\documentclass[a4paper,9pt]{article}
\usepackage[USenglish]{babel}
\usepackage[utf8]{inputenc}

\usepackage[left=3cm, right=3cm, top=3cm, bottom=3cm]{geometry}

\usepackage{authblk}

\usepackage[round]{natbib}
\usepackage{amsmath,amssymb,bm}

\usepackage{graphicx}
\usepackage{caption}
\usepackage{subcaption}
\usepackage{xcolor}
\usepackage{tikz}
\usetikzlibrary{shapes.misc,calc,angles}
\usetikzlibrary{shapes}

\usepackage{appendix}
\usepackage{comment}

\usepackage[normalem]{ulem}

\date{\today}
\title{On the spin-orbit problem for highly elliptical orbits and recursive excitation}

\author[1]{Erica Scantamburlo\thanks{erica.scantamburlo@polito.it}}
\author[2]{Davide Guzzetti\thanks{guzzetti@auburn.edu}}
\author[3]{Marcello Romano\thanks{marcello.romano@tum.de}}
\affil[1]{Department of Mechanical and Aerospace Engineering, Politecnico di Torino, Corso Duca degli Abruzzi, 24, 10129 Torino, Italy}
\affil[2]{Department of Aerospace Engineering, Auburn University, Auburn, Alabama}
\affil[3]{Department of Aerospace and Geodesy, Technical University of Munich, Germany}

\begin{document}

\maketitle
\begin{abstract}

Examining the spin-orbit coupling effects for highly elliptical orbits is relevant to the mission design and operation of cislunar space assets, such as the Lunar Gateway.  
In high-eccentricity orbits, the gravity-gradient moment is here modelled as an instantaneous excitation at each periapsis passage. By approximating the gravity-gradient moment through Dirac pulses, we derive a recursive discrete map describing the rotational state of the satellite at the periapsis passage. Thanks to the recursive map, we are able to find the initial attitude corresponding to an unbounded growth of angular velocity, and to identify initial conditions whose evolution is such that the pulses have the same sign (in-phase condition) or the alternate sign (counterphase condition) at successive periapsis passages. In the recursive map, we perform the numerical analysis up to ten periapsis passages. In order to justify the introduction of the discrete map, we compare the results of the discrete map with those found in the spin-orbit problem. Because of numerical errors due to the high eccentricity, we restrict the investigation up to three periapsis passages in the spin-orbit problem. Moreover, we apply the Fast Lyapunov Indicators method to draw the phase portrait and detect the initial conditions fulfilling the counterphase condition. 

\end{abstract}

\section{Introduction}

The Spin-Orbit Problem (SOP) concerns the dynamics of a rigid-body satellite performing a Keplerian motion around a pointlike celestial body and with the assumption that the satellite rotates solely around the axis perpendicular to the orbital plane \citep{danby_book,goldreich1966,wisdom1987,celletti1990_I,celletti1990_II}. This problem has been widely investigated in order to analyze the coupling effects between rotational and orbital dynamics in planetary systems, in particular in connection to the rotation-revolution resonances of the Moon and Mercury \citep{celletti1992,celletti2007}, and the Saturnian satellites Hypernion and Enceladus \citep{wisdom1984,wisdom2004}. Specifically, \citet{celletti2000}, and \citet{gkolias2016,gkolias2019} analytically investigate the primary and secondary spin-orbit resonances via the application of the Hamiltonian perturbation theory and normal forms for the definition of stability criteria for resonant spin rates, as well as for a mathematical justification of the reasons why some spin-orbit resonances in the Solar System are more frequent than others.
In \citet{shevchenko1999}, a two-dimensional area-preserving discrete map able to describe the motion in a vicinity of the $1$:$1$ resonance is constructed.
Previous works are largely focused on natural satellites or planets' rotational dynamics. Hence, the eccentricity values of interest are `relatively' small, (namely, the maximum value of the eccentricity is that characterizing Mercury, i.e., $\sim 0.2$). 
In \citet{misquero2020}, the capture into the primary $1$:$1$ resonance in the SOP problem, is analyzed, without constraints on the eccentricity,  but no attention is dedicated to the effects of high eccentricity on the attitude dynamics evolution for general initial conditions.  

The analysis of the spin-orbit dynamics for highly elliptical orbits was also conducted by investigating a second-order differential equation, called the Beletsky equation \citep{beletskii1965}. In particular, in \citet{zlatoustov1973} the stability of odd $2\pi$-periodic oscillations is examined for all the possible values of the inertia ratio and orbit eccentricity, and several papers are devoted to the examination Beletsky equation when the eccentricity converges to $1$ \citep[see for example][]{bruno1997,bruno2004}. In \citet{shevchenko1999}, the construction of a recursive map, combining the separatrix algorithmic map (SAM) for asymmetric perturbation and the regular projection algorithm (RPA), reproduce the motion of the spin-orbit problem if the terms beyond the first-order approximation are negligible. In particular, the SAM-RPA provides a good agreement for the spin-orbit motion of Phobos, Deimos, Amalthea, Janus, Epimetheus, Pandora and Prometheus. We notice that the greatest elliptical orbit considered for the aforementioned satellites is that of Phobos, whose orbital eccentricity is equal to $0.015$.

The investigation of the SOP with high eccentricity is relevant for several space mission scenarios, for example, for artificial satellites in Molnija orbits and for comets in highly elliptic orbits (such as Halley's comet), or Near-Rectilinar Halo Orbits (NRHOs) originating at $L_1$ and $L_2$ of the Earth--Moon system. 
In the last decade a growing interest has been dedicated to the investigation of the NRHOs originating at the $L_2$ point of the Earth--Moon circular restricted three-body problem (CR3BP). For further details on the numerical computation of the NRHOs, the reader is referred to \cite{howell1984}. The renewed interest on the NRHOs lie on their usage as reference orbit for the `Lunar Gateway', a space mission of the NASA Artemis program whose objectives are the exploration of the Moon's surface and of the deep space \citep[for example, see][]{guzzettiAAS2017,lee2018,lee2019,zimovan23}.
In connection to the attitude dynamics, the coupling effects between the orbital and rotational motion for Halo orbits in the vicinity of the Earth--Moon $L_1$ point is investigated in \citet{knutsonIAC12}, and then the calculation of librational attitude solutions as a function of the reference orbit, initial rotational configuration, and spacecraft inertia is provided in \citet{guzzetti_phdthesis}, and \citet{guzzetti2017} through the application of Floquet theory, multiple shooting and continuation algorithms. 

In \citet{colagrossi2017}, and \citet{guzzetti2018} the investigation of the attitude dynamics within the CR3BP of the NRHOs, originating at $L_1$ and $L_2$, reveals a sharp change of the angular velocity occurring at the periselene passage. It is shown that the gravity-gradient moment acts as an impulsive disturbance, and the effect of multiple close approaches on the spacecraft attitude dynamics is to vary the average angular rate. 
This phenomenon was explained by suggesting that this rotational evolution is present when a large reference orbit is considered (i.e., orbits that are not defined within the linearization approximation with respect to a Lagrangian point).
The effect of one periapsis passage on the rotational angular momentum of a body orbiting a central body is analyzed in \citet{scheeres1999}, \citet{scheeres2000_b}, \citet{scheeres2000_c}, and \citet{scheeres2001}, via the Lagrange planetary equations in connection to the change of orbit energy and angular momentum, but no attention is dedicated to the effect of multiple periapsis passages on the rotational dynamics in Highly Elliptical Orbits (HEOs). The pioneristic investigation of the attitude stability of a spacecraft located at the Lagrangian points is performed by \citet{robinson1974}. Then, the attitude dynamics close to the Lagrangian points for rigid satellites are investigated by \citet{brucker2007}, \citet{knutson2015} and \citet{guzzetti2017,guzzetti2018}.

In the present paper, we prove that the sharp change of the angular velocities at periapsis is not a peculiarity of the NRHOs, or of large reference orbits, but it is due to the high value of the osculating eccentricity associated to the spacecraft orbital motion. As a matter of fact, the NRHOs originating at the $L_2$ point of the Earth--Moon system are characterized by an osculating eccentricity with respect to the Moon, that ranges from $0.8$ to $0.95$. For this reason we investigate here the attitude dynamics within the Kepler problem and HEOs.

The high orbital eccentricity value complicates the application of analytical tools, such as those provided by the Hamiltonian perturbation theory. This is because for high values of eccentricity, the center equation, i.e., the equation expressing the true anomaly as a function of the mean anomaly through a power series of the eccentricity, does not always converge \citep[the reader is referred to ][pag. 46--47]{plummer1918}. 

In this investigation, the gravity-gradient moment is modeled as an impulsive moment acting instantaneously at the periapsis passage. 

After each periapsis passage, the satellite's angular velocity is affected by a sharp change, and its evolution can be described by a piecewise constant function. After suitable approximation, we derive the discrete map (DM) describing the evolution of the satellite's orientation and angular velocity. Even if the DM simplifies the SOP system, its investigation is useful for identifying initial conditions of interest. For example, we can identify initial conditions whereby the satellite rotates faster and faster. On the other hand, we can also define the initial conditions for which the evolution of the angular velocity remains bounded. Since the DM represents a simplified SOP model for HEO, we compare the results obtained in the DM and the SOP to provide a justification for its introduction. Moreover, in order to provide a graphical representation of the rotational initial conditions and the chaotic zones, we compute the Fast Lyapunov Indicator (FLI) as a numerical tool \citep[for further details, the reader is referred to][]{froeschle1997_a,froeschle1997_b,guzzo2023}, allowing us to distinguish between the regions of the initial conditions in which the angular velocity continues to grow or not, and those affected by exponential separations in the solution.

The paper is organized as follows: in Sec. \ref{sec:sop}, we describe the SOP and introduce its equations of motion. In Sec. \ref{ssec:sop_at_higheccentricity}, we justify the impulse approximation for high eccentricity and develop a recursive analysis of the effects of the gravity-gradient moment at each periapsis. Thanks to this approximation, we construct a recursive map able to identify initial conditions such that all the pulses are in-phase (i.e., have the same sign) or in counterphase (i.e., the sign of the pulse alternates at each periapsis passage). In Sec. \ref{ssec:comparison_DM_SOP}, we compare the results obtained in the discrete map and the SOP with high eccentricity. In Sec. \ref{ssec:SOP_FLI}, we apply the Fast Lyapunov Indicators (FLI) method to investigate the phase portrait of the SOP.

\section{The spin-orbit problem for elliptic orbits of the spacecraft}
\label{sec:sop}
\begin{figure}[htbp!]
\centering
\begin{tikzpicture}[scale=2]
    \draw[->] (-3.5,0) -- (1.8,0);
    \draw[->] (0,-2) -- (0,2.);   

    \draw[rotate around={0:(0,0)}] (-0.872,0) ellipse (2 and 1.8);
    \draw[fill,black] (0,0.) circle (0.25);
    \draw[fill,gray, rotate around = {60:(1,0.6)}] (0.8,0.5) rectangle (1.2,.7);
    
    \node at (-0.32,-0.2) {\normalsize{$P$}};
    \node at (1,0.95) {\normalsize{$S$}};

    \draw[black] (0,0) -- (1,0.6);
    \draw[black] (0.65,0) -- (1,0.6);

    \draw[black,domain=0:60] plot ({0.65+0.3*cos(\x)},{0.3*sin(\x)} );
    \draw[black,domain=0:30] plot ({0.4*cos(\x)}, {0.4*sin(\x)} );

    \node at (0.46,0.12) {\normalsize{\textcolor{black}{$f$}}};
    \node at (0.82,0.09) {\normalsize{\textcolor{black}{$\alpha$}}};
    \draw[fill,brown] (1,0.6) circle (0.005);

    \draw[->,brown] (1,0.6) -- ({1+0.5*cos(60)},{0.6+0.5*sin(60)});
    \draw[->,brown] (1,0.6) -- ({1+0.5*cos(150)},{0.6+0.5*sin(150)});
    \draw[fill,brown] (1,0.6) circle (0.025);

    \node at (1.33,0.95) {\normalsize{\textcolor{brown}{$i_1$}}};
    \node at (0.55,0.78) {\normalsize{\textcolor{brown}{$i_2$}}}; 
    \node at (1.1,0.55) {\normalsize{\textcolor{brown}{$i_3$}}}; 

    \node at (1.7,-0.15) {$\vec{e}$};
    \draw[fill,black] (1.128,0) circle (0.025);

    \draw[-,thin] (1.15,-0.05) -- (1.4,-0.45);
    \node at (1.5,-0.5) {\normalsize{periapsis}};
\end{tikzpicture}
\caption{Illustration of the SOP.}
\label{fig:sop_representation}
\end{figure}
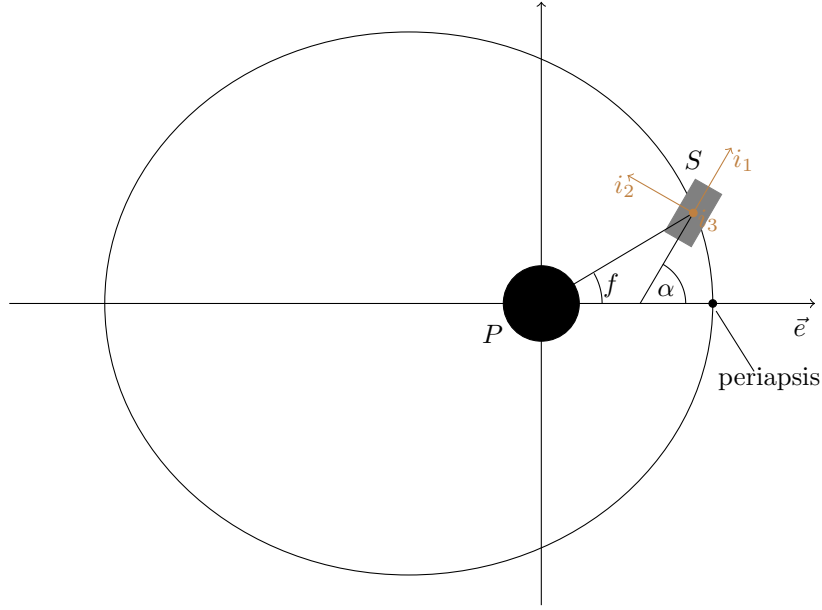

This section briefly introduces the well-known spin-orbit problem (SOP) for a satellite orbiting a primary body, and its governing equations. Let us assume that the center of mass of a rigid-body satellite $S$ is orbiting around a primary body $P$, having gravitational parameter $\mu$ (equal to the product of its mass and Newton's universal gravitation constant), under the modeling hypotheses of `Kepler restricted two-body problem'. Let us call $a$ the orbit semimajor axis, $e$ its eccentricity, $\omega$ the argument of periapsis, $f$ the true anomaly, and $\rho$ the radius-vector magnitude (i.e., the distance between the center of the primary and the center of mass of the satellite). See also Fig. \ref{fig:sop_representation}.

Let $i_1$, $i_2$, and $i_3$ be the principal axes of inertia of the satellite, with $i_1$ being the axis of minimum inertia and $i_3$ the axis of maximum inertia, and let $I_1$, $I_2$, $I_3$ be the three corresponding principal moments of inertia, satisfying the following inequality constraint:
\begin{equation} \label{I1lessthanI2lessthanI3}
I_1<I_2 \leq I_3.
\end{equation}

Let us assume that the satellite, in addition to orbiting the primary, is spinning about its third principal axis of inertia ($i_3$), which stays at all time normal to the orbital plane (i.e., the satellite is in a `flat-spin' about the axis of maximum inertia, which would be stable even for a `quasi-rigid-body' satellite).

Let us call $\alpha$ the attitude angle, between the orbital eccentricity vector ($\vec{e}$) and the satellite's first principal axis ($i_1$), and $\dot{\alpha}$ (where the dot symbol is used to indicate derivative with respect to time) the satellite's spin angular-rate, i.e., the magnitude of its absolute angular velocity, which by assumption is a vector always perpendicular to the orbital plane.  

By following common practice, for the sake of maximizing generality, the problem is normalized by scaling distances and time in what are here called `reference orbit’ (ro) units, as follows.
Distance (or, length) quantities are scaled by dividing them by the orbit semimajor axis $a$; i.e., the following unit of measurement is used for distance quantities,
\begin{equation}\label{defofDUro}
1 \, DU \triangleq  a.
\end{equation}
Periods of time are scaled by dividing them by the inverse of the orbit mean motion, i.e., the following unit of measurement is used for time-period quantities:
\begin{equation}\label{defofTUro}
1 \, TU \triangleq  \frac{1}{n}=\sqrt{\frac{a^3}{\mu}}.
\end{equation}
The value of the numerical conversion factors needed to convert from distances and periods expressed in meter and second, by following the international system of measurement SI, to the new `reference orbit' units, can be easily expressed as follows
\begin{align}
\label{meterinrounits}
1 \, DU \triangleq 1 \,a= N_{\text{SI}}\!\left(a\right)\,\text{m} &\rightarrow  1\, \text{m}= \left(\frac{1}{N_{\text{SI}}\!\left(a\right)}\right)\,DU,\\
1 \, TU   \triangleq \sqrt{\frac{a^3}{\mu}} = \sqrt{ \frac{\left(N_{\text{SI}}\!\left(a\right)\right)^3}{N_{\text{SI}}\!\left(\mu\right) \,}}\,s&\rightarrow 1\, \text{s}= \left(\sqrt{ \frac{N_{\text{SI}}\!\left(\mu\right)}{\left(N_{\text{SI}}\!\left(a\right)\right)^3}}\right)\,TU, \label{seinrounits}
\end{align}
where $N_{\text{SI}}$ indicate the numerical value in SI units of the variable within the parentheses,  
and, finally, the numerical value of the gravitational constant in reference orbit units yields to be conveniently equal to the scalar unit value, i.e., 
\begin{equation}\label{muinro}
\mu=1\,\frac{DU^3}{TU^2},
\end{equation}
which, by using Eqs.~\eqref{defofDUro} and~\eqref{defofTUro} yields
\begin{equation}\label{muinro2}
\frac{\mu}{n^2\,a^3}=1.
\end{equation}

Therefore, $N_{ro} (n) = 1, N_{ro} (\mu) = 1$.
Furthermore, the orbital period is 
\begin{equation}
T =  2\pi \,\, TU.
\end{equation}

The governing equations of motion of the above introduced SOP are, in dimensional form, 
\begin{equation}
\begin{split}
\dot{f} & = \frac{n(1+e \cos f)^2}{\sqrt{(1-e^2)^3}}  \cr 
\ddot{\alpha} &= A(f) \,\sin (2 ( f -\alpha + \omega)),
\end{split}
\label{eq:spinorbit_problem}
\end{equation}
where it has been defined 
\begin{equation}
    A(f) \triangleq \frac{3}{2} \frac{\kappa\,\mu}{\rho^3}, \qquad \kappa \triangleq \frac{I_3 - I_1}{I_3},
    \label{eq:A_SOP}
\end{equation}
and 
\begin{equation}
    \rho= \frac{a\,(1-e^2)}{1+e \cos f}.
\end{equation}
Because of the hypothesis in Eq.~\eqref{I1lessthanI2lessthanI3}, it is $0<\kappa<1$ \citep[see][]{hughes_book}.

Notice that the first of Eqs.~\eqref{eq:spinorbit_problem} governs the evolution in time of the true anomaly of the orbiting satellite, according to Kepler restricted two-body problem model, while the second equation governs the spin of the satellite about its axis of maximum inertia, coincident at any time with the normal to the orbital plane. In particular the second equation is the only non-trivial\footnote{The absolute angular velocity components along the first and second axis of inertia are at any time zero, by assumption.} Euler rotational equation for the problem at hand, governing the spinning motion under the effect of the gravity-gradient external moment with respect to the center of mass of the satellite \citep[for the expression of that moment see for instance][]{hughes_book}.    

Since the argument of periapsis $\omega$ in Eqs.~\eqref{eq:spinorbit_problem} introduces only an initial phase displacement, without loosing generality, we consider it to be zero, from now on.

Now, by applying the \emph{ro} units normalization introduced before, Eqs.~\eqref{eq:spinorbit_problem} become
\begin{equation}
\begin{split}
f' & = \frac{(1+e \cos f)^2}{\sqrt{(1-e^2)^3}}  \cr 
\alpha''  &= \overline{A}(f) \,\sin (2 (f-\alpha)),
\end{split}
\label{eq:spinorbit_problem_NORMZED}
\end{equation}
where it has been defined
\begin{equation}
    \overline{A}(f) \triangleq \frac{3}{2} \frac{\kappa}{\overline{\rho}^3}, \qquad \overline{\rho}= \frac{(1-e^2)}{1+e \cos f},
    \label{eq:A_SOP2}
\end{equation} 
and the prime symbol indicates the derivative with respect to the normalized time $\sigma=n\,t$. Moreover, it has been taken into account that
\begin{equation}
f'=n\,\dot{f}, \quad \alpha''=n^2\,\ddot{\alpha}.
\end{equation}

The second of Eq.~\eqref{eq:spinorbit_problem_NORMZED} can also be written as 
\begin{equation}
\alpha '' = \tau,
\label{eq:eqstau}
\end{equation}
where $\tau$ is the specific (i.e., per unit of $I_3$) gravity-gradient moment, 
\begin{equation}
    \tau \equiv \overline{A}(f) \sin (2 (f-\alpha)) .
    \label{eq:taucompleta}
\end{equation}

\subsection{Spin orbit problem for highly elliptical orbits: impulsive approximation}
\label{ssec:sop_at_higheccentricity}

From now on we add the hypothesis of HEOs (more specifically, we assume that $e \geq 0.8$). Under this hypothesis, in this and next two sections we introduce analytical and numerical results, that are original to the best knowledge of the authors. In particular, in this section and the next one we develop an approximate analysis of the effect of the gravity-gradient moment, modeling it as an impulsive force acting at the periapsis; furthermore, in Sec. \ref{ssec:SOP_FLI} we consider an application of the chaos indicator methods via the Fast Lyapunov Indicators (FLI).

In particular, in this Section we will prove three significant results:

\vspace{2mm}
\emph{Result 1.} For HEOs, the gravity-gradient moment can be modelled through a Dirac-delta moment acting at the periapsis.
\vspace{1mm}

\emph{Proof of Result 1.}
We note that $\overline{A}(f)$ describes the effects of the gravity-gradient moment and depends on the eccentricity $e$, and satellite moments of inertia ratio $\kappa$. 
Moreover, for $e \neq 0$ the minimum and maximum value of $\overline{A}(f)$ are

\begin{equation}
    \overline{A}_{m} \equiv \overline{A}(\pi) = \frac{3}{2} \frac{\kappa}{ (1+e)^3}, \quad
    \overline{A}_{M} \equiv \overline{A}(0) = \frac{3}{2} \frac{\kappa }{ (1-e)^3}.
\end{equation}
We note that $\overline{A}_M/\overline{A}_m = \left(\frac{1+e}{1-e} \right)^3$ which tend to infinity as $e \rightarrow 1$. The semi-maximum amplitude is reached for $f^{\ast}$ such that
\begin{equation}
\overline{A}(f^{\ast}) \equiv \overline{A}_m + \frac{\overline{A}_M - \overline{A}_m}{2},
\end{equation}
i.e., when 
\begin{equation}
2(1+e\cos f^{\ast})^3 - (1-e)^3 - (1+e)^3 = 0.
\end{equation}
It is immediate to prove that the previous equation has the following solutions
\begin{equation}
    f^{\ast} _{1,2} = \pm\arccos\left(\frac{(1+3e^2)^{1/3}-1}{e}\right).
    \label{eq:fast_ecc}
\end{equation}
We emphasize that $f^{\ast} _1$ and $f^{\ast} _2$ are symmetric with respect to the periapsis. Moreover, for $e=1$ it yields
\begin{equation}
    f^{\ast} _{1,2} \big|_{e=1} = \pm \arccos \left(4^{1/3}-1 \right) = \pm \left(\frac{\pi}{2} - \arcsin \left(4^{1/3} -1 \right) \right) \approx \pm 0.943 \, \text{rad},
\end{equation}
where we used the identity $\arcsin (x) = \pi/2 - \arccos(x)$, for $x \in [-1,1]$. 

Moreover, we notice that $f^{\ast} _{1,2}$ defined in Eq. \eqref{eq:fast_ecc} is a monotonic function of $e$, hence
$f^{\ast} _{1,2} |_{e=1}$ represent the lower and upper bound of $f^{\ast} _{1,2}$, for all $e$.
 
From the first equation of \eqref{eq:spinorbit_problem}, the time interval between the semi-maximum amplitudes is computed as

\begin{equation}
    \Delta \sigma^{\ast} = (1-e)^{3/2} \int _{f^{\ast}_1} ^{f^{\ast}_2} \frac{df}{(1+e \cos f)^2}.
    \label{eq:Deltasigmastar_def}
\end{equation}

It is immediate to prove that \citep[see formulas 2.554 and 2.553 in][]{gradshteynbook} 
\begin{equation}
    \int  \frac{df}{(1+e \cos f)^2} 
    = -\frac{e \sin f}{(1-e^2) (e \cos f + 1)} + \frac{{4}}{(1-e^2)^{3/2}} \arctan \left(  \sqrt{\frac{1-e}{1 + e}} \tan(f/2) \right) .
    \label{eq:indefinite_int}
\end{equation}

We recall that the period of the orbit is $T = 2\pi$, and by inserting Eq. \eqref{eq:fast_ecc} into \eqref{eq:Deltasigmastar_def}, and taking into account Eq. \eqref{eq:indefinite_int}, the portion of orbital time in which $\overline{A}(f)$ is larger than its semi-maximum, yields to be
\begin{equation}
    \begin{split}
    \frac{\Delta {\sigma^{\ast}}  }{T} & = -\frac{1}{\pi}\left\{  \frac{\sqrt{1-e^2} \sqrt{e^2 + 2 \eta - \eta^2 -1}}{\eta} \right. \cr
    & \left. \qquad \qquad - 2 \arctan  \left[\sqrt{\frac{1-e}{1+e}} \tan \left( \frac{\arccos((\eta-1)/e)}{2} \right)   \right] \right\}, \qquad \eta \equiv (3e^2 +1)^{1/3}.
    \end{split}
    \label{eq:delta_sigmaast}
\end{equation}
 
In Table \ref{tb:rate_Deltat} we show the values of $\overline{A}_{M}/\overline{A}_{m}$ and  $\Delta{\sigma^{\ast}}/T$ for different values of the orbital eccentricity.

\begin{table}[htbp!]
\centering
\begin{tabular}{|c|c|c|}
\hline
\rule{0pt}{3ex}
\rule[-1.2ex]{0pt}{3ex}
$\bm{e}$ & $\overline{\bm{A}}_{\bm{M}}/\overline{\bm{A}}_{\bm{m}}$ & $\bm{\Delta}  \bm{\sigma}^{\ast}/\bm{T}$ \cr
\hline
$0.2$ & $3$ & $0.31$ \cr
$0.4$ & $13$ & $0.17$ \cr 
$0.6$ & $64$ & $0.08$ \cr
$0.8$ & $729$ & $0.03$ \cr 
$0.9$ & $6859$ & $0.0083$ \cr
$0.95$ & $59319$ & $0.0029$ \cr
\hline
\end{tabular}
\caption{Values of the ratio between maximum and minimum amplitude $(\overline{A}_{M}/\overline{A}_{m})$, and the portion of orbital time in which $\overline{A}(f)$ is larger than its semi-maximum ($\Delta \sigma^{\ast}/T$), for different values of the orbital eccentricity. }
\label{tb:rate_Deltat}
\end{table}
\begin{figure}[htbp!]
\centering
\includegraphics[scale=0.6]{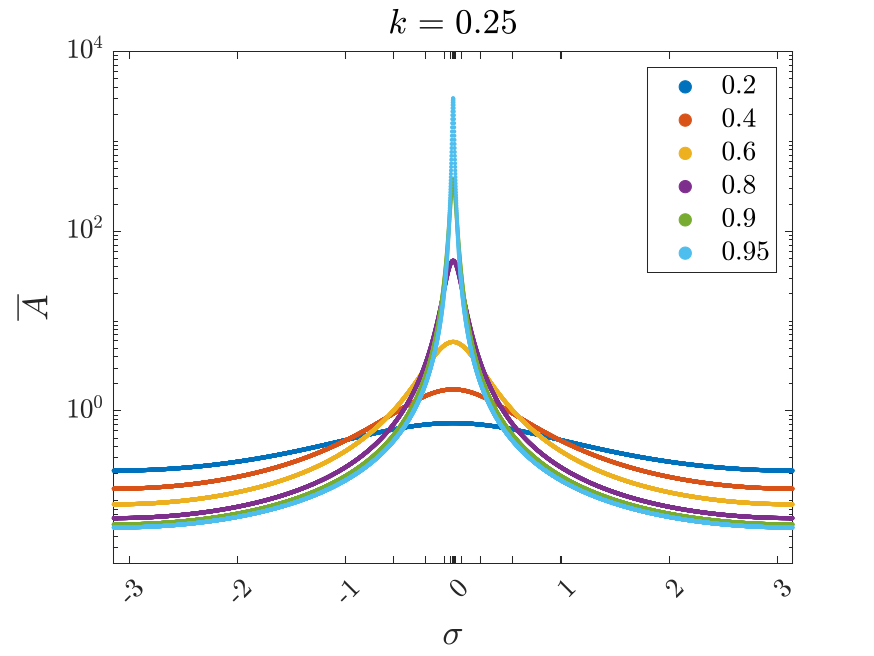}
\includegraphics[scale=0.6]{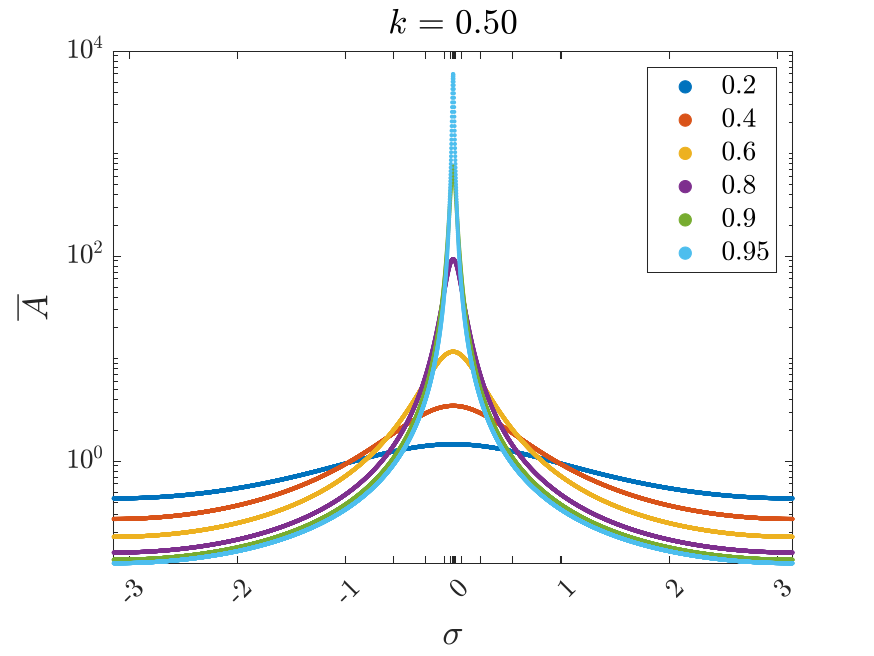}
\includegraphics[scale=0.6]{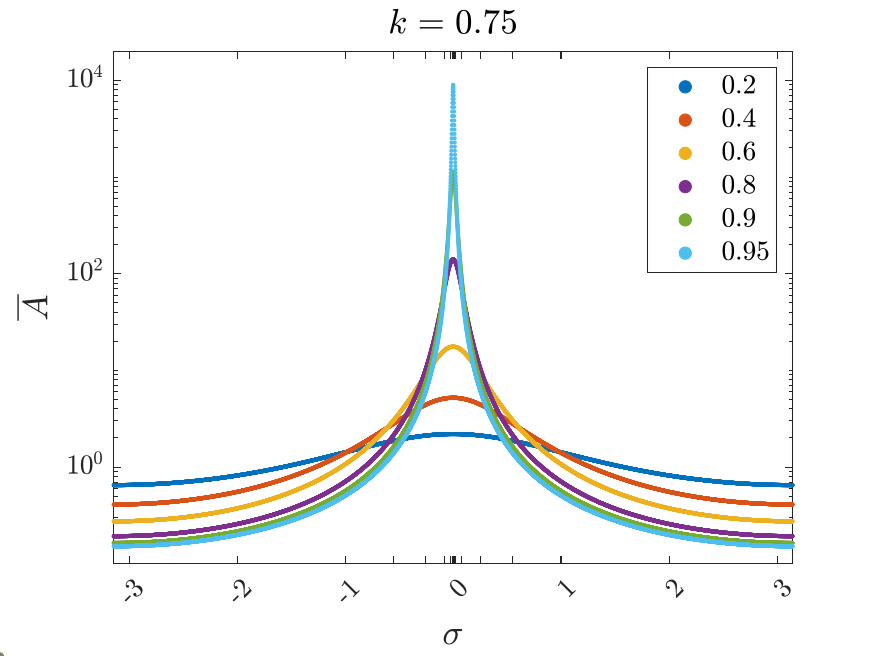}
\caption{Representation of $\overline{A}(\sigma)$ for different orbital eccentricities and for one orbit (at $\sigma=0$ the satellite is at the periapsis). The values of the inertia ratio used are $\kappa = 0.25$ (top panel), $\kappa = 0.5$ (central panel), and $\kappa = 0.75$ (bottom panel).  }
\label{fig:diffe_t_A}
\end{figure}

Having the goal of performing an approximate analysis of the coupling effects of spin-orbit dynamics, by considering Table \ref{tb:rate_Deltat} and Fig. \ref{fig:diffe_t_A},  we notice that for HEOs ($e \geq 0.8$), $\overline{A}(f)$ can be approximated by a pulse. For the mathematical description, we decided to model the gravity-gradient through  Dirac deltas, whose maximum amplitude impulse is given by the parameter $K$: i.e., having a zero value everywhere, except than at the periapsis, where the true anomaly and the adimensional time have values
\begin{equation}
f_p=f_r \triangleq \pm 2 r \pi, \qquad \sigma_p=\sigma_r \triangleq \pm 2 r \pi \equiv f_r,
\end{equation}
where the $p$ at the subindex indicates `at periapsis', and $r \in \mathbb{N}_0$ (a natural number, including $0$) is used as an index enumerating each orbit; it has been taken into account that the orbital period in adimensional time $\sigma$ is equal to $2 \pi$, and therefore it yields $\sigma_r\equiv f_r$.

Finally, the gravity-gradient moment for $R$ periapsis passages is  
\begin{equation}
    \tau =  -K \sum _{r=1} ^R \sin (2 \alpha (\sigma_r)) \delta (\sigma - \sigma_r),
    \label{eq:tau}
\end{equation}
where 
\begin{equation}
K \equiv \overline{A}_M \Delta \sigma ^{\ast},
\label{eq:definition_K}
\end{equation} 
and $\delta$ indicates the Dirac delta, such that
\begin{equation}
    \delta (\sigma-\sigma_r) = \begin{cases}
        \infty \quad \text{when } \sigma = \sigma_r  \cr
        0 \quad \text{otherwise}
    \end{cases},
\end{equation}
and it has been taken into account that, because of the sine periodicity, it is 
\begin{equation}
\sin(2(\pm2 r \pi-\alpha (\sigma_r))) \equiv - \sin(2 \alpha (\sigma_r)).
\end{equation}

\vspace{2mm}
\emph{Result 2.} With the Dirac delta approximation, a recursive analytical discrete map is computed, providing the spacecraft angular velocity, and orientation at the periapsis passage. 
\vspace{1mm}

\emph{Proof of Result 2.}
For HEOs, the rotational dynamics in Eq.~\eqref{eq:eqstau}, by inserting Eq. \eqref{eq:tau}, is approximated by the differential equation 

\begin{equation}
\alpha''(\sigma) = - K \, \sum _{r=0} ^R \sin \left(2 \alpha (\sigma_r)\right) \,\delta (\sigma-\sigma_r).
\label{eq:approx_spinorbit}
\end{equation}

Now, by assuming the following initial conditions at epoch $\sigma_0 \equiv \sigma= 0^{-}$:
\begin{equation} \label{SpinOrbitInitCond}
f_0 \equiv f (\sigma_0) =0, \quad  \alpha_0 \equiv \alpha (\sigma_0) ,\quad \alpha_0 ' \equiv \alpha'( \sigma_0 ) , 
\end{equation}
where $\alpha_0, \alpha_0 ' \in \mathbb{R}$ are considered \emph{known constants}, let us integrate forward, in the adimensional time, the differential equation \eqref{eq:approx_spinorbit}, via Laplace transform\footnote{We denote the Laplace trasform of a function $F (\sigma)$ by $\mathcal{L}\{F\}(s)$. We will omit for simplicity the independent variable $s$ in $\mathcal{L}\{F\}(s)$.}. Then, by recalling that
\begin{equation}
    \mathcal{L} \{\alpha '' \} = s^2 \mathcal{L} \{\alpha\} -s \alpha_0 - \alpha_0 ' ,
\end{equation}
and that
\begin{equation}
    \mathcal{L} \left\{ \alpha'' \right\} = -K \mathcal{L} \left\{ \sum_{r=1} ^R \sin (2\alpha (\sigma_r) ) \delta (\sigma - \sigma_r) \right\} = -K \sum _{r=1} ^R \sin (2 \alpha (\sigma_r)) e^{-\sigma_r s},
\end{equation}
it yields
\begin{equation}
     \mathcal{L}\{ \alpha \} = \frac{\alpha_0}{s} + \frac{\alpha_0 '}{s^2} - K \sum_{r=1} ^R \sin (2 \alpha (\sigma_r)) \frac{e^{-\sigma_r s}}{s^2},
     \label{eq:eqtoantitrasf}
\end{equation}

Finally, by antitrasforming Eq. \eqref{eq:eqtoantitrasf}, the solution of the differential equation \eqref{eq:approx_spinorbit} yields
\begin{equation}
    \quad 0 \leq \sigma \leq \sigma_R\qquad \rightarrow \qquad  \left\{\begin{split}
        \alpha (\sigma) &= \alpha_0 +  \alpha '_0 \sigma - K \,\sum _{r=1} ^R \left[ \sin (2 \alpha (\sigma_r) )  (\sigma -\sigma_r) H (\sigma- \sigma_r) \right] \cr 
        \alpha' (\sigma) & = \alpha'_0 - K \sum _{r=1} ^R \sin (2 \alpha (\sigma_r) ) H (\sigma - \sigma_r) , \cr        
    \end{split}\right. 
    \label{eq:solution_approx}
\end{equation}
where $H$ denotes the Heaviside step function, i.e.,
\begin{equation}
    H(\sigma-\sigma_r ) \triangleq \begin{cases}
        0 \quad \sigma < \sigma_r \cr 
        1 \quad \sigma \geq \sigma_r
    \end{cases}.
\end{equation}
Since we are interested in the effect of the periapsis passages, we rewrite system \eqref{eq:solution_approx} as a recursive map here below. 

Let us define
\begin{equation}
    \alpha_r \equiv \alpha_r (\sigma_r), \qquad \alpha_r ' \equiv \alpha' (\sigma_r).
\end{equation}

At the first periapsis passage ($r=1$), occuring at $\sigma_1\equiv \sigma=0$, from Eq. \eqref{eq:solution_approx} it yields
\begin{equation}
    \begin{split}
        \alpha_1 & = \alpha_0 \\
        \alpha_1 ' & = \alpha_0 ' - K \sin (2 \alpha_1).
    \end{split}
    \label{eq:step1}
\end{equation}

At the second periapsis passage ($r=2$), occuring at $\sigma_2 \equiv \sigma = 2\pi$,  from Eq. \eqref{eq:solution_approx} it yields
\begin{equation}
    \begin{split}
        \alpha_2 & = \alpha_0 + \alpha_0 ' 2\pi -K \sin ( 2\alpha_1) (2\pi-2\pi) = \alpha_1 + \alpha_1 '  2\pi \\
        \alpha_2 ' & = \alpha_0 ' -K \left[ \sin (2 \alpha_1 ) 
         + \sin (2 \alpha_2)) \right] = \alpha_1 ' -K \sin ( 2\alpha_2).
    \end{split}
    \label{eq:step2}
\end{equation}

At the third periapsis passage ($r=3$), occuring at $\sigma_3 \equiv \sigma = 4 \pi$, from Eq. \eqref{eq:solution_approx} it yields
\begin{equation}
    \begin{split}
        \alpha_3 & = \alpha_0 + \alpha_0 ' 4 \pi - K \left[ \sin (2 \alpha_1) 4\pi + \sin ( 2\alpha_2) 2\pi\right] \\
        & = \alpha_0 + \left[ \alpha_0 ' -K \sin (2\alpha_1) \right]4\pi - K \sin ( 2\alpha_2) 2 \pi \\
        & = \alpha_1 + \alpha_1 ' 2\pi + \left[ \alpha_1 ' -K \sin (2\alpha_2) \right] 2\pi \\
        & = \alpha_2 + \alpha_2 ' 2\pi \\
        \alpha_3 ' & = \alpha_0 ' - K \left[ \sin (2\alpha_1) + \sin (2 \alpha_2) + \sin ( 2 \alpha_3) \right]  \\
        & = \alpha_2' - K \sin (2 \alpha_3).
    \end{split}
    \label{eq:step3}
\end{equation}

Therefore, the following recursive map can be written regarding the values of the adimensional spin angular-rate $\alpha'$ and spin angle $\alpha$ at a generic periapsis passage, indicated by the value of the subscript index $r$:
\begin{equation}\label{eq:recursivemapnew}
\begin{cases}
    \alpha_1 &= \alpha_0 \\
    \alpha_1 ' &= \alpha_0 ' - K \sin (2\alpha_1) \\
\end{cases}; \qquad 
\begin{cases}
\alpha_r &= \alpha_{r-1}+\alpha'_{r-1} \,2\,\pi  \\
\alpha'_r &= \alpha'_{r-1} - K \sin (2\alpha_{r}) \\
\end{cases}, \quad r \geq 2.
\end{equation}

\vspace{2mm}
\emph{Result 3.} With the Dirac delta approximation, it is possible to identify the initial state leading to an uncontrolled growth of the spacecraft angular velocity.

\vspace{1mm}

\emph{Proof of Result 3.}
Let us consider the following two questions. \emph{Question 1:} given the orbit eccentricities and the inertia ratio $\kappa$, is there a range of values of the initial conditions ($\alpha_0$, and $\alpha'_0$) such that all impulses are `in-phase' (i.e., each successive impulse has  the same sign of the previous one)? \emph{Question 2:} is there  a range of values of the initial conditions such that all impulses are `in counterphase' (i.e., each successive impulse has  the opposite sign with respect to the previous one)? We would like to answer those questions for a series of two, three, or $R$ periapsis passages. 

Notably, knowing any initial conditions fulfilling the in-phase behavior is important for astrodynamics applications, since they would identify the initial states that cause the satellite to rotate faster and faster. In particular, a sequence of such periapsis passages guarantees spin-up. This is because, the only sequence values of sines that, while remaining positive (or negative), quickly tend to zero, are those converging to the fixed points. Since the fixed points are linearly unstable for values of $K$ of interest in our application (i.e., $K>1/\pi$), then there are no initial data $\alpha_0$ and $\alpha_0 '$ converging to null values of the sine (see Appendix \ref{app:fixedpoints}).  
On the other hand, the initial conditions leading to a bounded growth of the rotation speed can be found in the set of initial conditions fulfilling the counterphase condition.

\vspace{5mm}
\noindent\textbf{Initial conditions for in-phase gravity-gradient moment impulses}

Now, we consider the initial states fulfilling the in-phase condition of the gravity-gradient moment impulse up to an arbitrary number $R$ of periapsis passages. This condition is fulfilled by values of $\alpha_0$ and $\alpha_0 '$ such that
\begin{equation}
    \text{sign}(\sin ( 2 \alpha_{R})) = \cdots = \text{sign}(\sin ( 2 \alpha_2)) = \text{sign}(\sin ( 2 \alpha_1)) .
    \label{eq:gen_incase}
\end{equation}
Since $\alpha_1 = \alpha_0$ (see the first Eq. of \eqref{eq:step1}), we stress that 
\begin{equation}
    \text{sign}(\sin (2 \alpha_1)) = \text{sign}(\sin (2 \alpha_0)).
\end{equation}

We stress that in this section we do not provide a formal proof by induction of the existence of initial data satisfying the in-phase condition up to an arbitrary number of periapsis. We proceed analytically for the first two passages, and then we perform a numerical computation.
Hence, two subcases can occur in the investigation of the in-phase condition: i) $\text{sign}(\sin (2 \alpha_0)) = 1$, and ii) $\text{sign}(\sin (2 \alpha_0)) = -1$.

We emphasize that, writing 
\begin{equation}
    \text{sign}(\sin (2\alpha_0)) = 1
\end{equation}
is equivalent to
\begin{equation}
    \text{mod}(2 \alpha_0, 2\pi) \in (0,\pi).
    \label{eq:pos_cond}
\end{equation}
Noticing that, for given $a,m\in \mathbb{R}$, it yields
\begin{equation}
    \text{mod}(2a,2m) = 2a - 2m \Bigl\lfloor \frac{2a}{2m} \Bigl\rfloor = 2 \left( a -m  \Bigl\lfloor \frac{a}{m} \Bigl\rfloor \right) = 2 \, \text{mod}(a,m),
\end{equation}
then Eq. \eqref{eq:pos_cond} is equivalent to
\begin{equation}
    \text{mod}(\alpha_0, \pi) \in (0,\pi/2) \rightarrow \alpha_0 \in \left(n_0 \pi, \frac{1+2n_0}{2} \pi \right), \quad n_0 \in \mathbb{Z}.
\end{equation}

On the other hand, the condition
\begin{equation}
    \text{sign}(\sin ( 2\alpha_0)) = -1
\end{equation}
is equivalent to 
\begin{equation}
    \text{mod}(\alpha_0,\pi) \in (\pi/2,\pi) \rightarrow \alpha_0 \in \left( \frac{1+2n_0}{2} \pi, (1+n_0) \pi \right), n_0 \in \mathbb{Z}.
\end{equation}

We start our investigation by considering analytic results on the initial states fulfilling the in-phase condition for the first two impulses with $\text{sign}(\sin (2\alpha_0)) = 1$. Later, we will consider the subcase with $\text{sign}(\sin (2\alpha_0)) = -1$. Finally, we will show numerical results for initial conditions satisfying the in-phase condition up to ten periapsis passages.
 
\vspace{1mm}
\emph{Analytic results for the first two impulses.} By analysing Eq.~\eqref{eq:step1} and \eqref{eq:step2}, it is straightforward to see that the gravity-gradient moment impulse at the second periapsis passage ($r=2$) has the same sign of the first gravity-gradient moment impulse ($r=1$) only if
\begin{equation}
\text{sign}(\sin(2 \alpha_2)) =\text{sign}(\sin(2\alpha_1)),
\end{equation}
where, from Eq. \eqref{eq:step2}, and \eqref{eq:step1}, it is
\begin{equation}
    \alpha_2 = \alpha_1 +\alpha_1 '  2\pi  =\alpha_0 + 2\pi \left[ \alpha_0 ' -K \sin (2\alpha_0) \right] .
    \label{eq:twoimp_inphasepos}
\end{equation}

Since the first of Eq. \eqref{eq:step1}, the previous relation can be written as
\begin{equation}
\text{sign}(\sin(2 \alpha_2)) =\text{sign}(\sin(2\alpha_0)).
\end{equation}

Let us consider the following \emph{first subcase}: the initial spin-angle $\alpha_0$ is such that 
\begin{equation}
    \text{sign}(\sin(2 \alpha_0)) = 1 \leftrightarrow \sin (2\alpha_0) > 0 \rightarrow \mod (2 \alpha_0, 2\pi) \in (0,\pi).
\end{equation}
Hence, the condition that needs to be satisfied by $\alpha_0$ in order for the two impulses $r=0$ and $r=1$ to be in-phase is
\begin{equation}
    \sin (2\alpha_0) >0 \rightarrow \text{mod} (\alpha_0,\pi) \in (0,\pi/2).
    \label{eq:alpha0_inphase}
\end{equation}

The condition that needs to be satisfied in order for the two impulses $r=1$ and $r=2$ to be in-phase, is
\begin{equation}
    \text{mod}(\alpha_2,\pi) \in (0,\pi/2) \rightarrow \text{mod} (\alpha_0 + 2\pi \left[ \alpha_0 ' -K \sin (2\alpha_0) \right],\pi) \in (0,\pi/2) .
\end{equation}
We emphasize that the previous relation is equivalent to 
\begin{equation}
    \alpha_0 + 2\pi \left[ \alpha_0 ' -K \sin (2\alpha_0) \right] \in \left( n_2 \pi, \frac{\pi}{2} + n_2 \pi\right), \quad \forall n_2 \in \mathbb{Z}.
\end{equation}

Hence, the condition that needs to be satisfied by the initial spin angle $\alpha_0$ and the initial spin-angle rate $\alpha'_0$, in order for the two impulses $r=1$ and $r=2$ to be in-phase is
\begin{equation}
    \begin{split}
    \alpha_0 \in &\left(n_0 \pi, (1+2 n_0)\frac{\pi}{2} \right) , \\
    \alpha_0 ' \in & \left(\frac{n_2}{2} - \frac{\alpha_0}{2\pi} + K \sin (2\alpha_0), \frac{1+2n_2}{4} - \frac{\alpha_0}{2\pi} + K \sin (2\alpha_0) \right)  ,  
    \end{split}
    \quad  \forall n_0,n_2 \in \mathbb{Z}.
    \label{eq:alpha_three_inphase} 
\end{equation}
We notice that for a given values of $\alpha_0$ satisfying the first of Eq. \eqref{eq:alpha_three_inphase}, the length of the interval
\begin{equation}
    \left(\frac{n_2}{2} - \frac{\alpha_0}{2\pi} + K \sin (2\alpha_0), \frac{1+2n_2}{4} - \frac{\alpha_0}{2\pi} + K \sin (2\alpha_0) \right) 
\end{equation}
is independent of $K$, and equal to $1/4$. Hence, for an arbitrary $K$ and $\alpha_0 \in \left(n_0 \pi, (1+2 n_0)\frac{\pi}{2} \right)$, it is always possible to find a spin-angle rate $\alpha_0 '$ value fulfilling Eq. \eqref{eq:alpha_three_inphase}.

\begin{figure}[h!]
\centering
\includegraphics[scale=0.6]{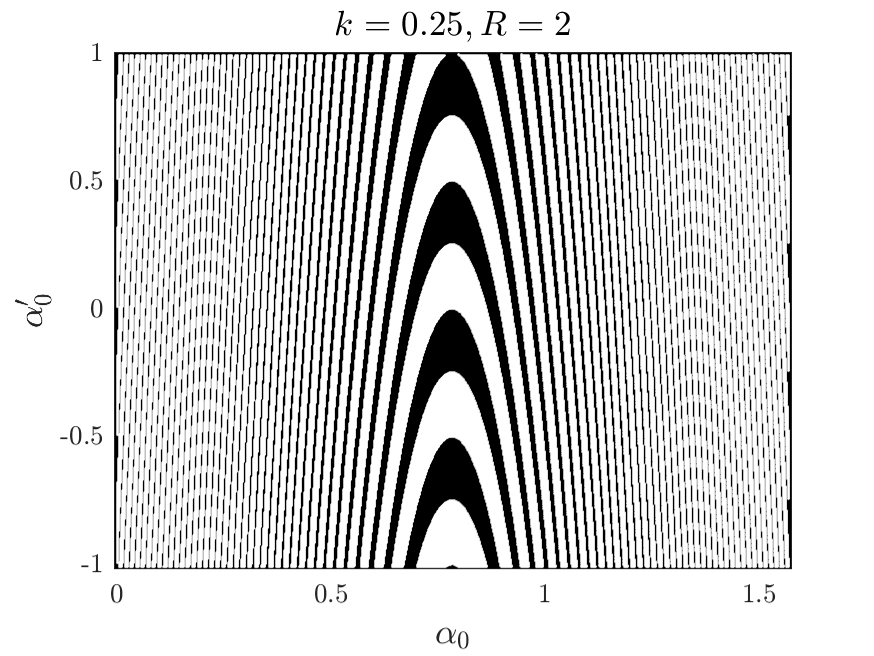}
\includegraphics[scale=0.6]{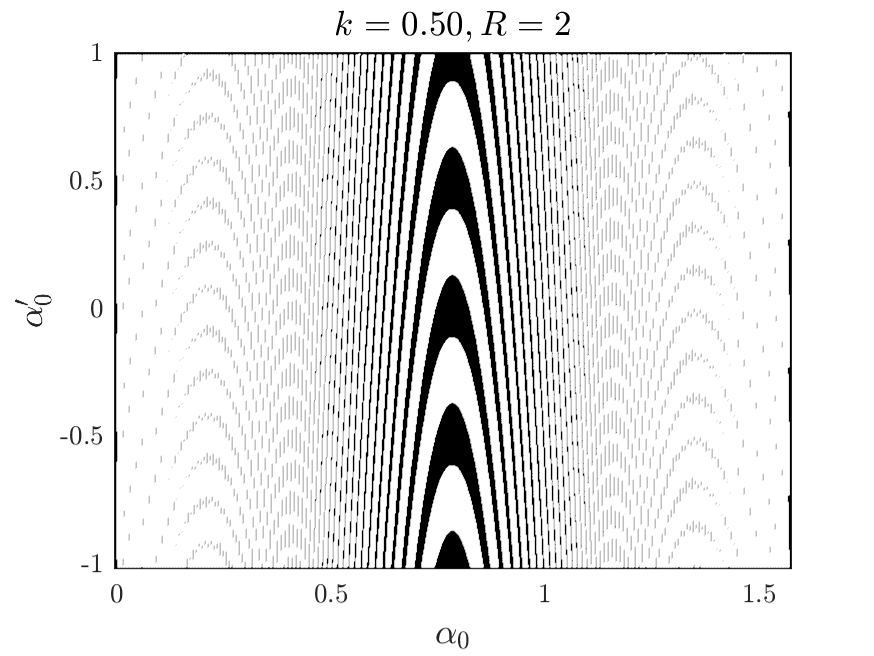}
\includegraphics[scale=0.6]{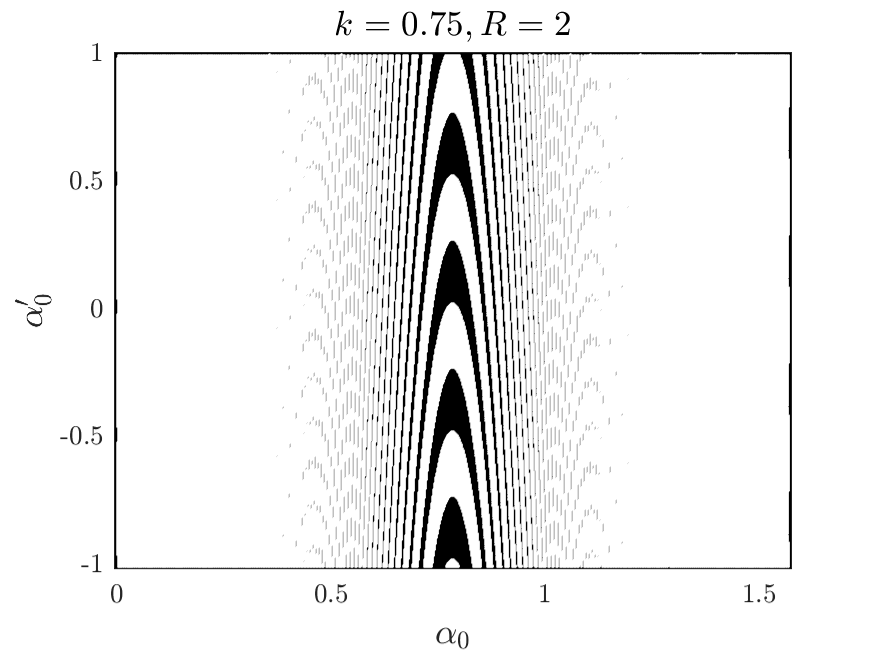}
\caption{Set of initial conditions ($\alpha_0$, $\alpha_0'$), represented in white color, for which the in-phase impulse condition (with $\sin(2\alpha_{0}) > 0$) is satisfied for two successive periapsis passages ($r=1,2$) for $e= 0.9$ and $\kappa = 0.25$ (top panel), $\kappa = 0.5$ (central panel), $\kappa = 0.75$ (bottom panel). }
\label{fig:SOP_DiscMap_R_2}
\end{figure}

Fig. \ref{fig:SOP_DiscMap_R_2} reports the initial conditions $\alpha_0, \alpha_0 '$ fulfilling the in-phase condition for the first two periapsis passages (represented with the white color) that we obtain via a numerical implementation of the DM. For the definition of $K = \frac{3}{2} \frac{\kappa}{(1-e)^3} \Delta \sigma$, we select $e=0.9$, and the three values of the inertia ratios $\kappa = 0.25,~0.5,~0.75$. Since the in-phase condition is periodic of period $\pi$ in $\alpha_{r}$, we restrict the numerical investigation for $\alpha_0 \in [0,\pi/2)$. As initial angular velocity, we consider a grid in the interval $\alpha_0 ' \in [-1,1]$. 
From the figure, we notice that for higher values of the inertia ratios $\kappa$, the regions of the initial conditions fulfilling the in-phase condition tend to expand. Moreover, for all of the three inertia-ratio values, the regions of the initial conditions not fulfilling the in-phase condition accumulate nearby $\alpha_0 = \pi/4$. We emphasize that this behavior is periodic in $\alpha_0'$. 

\vspace{1mm}
Let us now consider the following \emph{second subcase}: the initial spin-angle $\alpha_0$ is such that
\begin{equation}
    \text{sign}(2 \alpha_0) = -1 \leftrightarrow \sin ( 2\alpha_0) < 0 \rightarrow \text{mod}(\alpha_0,\pi) \in (\pi/2,\pi).
\end{equation}
By following an analogous procedure of the subcase $\text{sign}(\sin(2 \alpha_0)) = 1$, we obtain that the condition that needs to be satisfied in order for the two impulses $r=1$ and $r=2$ to be in-phase, is
\begin{equation}
    \text{mod}(\alpha_0 + 2\pi \left[ \alpha_0 ' -K \sin (2\alpha_0) \right],\pi) \in (\pi/2,\pi).
\end{equation}
We emphasize that the previous relation is equivalent to
\begin{equation}
    \alpha_0 + 2\pi \left[ \alpha_0 ' -K \sin (2\alpha_0) \right] \in \left( (1+ 2n_2) \frac{\pi}{2} , (1+n_2) \pi \right), \quad \forall n_1 \in \mathbb{Z}.
\end{equation}

Finally, the condition that needs to be satisfied by the initial spin angle $\alpha_0$ and the initial spin-angle rate $\alpha_0 '$, in order for the two impulses $r=1$ and $r=2$ to be in-phase is
\begin{equation}
    \begin{split}
        \alpha_0 &\in \left( (1+2n_0)\frac{\pi}{2}, (1+n_0)\pi \right) \cr 
        \alpha_0 ' &\in \left( - \frac{\alpha_0}{2\pi} + K \sin (2\alpha_0) + \frac{1+2n_2}{4},  - \frac{\alpha_0}{2\pi} + K \sin (2\alpha_0) + \frac{1+n_2}{2}  \right) 
    \end{split} \quad \forall n_0,n_2 \in \mathbb{Z}.
\end{equation}

\vspace{1mm}
\emph{Numeric results for any number of impulses.} 
The gradient-gradient moment impulse at the third periapsis passage ($r=3$) has the same sign of the gravity-gradient impulse at the second periapsis passage ($r=2$) only if
\begin{equation}
    \text{sign}(\sin (2\alpha_3)) = \text{sign}(\sin (2\alpha_2)),
\end{equation}
where 
\begin{equation}
    \alpha_3 = \alpha_2 + \alpha_2 ' 2\pi = \alpha_0 + \left[ \alpha_0 ' - K \sin (2\alpha_0) \right] 4\pi - K \sin \left( 2 (\alpha_0 + (\alpha_0 ' -K \sin (2\alpha_0)) 2\pi) \right)  .
\end{equation}

For a number of impulses larger than or equal to three, it does not seem possible to obtain analytic results about the existence of initial conditions fulfilling the in-phase condition. Therefore, we proceed numerically.

\begin{figure}[h!]
\centering
\includegraphics[scale=0.47]{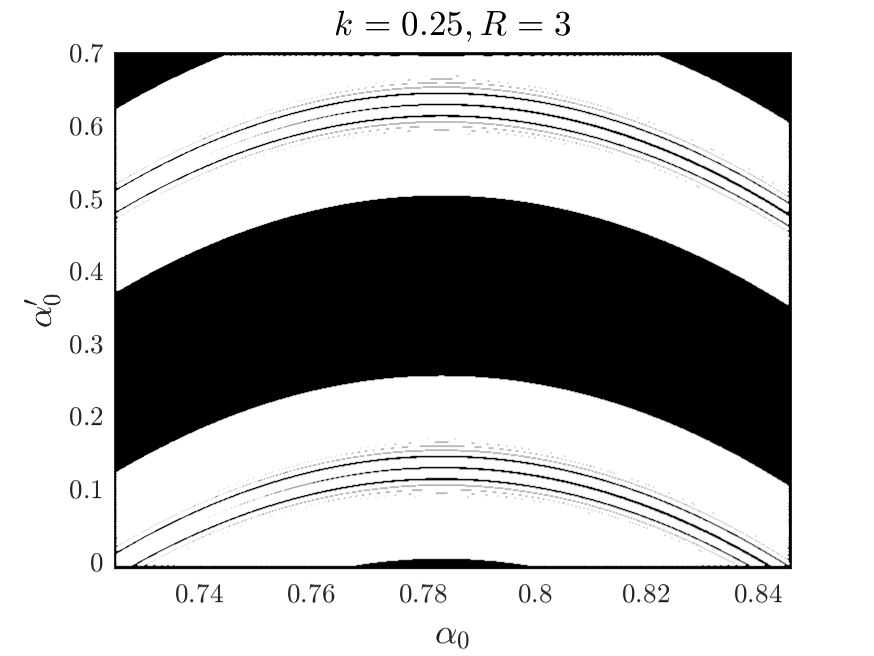}
\includegraphics[scale=0.47]{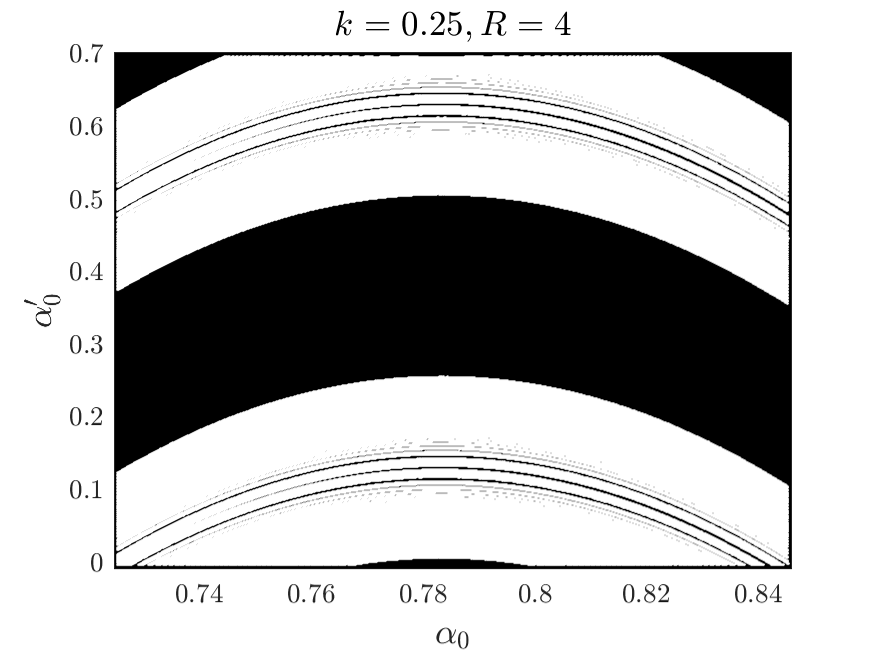}

\includegraphics[scale=0.47]{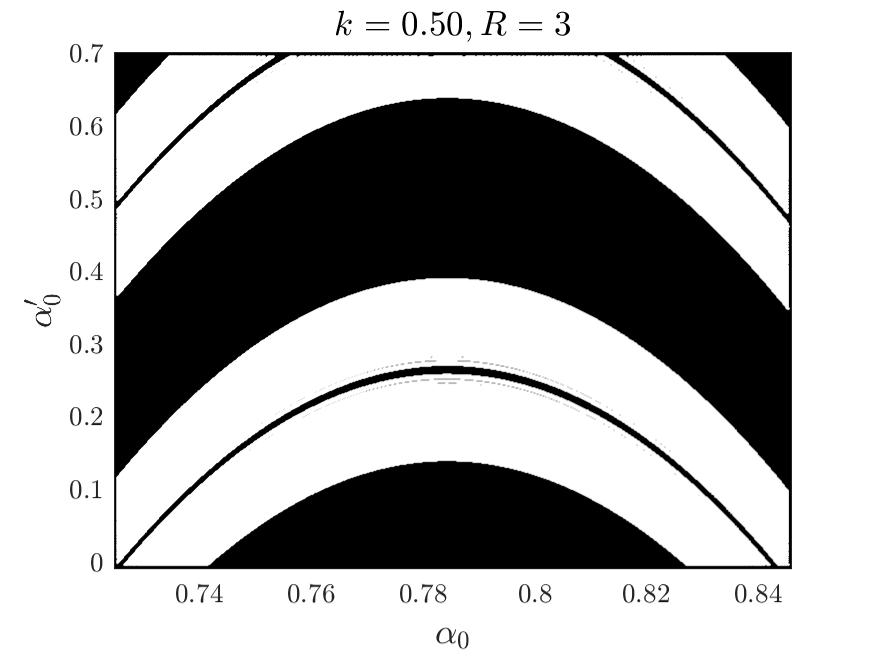}
\includegraphics[scale=0.47]{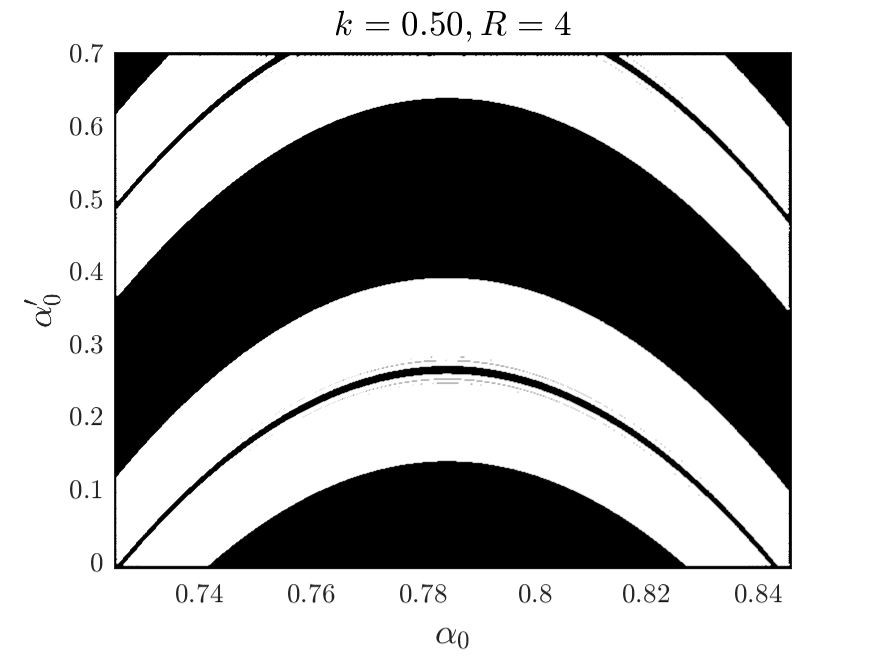}

\includegraphics[scale=0.47]{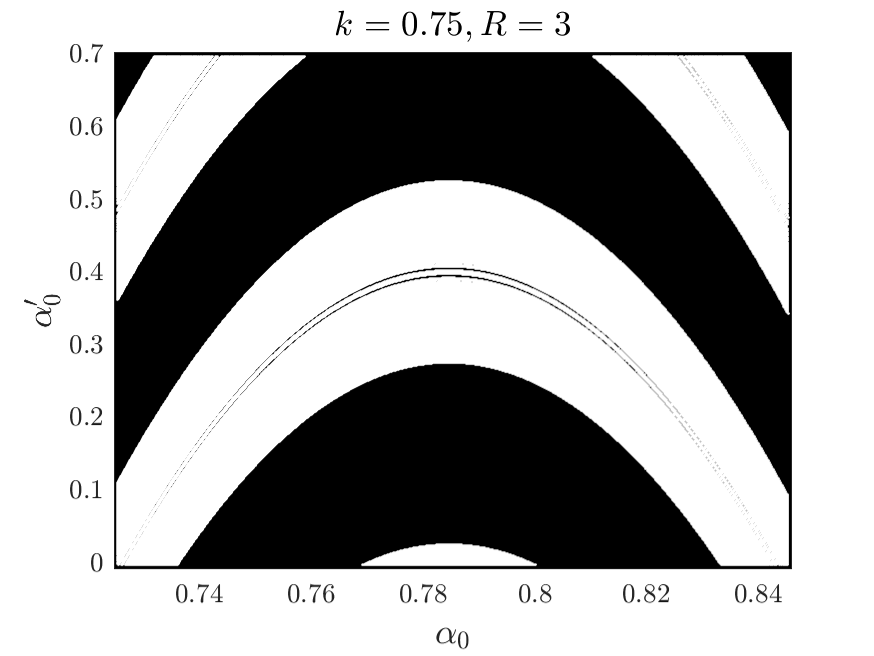}
\includegraphics[scale=0.47]{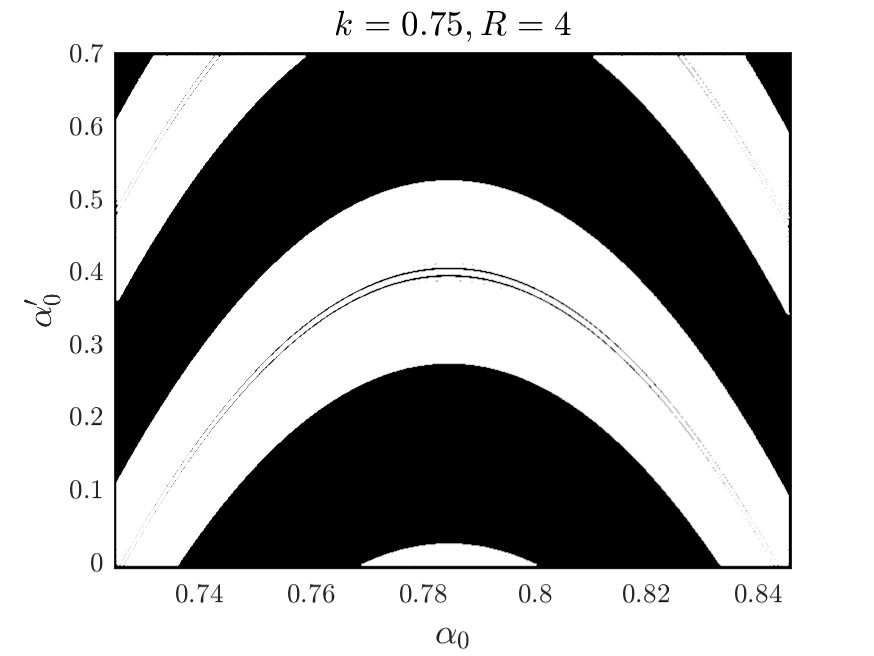}
\caption{Set of initial conditions ($\alpha_0$, $\alpha_0'$), represented in white color, for which the in-phase impulse condition (with $\sin(2\alpha_0) > 0$) is satisfied for three (left panels) and four (right panels) periapsis passages for $e= 0.9$ and $\kappa =0.25$ (top panels), $\kappa = 0.5$ (central panels), and $\kappa = 0.75$ (bottom panels). 
}
\label{fig:SOP_DiscMap_R_3_4_zoom1}
\end{figure}

In Fig. \ref{fig:SOP_DiscMap_R_3_4_zoom1} we plot the initial conditions fulfilling the in-phase condition (with $\sin (2 \alpha_{0} ) > 0$) in white color for three (left panels), and four periapsis passages (right panels), and for the three inertia ratio values $\kappa = 0.25,\,0.5,\,0.75$. We notice that the regions of initial conditions fulfilling the in-phase condition at four periapsis passages seems to be similar to the regions of initial conditions fulfilling the in-phase condition up to three periapsis passages.

\begin{figure}[h!]
\centering
\includegraphics[scale=0.47]{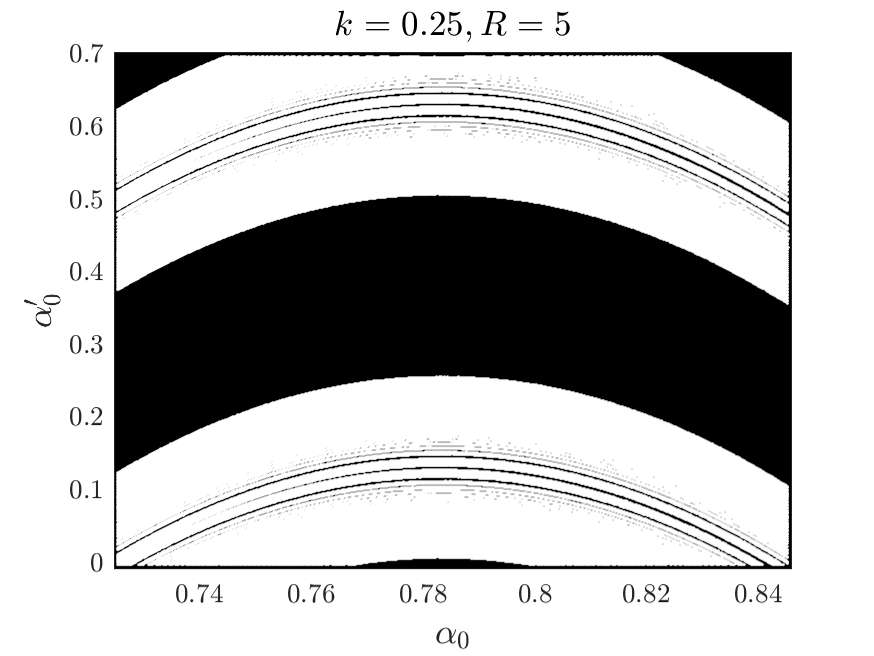}
\includegraphics[scale=0.47]{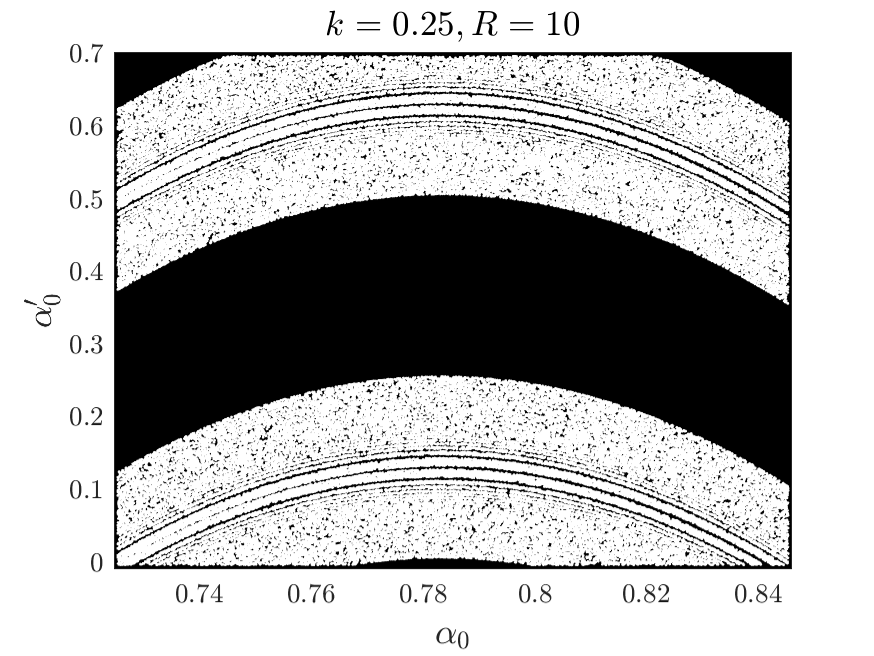}

\includegraphics[scale=0.47]{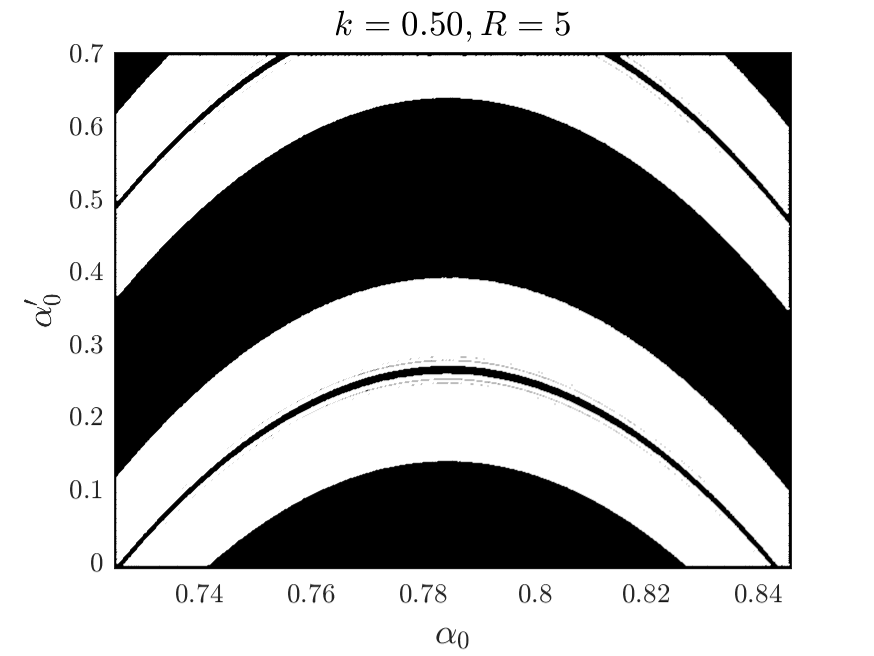}
\includegraphics[scale=0.47]{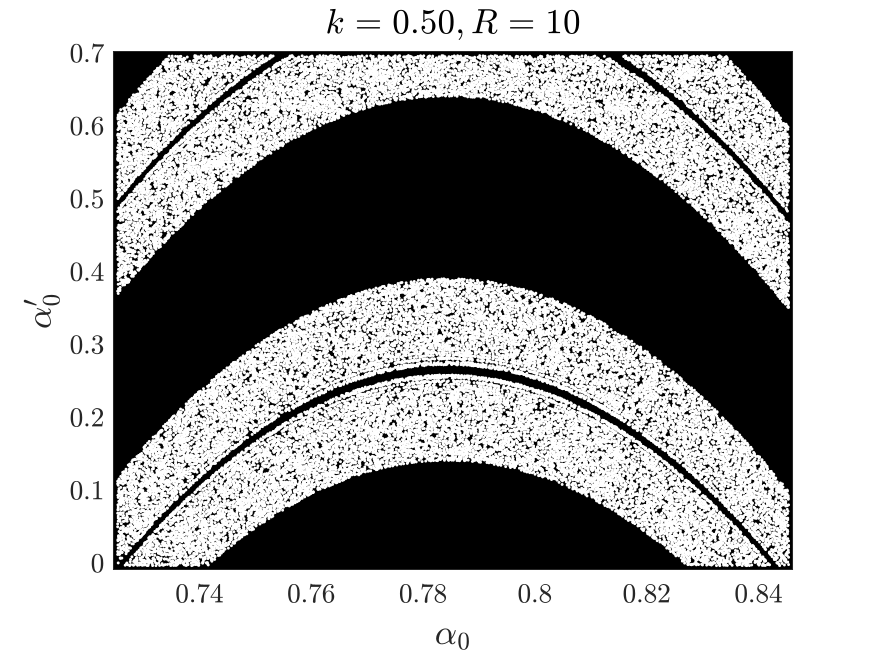}

\includegraphics[scale=0.47]{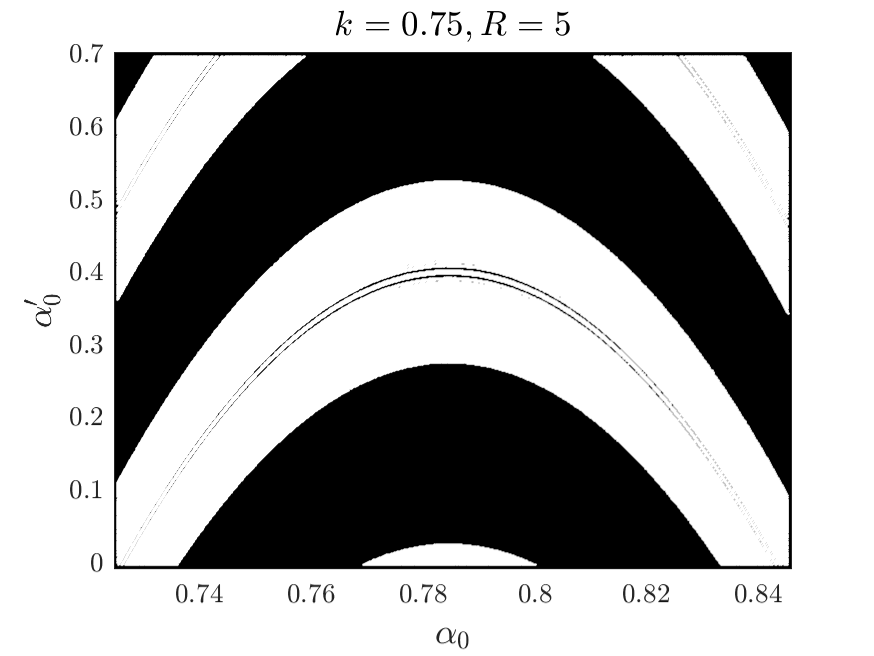}
\includegraphics[scale=0.47]{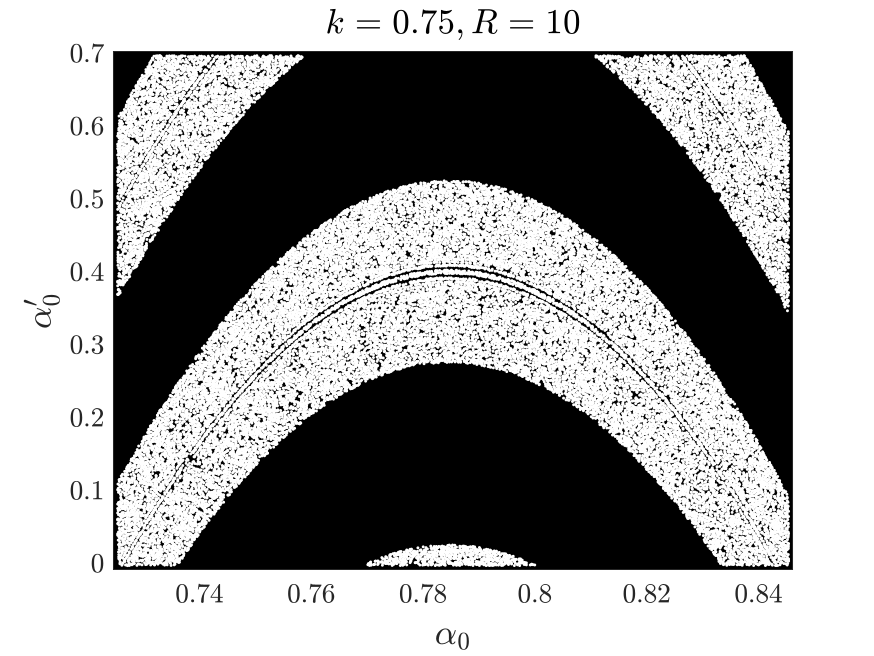}
\caption{Set of initial conditions ($\alpha_0$, $\alpha_0'$), represented in white color, for which the in-phase impulse condition (with $\sin(2\alpha_0) > 0$) is satisfied for five (left panels) and ten (right panels) periapsis passages for $e= 0.9$ and $\kappa =0.25$ (top panels), $\kappa = 0.5$ (central panels), and $\kappa = 0.75$ (bottom panels). 
}
\label{fig:SOP_DiscMap_R_5_10_zoom1}
\end{figure}

In Fig. \ref{fig:SOP_DiscMap_R_5_10_zoom1} we plot the initial conditions fulfilling the in-phase condition (with $\sin (2 \alpha_{0} ) > 0$) in white color up to five (left panels), and ten periapsis passages (right panels), and for the three inertia ratio values $\kappa = 0.25,\,0.5,\,0.75$. We notice that the regions of initial conditions fulfilling the in-phase condition up to ten periapsis passages is less dense and more disrupted with respect to those obtained up to five passages.

In Table \ref{tab:DMinphase} we show the evolution of $\alpha_r$, and $\sin (2 \alpha_{r})$ with $1 \leq r \leq 10$, by considering an initial datum fulfilling the in-phase condition with $e=0.9$, and $\kappa = 0.75$. More precisely, we consider $\alpha_0 = 0.83524816339744833993$, and $\alpha_0 ' = 0.054550000000000001266$.
\begin{table}[htbp!]
    \centering
    \begin{tabular}{|c|c|c|}
        \hline
        \rule{0pt}{3ex}
        \rule[-1.2ex]{0pt}{3ex}
        $\bm{r}$ & $\bm{\alpha_r}$ & $\bm{\sin (2 \alpha_{r})}$ \\
        \hline
        $1$  & $0.8352$ & $0.9950$ \\
        $2$  & $3.6222$ & $0.8199$ \\
        $3$  & $4.5638$ & $0.2928$ \\
        $4$  & $3.9364$ & $0.9998$ \\
        $5$  & $3.9801$ & $0.9944$ \\
        $6$  & $0.4340$ & $0.7630$ \\
        $7$  & $3.5269$ & $0.6965$ \\
        $8$  & $0.1593$ & $0.3132$ \\
        $9$  & $0.2620$ & $0.5003$ \\
        $10$ & $3.6933$ & $0.8927$ \\
        \hline
    \end{tabular}
    \caption{Evolution of $\alpha_r$, and $\sin (2\alpha_{r})$ for the discrete map with $e=0.9$, $\kappa = 0.75$, and the initial datum $\alpha_0 = 0.83524816339744833993$, and $\alpha_0 ' =  0.054550000000000001266$ fulfilling the in-phase condition. We use $4$ digits for the representation of the values.}
    \label{tab:DMinphase}
\end{table}

\vspace{2mm}
\noindent \textbf{Initial conditions for counterphase gravity-gradient moment impulses}

Now, we consider the initial states fulfilling the counterphase condition of the gravity-gradient moment impulse up to an arbitrary number $R$ of periapsis passages. This condition is fulfilled by values of $\alpha_0$ and $\alpha_0 '$ such that
\begin{equation}
    \text{sign}(\sin ( 2 \alpha_{R})) = \cdots = +\text{sign}(\sin (2\alpha_3) = -\text{sign}(\sin ( 2 \alpha_2)) = +\text{sign}(\sin ( 2 \alpha_1)) .
    \label{eq:gen_incase}
\end{equation}

We start our investigation by considering analytic results on the initial states fulfilling the counterphase condition for the first two impulses with $\text{sign}(\sin (2\alpha_0)) = 1$. Later, we will consider the subcase with $\text{sign}(\sin (2\alpha_0)) = -1$. Finally, we will show numerical results for initial conditions satisfying the in-phase condition up to ten periapsis passages.

\vspace{1mm}
\emph{Analytic results for the first two impulses.} By analysing Eq. \eqref{eq:step1} and \eqref{eq:step2}, it is straighforward to see that the gravity-gradient moment impulse at the second periapsis passage ($r=2$) has the opposite sign of the first gravity-gradient moment impulse ($r=1$), only if
\begin{equation}
    \text{sign}(\sin ( 2\alpha_2) ) = -\text{sign} (\sin ( 2 \alpha_1)).
\end{equation}
Let us consider the following \emph{first subcase}: the initial spin-angle $\alpha_0$ is such that
\begin{equation}
    \text{sign}(\sin ( 2\alpha_0)) = 1 \leftrightarrow \alpha_0 \in (n_0 \pi, (1+2n_0)\pi/2).
\end{equation}

The condition that needs to be satisfied in order for the two impulses $r=1$ and $r=2$ to be in counterphase, is
\begin{equation}
    \text{mod}(\alpha_2,\pi) \in (\pi/2,\pi) \rightarrow \text{mod}( \alpha_0  + \left[ \alpha_0 ' -K \sin (2 \alpha_0) \right] 2\pi,\pi) \in (\pi/2,\pi)
\end{equation}
We emphasize that the previous relation is equivalent to
\begin{equation}
    \alpha_0 + \left[ \alpha_0 ' -K \sin (2\alpha_0) \right] 2\pi 0' \in \left( \frac{1+2n_2}{2} \pi, (1+2n_2)\pi \right), \qquad n_2 \in \mathbb{Z}.
\end{equation}

Finally, the condition that needs to be satisfied by the initial spin angle $\alpha_0$ and the initial spin-angle rate $\alpha_0 '$, in order for the two impulses $r=1$ and $r=2$ to be in counterphase is
\begin{equation}
    \begin{split}
        \alpha_0 & \in \left( n_0\pi, \frac{1+2 n_0}{2} \pi \right) \\
        \alpha_0 ' & \in \left( -\frac{\alpha_0}{2\pi} + K \sin (2\alpha_0) + \frac{n_2}{2}, -\frac{\alpha_0}{2\pi} + K \sin (2\alpha_0) + \frac{1+2 n_2}{4}\right)
    \end{split} \quad n_0,n_2 \in \mathbb{Z}.
\end{equation}

As before, let us assume that 
\begin{equation}
    \text{sign}(\sin (2\alpha_1)) = 1 \leftrightarrow \sin ( 2\alpha_0) > 0 \rightarrow \text{mod}(\alpha_0,\pi) \in (0,\pi/2),
\end{equation}
or equivalently,
\begin{equation}
    \alpha_0 \in (n_0 \pi, (1+2n_0) \pi/2)), \quad n_0 \in \mathbb{Z}.
\end{equation}

\vspace{1mm}
\emph{Numeric results for any number of impulses.} 
As for the in-phase condition, for a number of impulses larger than or equal to three, it does not seem possible to obtain analytic results. Therefore, we proceed numerically. 

\begin{figure}[h!]
\centering
\includegraphics[scale=0.47]{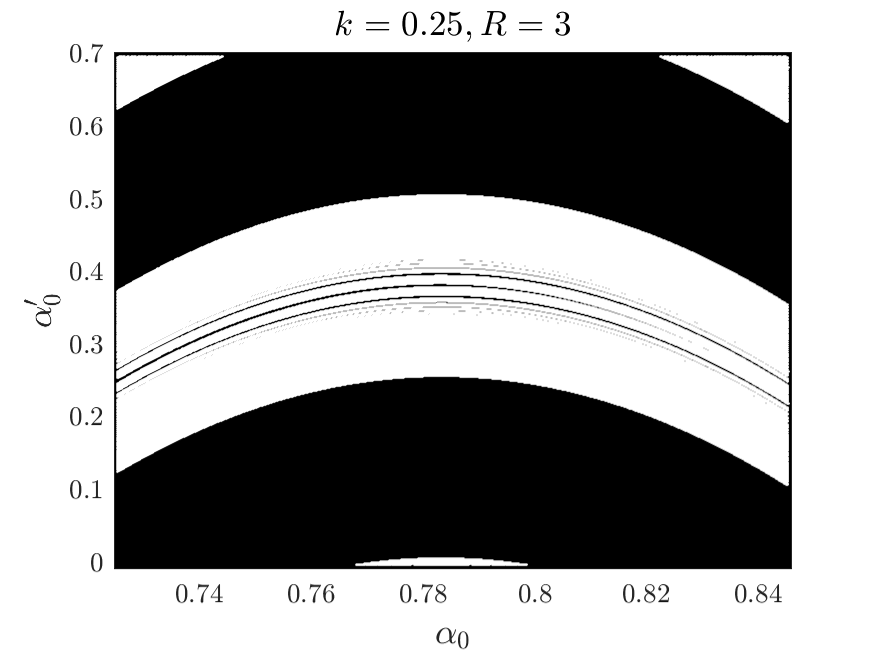}
\includegraphics[scale=0.47]{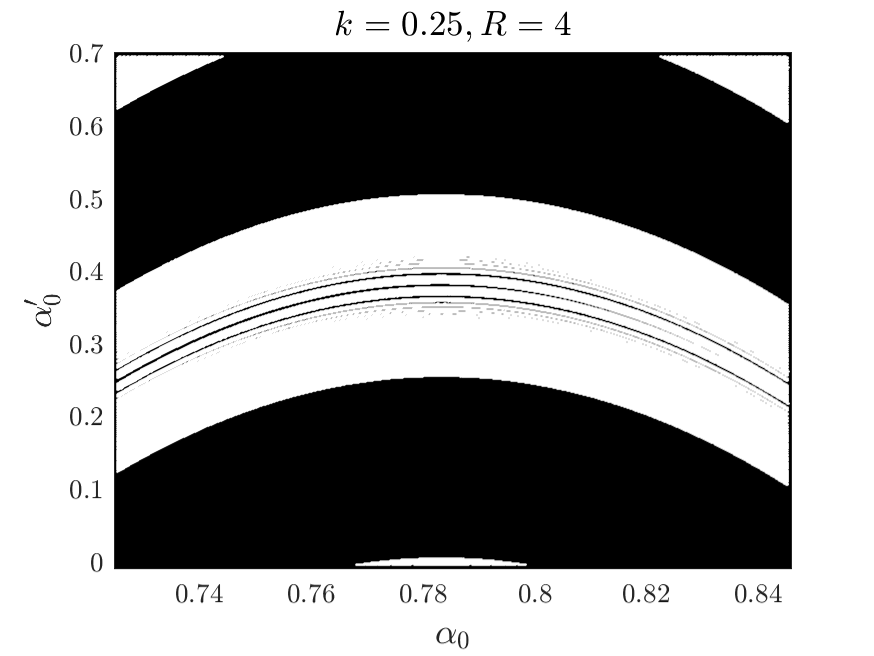}

\includegraphics[scale=0.47]{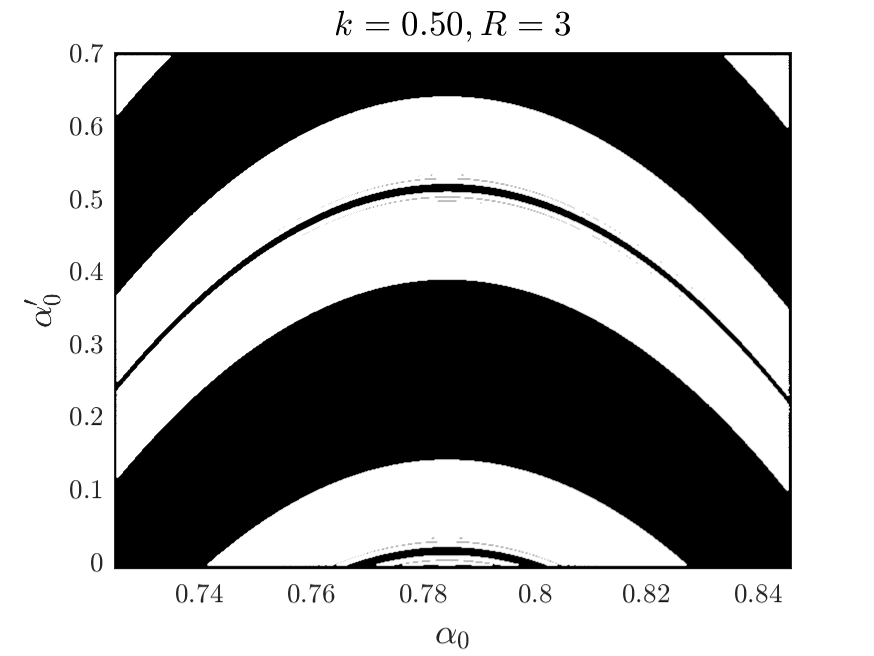}
\includegraphics[scale=0.47]{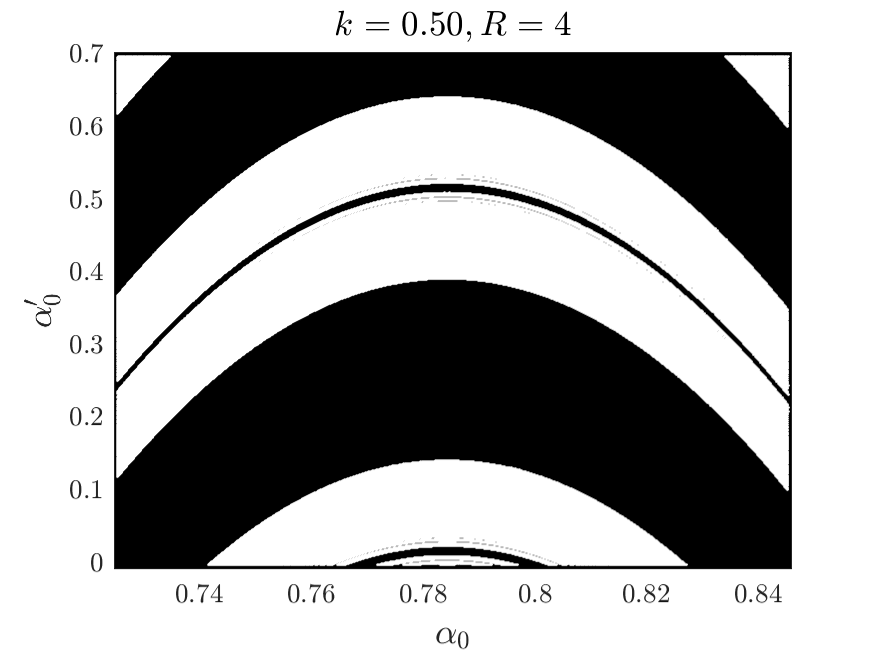}

\includegraphics[scale=0.47]{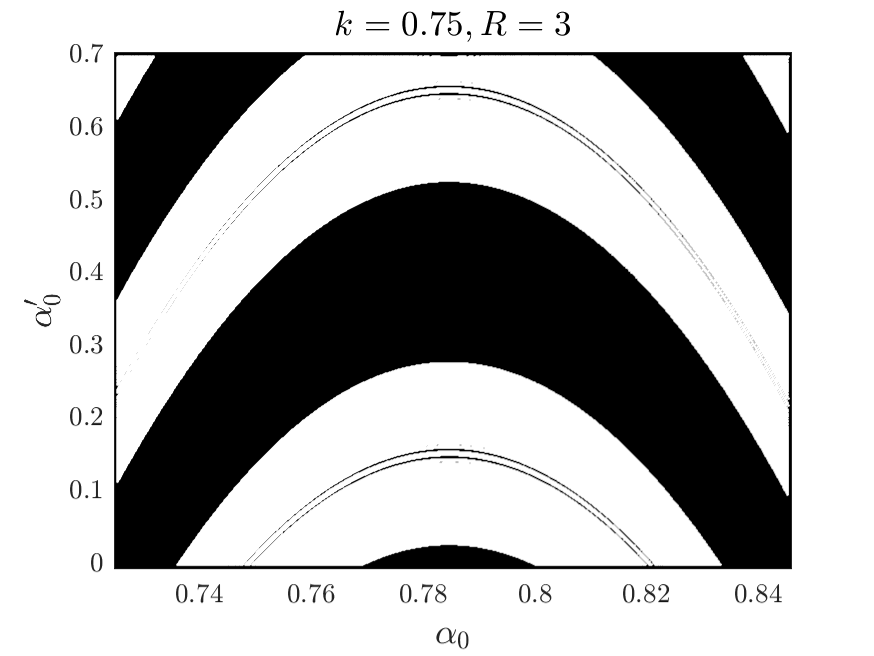}
\includegraphics[scale=0.47]{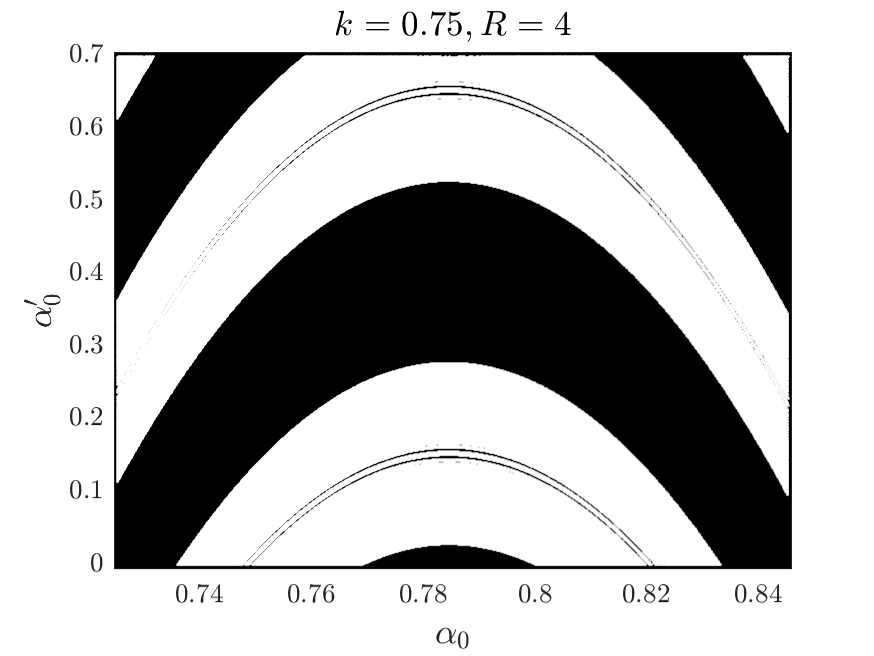}
\caption{Set of initial conditions $(\alpha_0,\alpha_0 ')$, represented in white color, for which the in counterphase impulse condition (with $\sin (2\alpha_0)>0$) is satisfied for three (left column), and four (right column) periapsis passages for $e=0.9$, and $\kappa = 0.25$ (top row), $\kappa =0.5$ (central row), and $\kappa = 0.75$ (bottom row).}
\label{fig:DM_counterphase_R_3_4}
\end{figure}

In Fig. \ref{fig:DM_counterphase_R_3_4} we plot in white color the initial conditions $\alpha_0$, $\alpha_0 '$ fulfilling the counterphase condition with ($\sin ( 2\alpha_0) > 0$) up to three periapsis passages (left panels) and four periapsis passages (right panels). 

\begin{figure}[h!]
\centering
\includegraphics[scale=0.47]{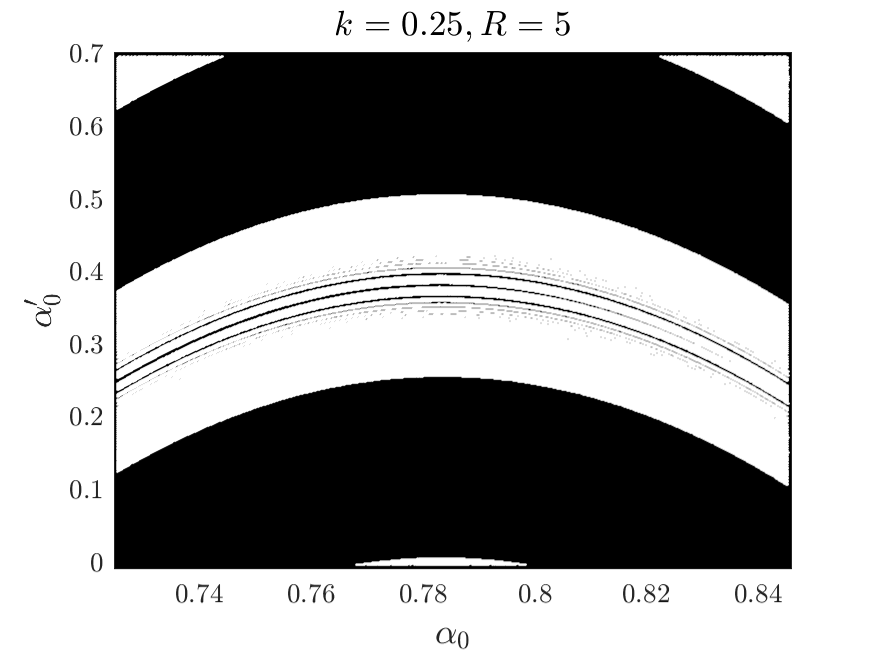}
\includegraphics[scale=0.47]{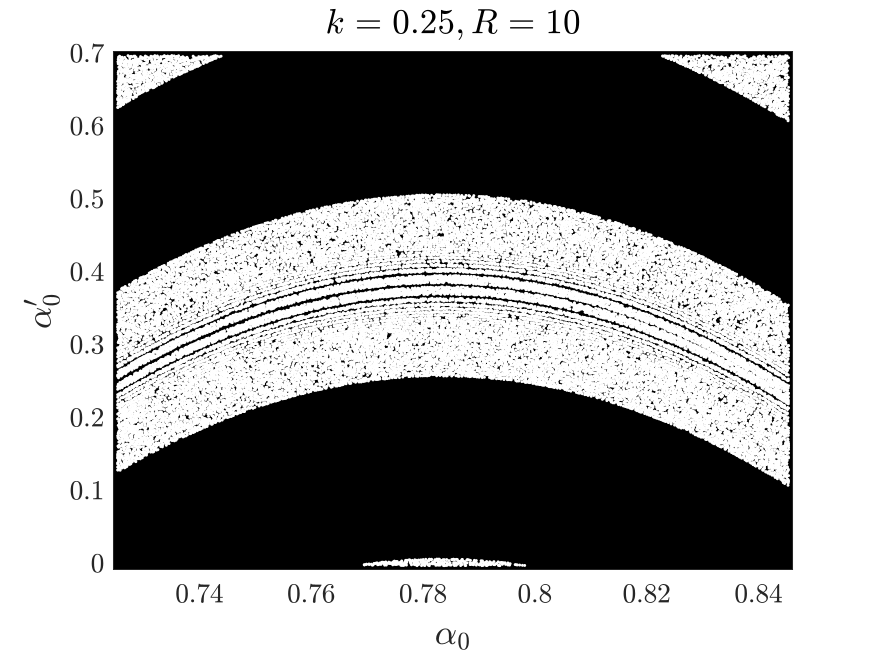}

\includegraphics[scale=0.47]{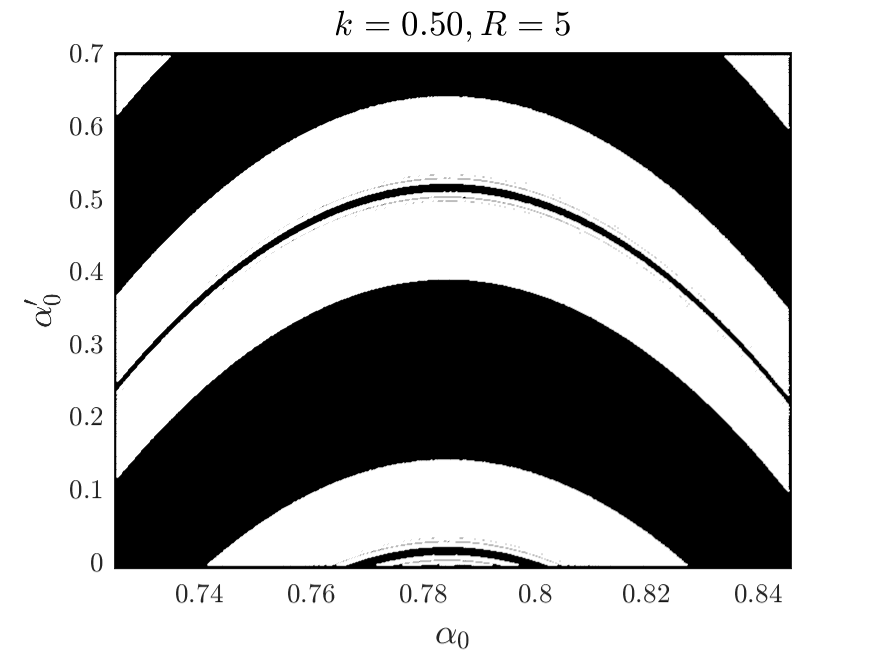}
\includegraphics[scale=0.47]{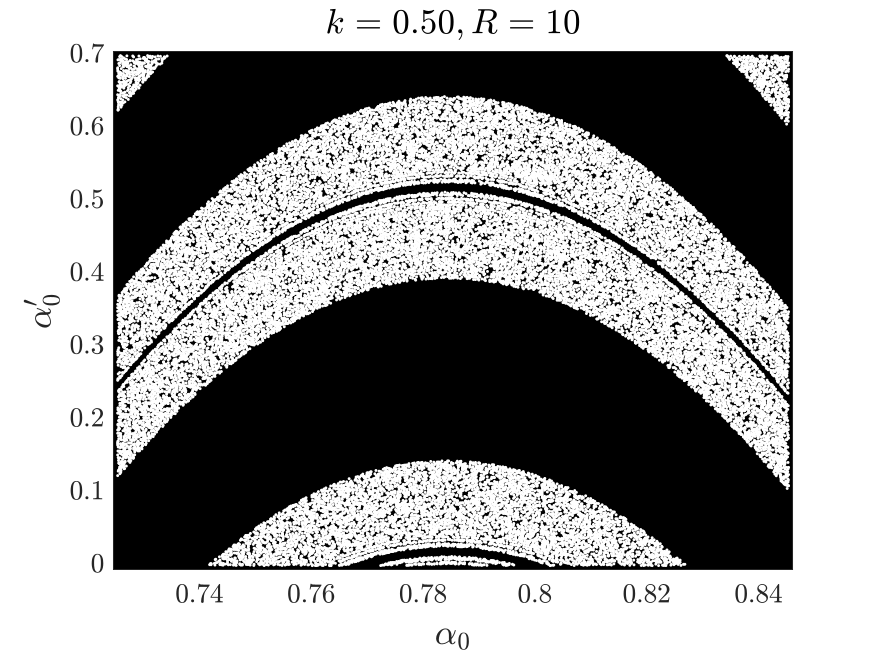}

\includegraphics[scale=0.47]{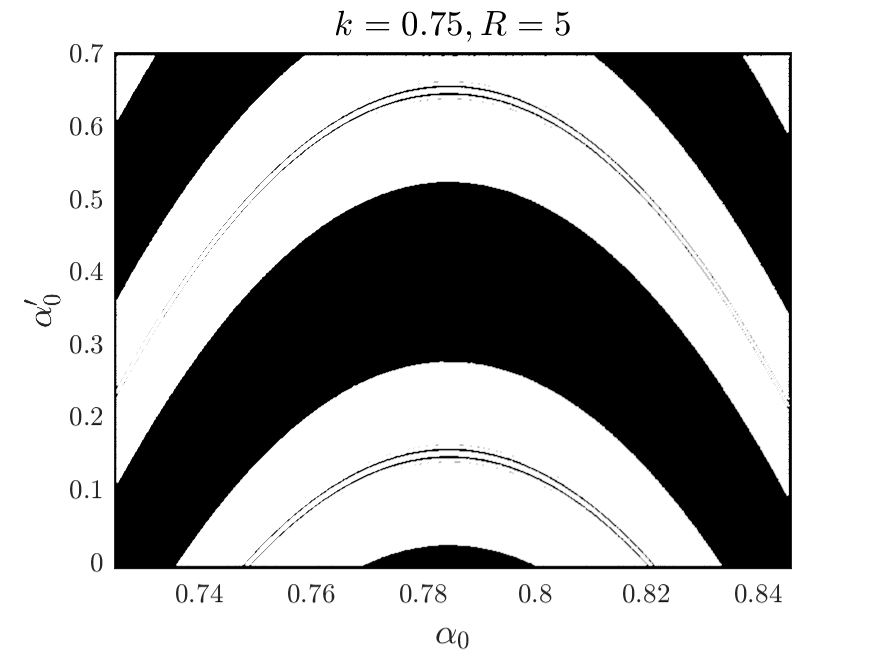}
\includegraphics[scale=0.47]{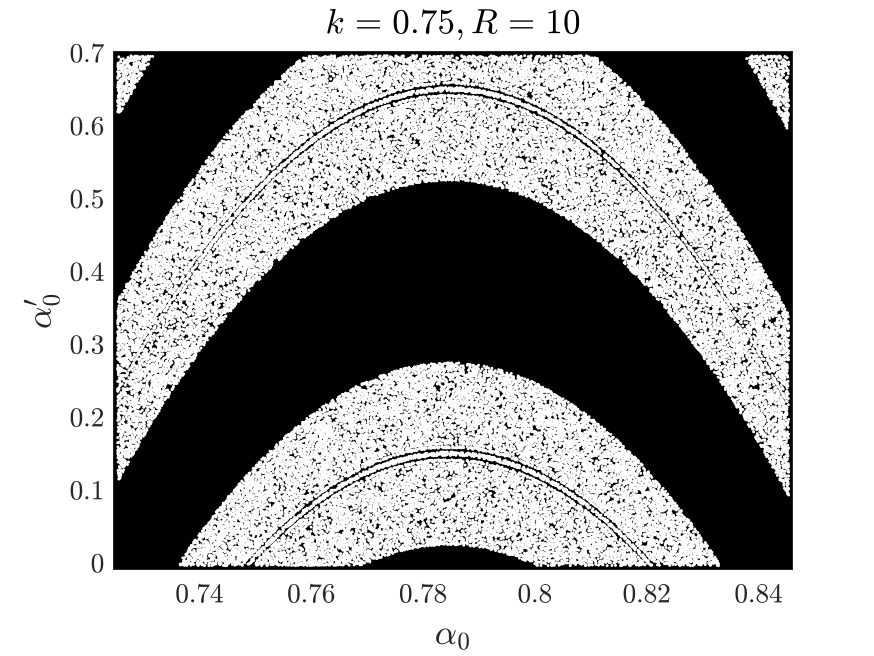}
\caption{Set of initial conditions $(\alpha_0,\alpha_0 ')$, represented in white color, for which the in counterphase impulse condition (with $\sin (2\alpha_0)>0$) is satisfied for five (left column), and ten (right column) periapsis passages for $e=0.9$, and $\kappa = 0.25$ (top row), $\kappa =0.5$ (central row), and $\kappa = 0.75$ (bottom row).}
\label{fig:DM_counterphase_5_10}
\end{figure}

In Fig. \ref{fig:DM_counterphase_5_10} we plot in white color the initial conditions $\alpha_0$, $\alpha_0 '$ fulfilling the counterphase condition with ($\sin ( 2\alpha_0) > 0$) up to five periapsis passages (left panels) and ten periapsis passages (right panels).
It is evident a disruption of the white regions for ten periapsis passages.

In Table \ref{tab:DMcounter} we show the evolution of $\alpha_r$, and $\sin (2 \alpha_{r})$ with $ 1 \leq r \leq 10$, by considering an initial datum fulfilling the counterphase condition with $e=0.9$, and $\kappa =0.75$. More precisely, we consider $\alpha_0 = 0.81234816339744830849$, and $\alpha_0 ' = 0.11080000000000000959$.
\begin{table}[htbp!]
\centering
\begin{tabular}{|c|c|c|}
\hline
\rule{0pt}{3ex}
\rule[-1.2ex]{0pt}{3ex}
$\bm{r}$ & $\bm{\alpha}_{\bm{r}}$ & $\bm{\sin} \bm{(2 \alpha_{r})}$ \\
\hline
$1$  & $0.8123$ & $+0.9985$ \\
$2$  & $2.6523$ & $-0.8297$ \\
$3$  & $3.6912$ & $+0.8908$ \\
$4$  & $1.7626$ & $-0.3743$\\
$5$  & $0.1457$ & $+0.2873$ \\
$6$  & $5.2797$ & $-0.9064$ \\
$7$  & $0.2868$ & $+0.5427$ \\
$8$  & $1.7996$ & $-0.4418$ \\
$9$  & $3.4496$ & $+0.5778$ \\
$10$ & $4.8939$ & $-0.3550$ \\
\hline    
\end{tabular}
\caption{Evolution of $\alpha_r$, and $\sin (2 \alpha_{r})$ for the discrete map with $e=0.9$, $\kappa =0.75$, and the initial datum $\alpha_0 = 0.81234816339744830849$, $\alpha_0 ' = 0.11080000000000000959$ fulfilling the counterphase condition.  We use $4$ digits for the representation of the values. }
\label{tab:DMcounter}
\end{table}

\subsection{Comparison between the impulsive approximation and the spin orbit problem for highly elliptical orbits}
\label{ssec:comparison_DM_SOP}

In this section we justify the introduction of the discrete map in \eqref{eq:recursivemapnew} as a preliminary study of the SOP model in Eq. \eqref{eq:eqstau} and \eqref{eq:taucompleta}, for HEOs. As a matter of fact, for $\overline{A}(f)$ to be suitably approximated by a Dirac pulse, we expect that at each periapsis passage the slope of the linear dependence in $\alpha (\sigma)$ changes, while $\alpha' (\sigma)$ is affected by a jump, and it is reasonable that the DM can be treated as a Poincar\'e map of the SOP for $f=r 2\pi$, with $r \in \mathbb{Z}$.
Obviously, there will be some difference between the DM and the SOP, because $\overline{A}(f)$ is not an ideal Dirac pulse acting instantaneously, and the gravity-gradient moment is distributed in a time interval that it is not infinitesimal (even it is really small, as seen in Table \ref{tb:rate_Deltat}). One of the main difference between the DM and the SOP is that in the DM the initial conditions $\alpha_0,\alpha_0'=0$ is an equilibrium state, while this is not true in the SOP \citep[for example, the reader is referred to][for further details]{celletti2000,misquero2020}. As a matter of fact, it is well known that after the introduction of the new variable $\Theta = 2(f-\alpha)$, the SOP equations assume the form
\begin{equation}
    2\ddot{f}-\ddot{\Theta} = A(f) \sin \Theta,
\end{equation}
and for $e=0$, the previous equations is pendulum-like, i.e.,
\begin{equation}
    \ddot{\Theta} + A(f) \sin \Theta = 0.
    \label{eq:pendulumlike}
\end{equation}
We recall that the equilibrium of \eqref{eq:pendulumlike} are 
$(\Theta^{\ast},\dot{\Theta}^{\ast}) = (0,0)$ and $(\Theta^{\ast},\dot{\Theta}^{\ast}) = (\pi,0)$; hence the resonance for the SOP with $e=0$ are found for $\alpha'_{\ast} = 1$, $f(\sigma)-\alpha(\sigma) = 0$ and $f(\sigma)-\alpha(\sigma) = \pi/2$. The latter resonances are known as $1$:$1$ spin-orbit resonances. When $e\neq 0$, it is possible to prove that the SOP equations can be written in Fourier series as (for values of eccentricity in which the series converges)
\begin{equation}
    \alpha'' + \frac{3 \kappa}{2} \sum _{m\neq 0, m = -\infty} ^{+\infty} W \left (\frac{m}{2},e \right) \sin (2 \alpha - m \sigma) = 0, 
\end{equation}
where $W$ are suitable coefficients of the Fourier series \citep[for an explicit representation, the reader is referred to][]{cayley1859}. Hence for $e \in (0,1)$ there may occur other spin-orbit resonances, such as the famous $3$:$2$ resonance appearing in the Sun--Mercury system. For recent research works on the spin-orbit resonances, the reader is referred to \citet{celletti2000}, \citet{wisdom2004} and \citet{gkolias2016}.

Another difference between the DM and the SOP is that the gravity-gradient moment is not an ideal Dirac pulse and it does not act instantaneously. Hence, the term $\overline{A}(f) \sin(2 (f-\alpha))$ can change in sign several times during a single periapsis passage, depending on the evolution of $\alpha(\sigma)$. For this reason, in the SOP we expect two main scenarios close to each periapsis passage: the gravity-gradient moment can be described i) through a single Dirac pulse, or ii) through two or more Dirac pulses with opposite sign.

Moreover, we stress that since we are interested in HEOs, at every periapsis passage the solution is really close to the singularity (i.e., the major body) and the numerical integration are affected by errors. Hence, up to now, there are no regularization techniques that are able to `remove' the singularities both in the orbital and rotational dynamics. 
Therefore, we can numerically propagate the SOP only for a few periapsis passages. In this works, we consider a maximum of three periapsis passages. Furthermore, for the simulations we use the Matlab ode solver \texttt{ode89} with $10^{-13}$ as relative and absolute tolerance.

\begin{figure}[h!]
    \centering
    \includegraphics[scale=0.56]{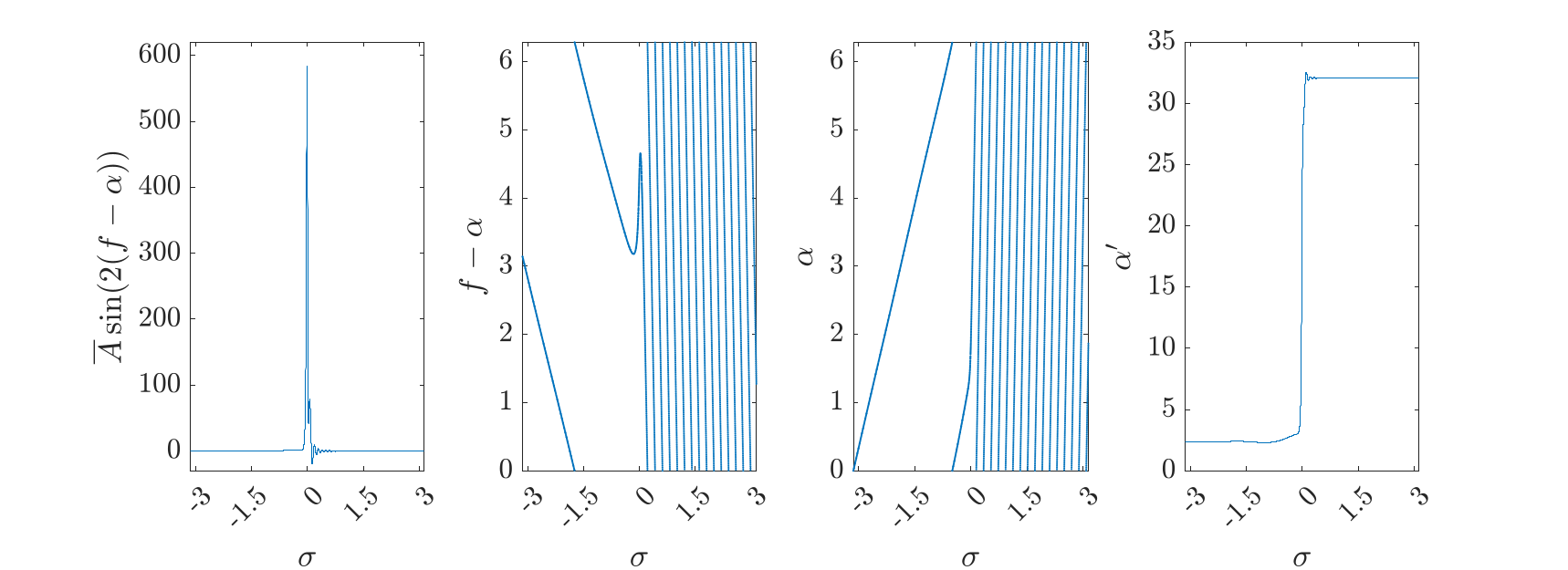}

    \includegraphics[scale=0.56]{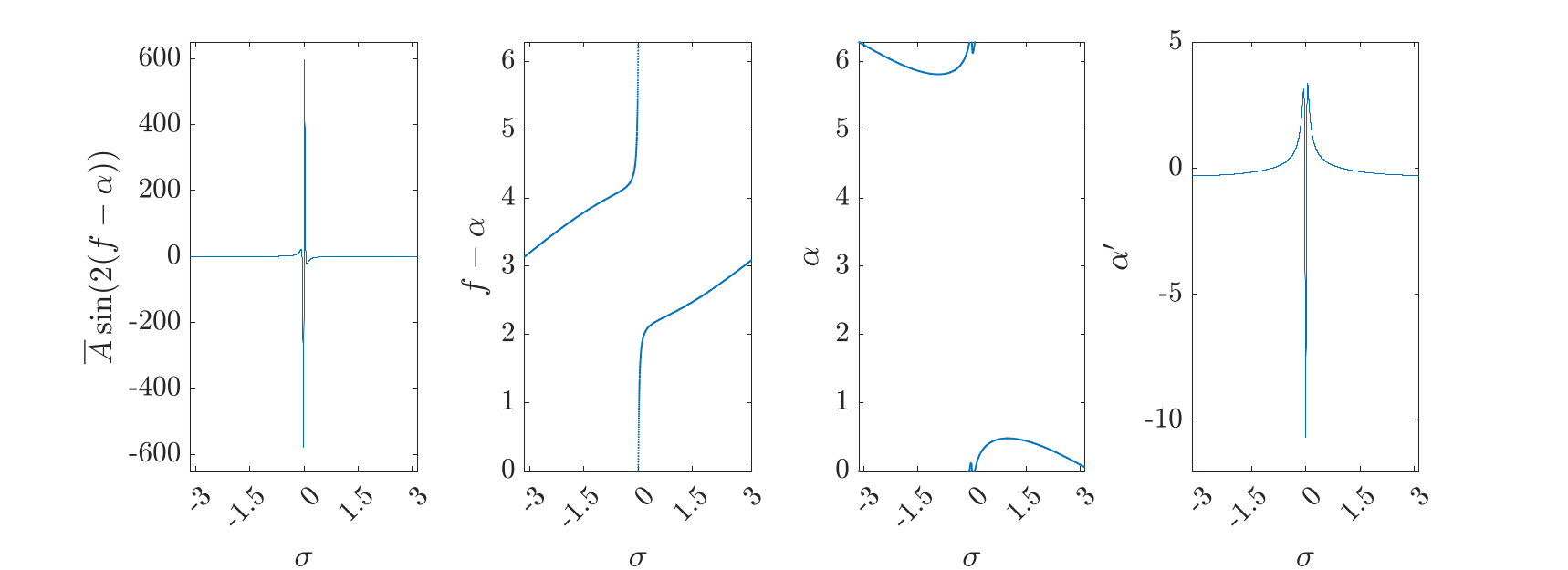}

    \includegraphics[scale=0.56]{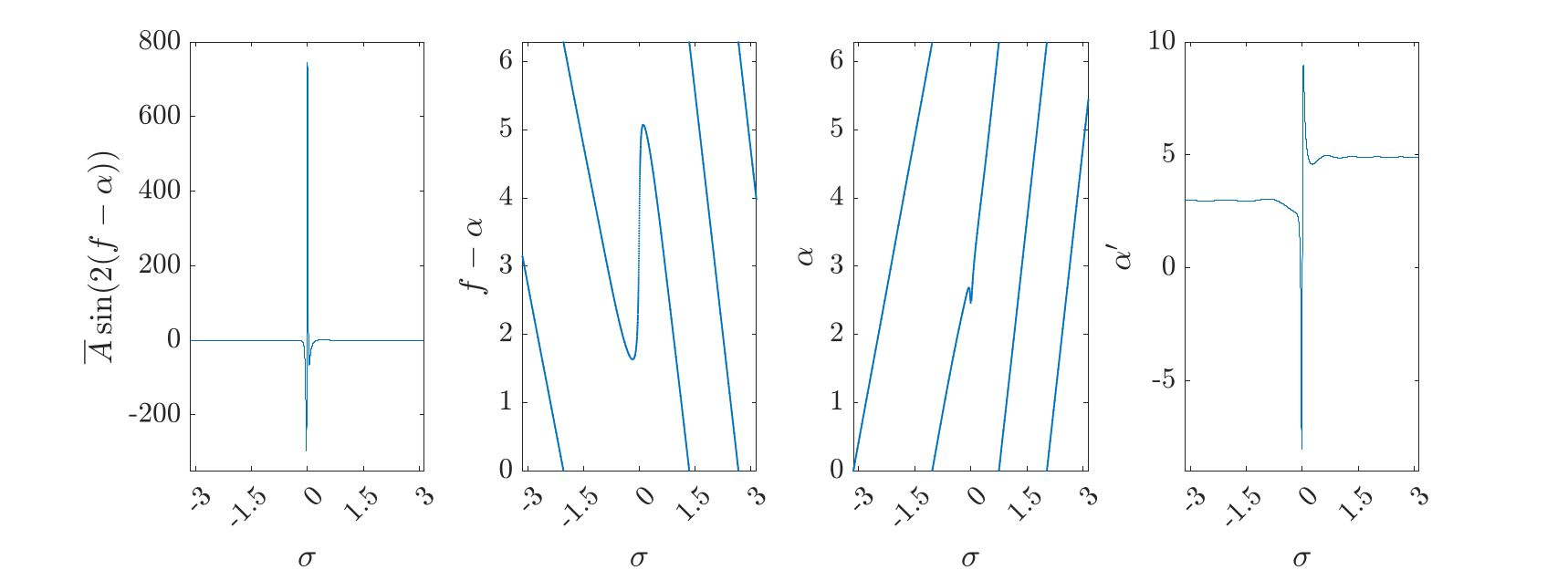}

    \caption{Numerical solutions of the SOP with $e=0.9$ and $\kappa = 0.5$ associated to the initial conditions $(f,\alpha,\alpha_0 ')= (-\pi,0,2.4)$ (top row), $(f,\alpha,\alpha_0 ')= (-\pi,0,-0.3)$ (middle row) and  $(f,\alpha,\alpha_0 ')= (-\pi,0,3)$ (bottom row). }
    \label{fig:spinorbit_solution_example}
\end{figure}

In Fig. \ref{fig:spinorbit_solution_example} we show the numerical solution of the SOP with $e=0.9$ and $\kappa = 0.5$ associated to the integration of three different initial conditions. 
The panels of the first, second, third and fourth
column show the values of $\overline{A}(f) \sin(2(f-\alpha))$, $f-\alpha$, $\alpha$ and $\alpha ' $ respectively. 
In the case represented in the top row of Fig. \ref{fig:spinorbit_solution_example}, the gravity-gradient moment can be described by a single Dirac pulse. We note that the angular velocity is affected by a jump, whose amplitude is not equal to the amplitude of the pulse, while the simple rotation angle grows linearly in time with a change of slope after the periapsis passage. We note that the greatest peak of the gravity-gradient moment is positive, because $f-\alpha \in \left(\pi,\frac{3}{2}\pi \right)$ in the time interval in which the gravity-gradient moment is characterized by the peak.
The middle and bottom rows in Fig. \ref{fig:spinorbit_solution_example} represent cases in which the gravity-gradient moment can be described by two or more Dirac pulses having opposite sign in the amplitudes. In particular, the central panels of the figure show the case in which the gravity-gradient moment can be modelled by two consecutive and almost equal in amplitude Dirac deltas. In such a case, around the maximum value of the gravity-gradient moment, $\alpha$ passes from the interval $\left(\frac{3}{2}\pi,2\pi \right)$ to $\left(0,\frac{\pi}{2}\right)$, and then $\sin (2(f-\alpha))$ changes sign. We note that since the gravity-gradient moment present two peaks with almost opposite amplitude in a very short time-span, then the effects of these two peaks on the dynamics almost cancel out. This effect can be seen in the evolution of $\alpha'$; as a matter of fact $\alpha'$ is affected by a sharp variation close to the periapsis passage, but its value before and after the periapsis passage are almost the same. We stress that when the gravity-gradient moment can be modelled as two Dirac pulse with opposite amplitude (and both pulses are distributed in the same time-span), then we expect that the evolution of $\alpha$ and $\alpha'$ are periodic. 
In the bottom panels, a case in which the gravity-gradient moment  changes sign twice, but with a different value of the amplitude is presented. We note that the greatest peak is positive in amplitude and the angular velocity increases its value after the periapsis passage.

Since in the SOP multiple peaks may occur during a periapsis passages, in order to distinguish if the gravity-gradient moment can be modelled as a positive, negative, or null pulse during a periapsis passage, we compute the integral 
\begin{equation}
    \int _{\overline{\sigma} - \Delta \sigma^{\ast}/2} ^{\overline{\sigma} + \Delta \sigma^{\ast}/2 }\overline{A} \sin (2 (f-\alpha) ) d \sigma ,
\end{equation}
where $\overline{\sigma}$ indicates the time of periapsis passage, and $\Delta \sigma^{\ast}$ is calculated through \eqref{eq:delta_sigmaast}.
This integral helps to discriminate if the sharp change in $\alpha'$ before and after the periapsis passage is positive, negative or null. Hence, to plot the initial conditions fulfilling the in-phase condition in the SOP, we decided to consider a set of initial conditions $f=0$, $\alpha_0$ and $\alpha_0 '$, and to numerically propagate Eq. \eqref{eq:eqstau} and \eqref{eq:taucompleta} in the time interval $[-\pi,2\pi R-\pi)$ (i.e., we consider $R$ periapsis passages). For each solution, we calculate the gravity-gradient moment evolution, and the integrals 
\begin{equation}
    I_0 = \int _{-\pi} ^{-\pi+\Delta \sigma^{\ast}/2} \overline{A} \sin (2(f-\alpha)) d \sigma , \quad 
    I_r = \int _{-\pi(1-2r) - \Delta \sigma^{\ast}/2} ^{-\pi(1-2r)+\Delta \sigma^{\ast}/2} \overline{A} \sin (2(f-\alpha)) d \sigma \quad r \neq 1, R.
\end{equation} 
We remark that we do not compute $I_R$, because for $r=R$ the periapsis passage is not complete (but we are considering only `half periapsis passage').

The values of the integral
\begin{equation}
    P = 100 \times \frac{\int _{-\Delta \sigma^{\ast}/2} ^{\Delta \sigma^{\ast}/2} \overline{A} \sin(2 (f-\alpha)) d \sigma}{ \int _{-\Delta \sigma^{\ast}/2} ^{\Delta \sigma^{\ast}/2}\overline{A}  d \sigma }
    \label{eq:P_formula}
\end{equation}
associated to the periapsis passages in the three orbits plotted in Fig. \ref{fig:spinorbit_solution_example} are written in Table \ref{tb:3SOP_integrals}. We decided to display the rate defined in Eq. \eqref{eq:P_formula}, instead of $I_r$, to quantify the difference in the integral values due to $\sin(2(f-\alpha))$. The sign of the values written in the table show that the gravity-gradient moment acts at the periapsis passages as a positive pulse. This beheaviour is confirmed by the evolution of $\alpha$ and $\alpha'$ plotted in Fig. \ref{fig:spinorbit_solution_example}. 
\begin{table}[htbp!]
\centering
\begin{tabular}{|c|c|}
\hline
\rule{0pt}{3ex}
\rule[-1.2ex]{0pt}{3ex}
\textbf{Panel} & $\bm{P}$\cr
\hline
Top & $59$\cr 
Middle & $1.8$ \cr 
Below & $47$ \cr
\hline
\end{tabular}
\caption{Value of $P$ associated to the $3$ periapsis passage in the three scenarios represented in Fig. \ref{fig:spinorbit_solution_example}. }
\label{tb:3SOP_integrals}
\end{table}

\begin{figure}[h!]
\centering
\includegraphics[scale=0.35]{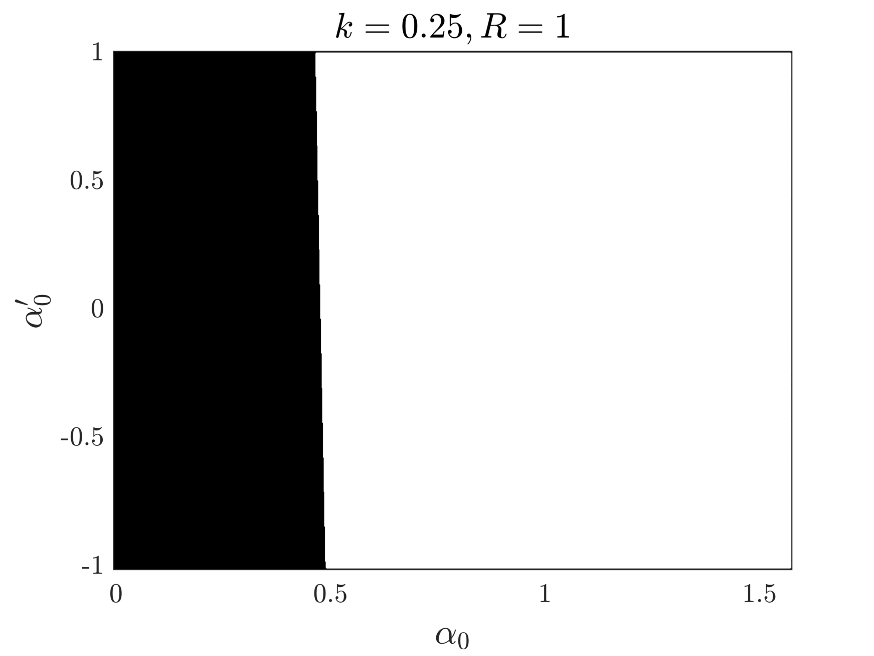} \hspace{-5mm}
\includegraphics[scale=0.35]{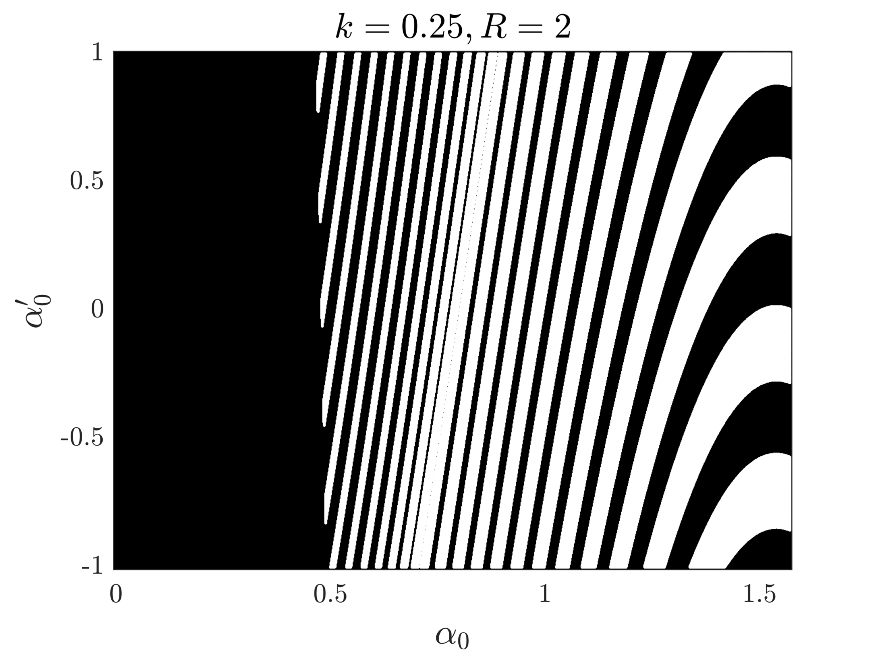} \hspace{-5mm}
\includegraphics[scale=0.35]{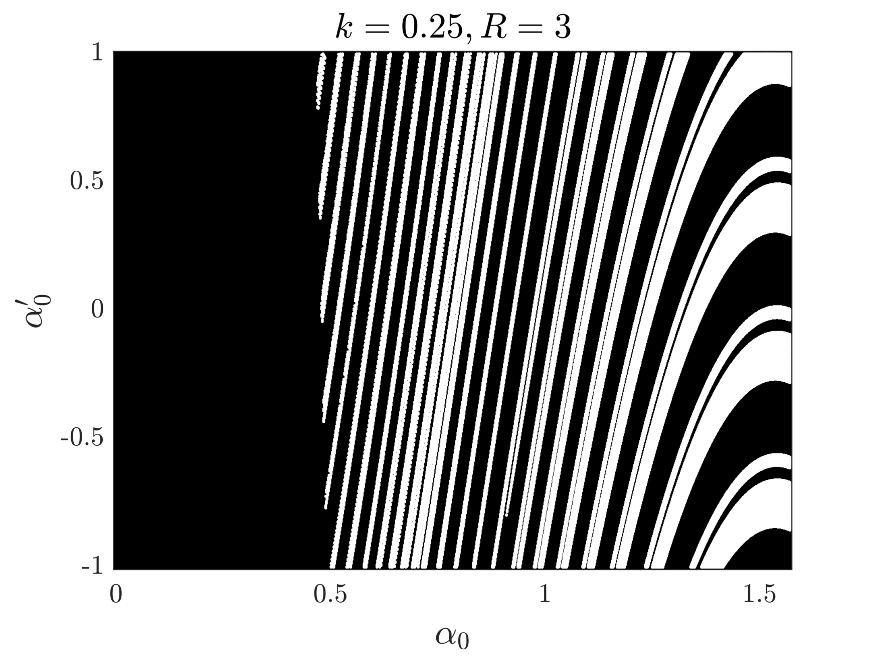}

\includegraphics[scale=0.35]{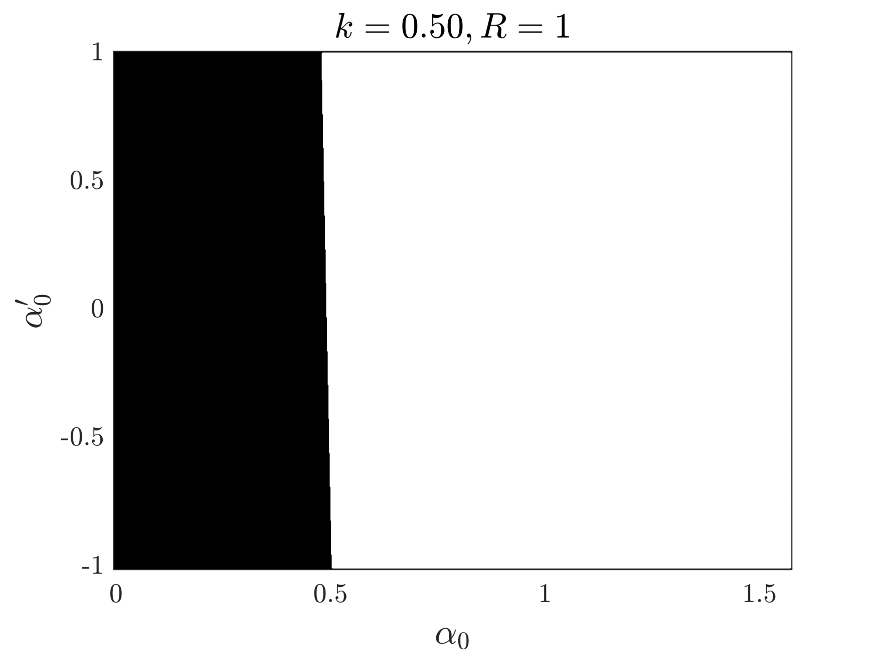} \hspace{-5mm}
\includegraphics[scale=0.35]{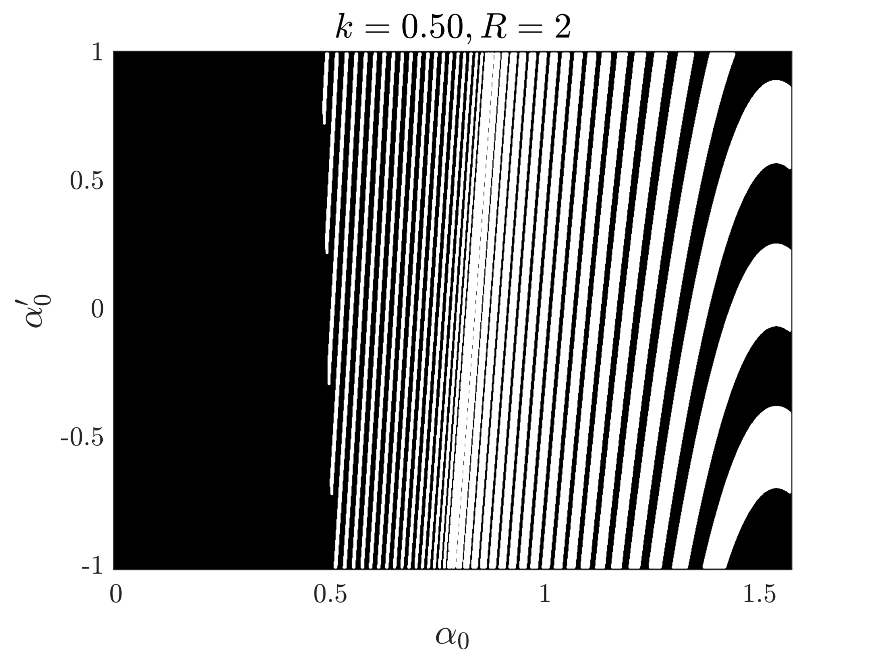} \hspace{-5mm}
\includegraphics[scale=0.35]{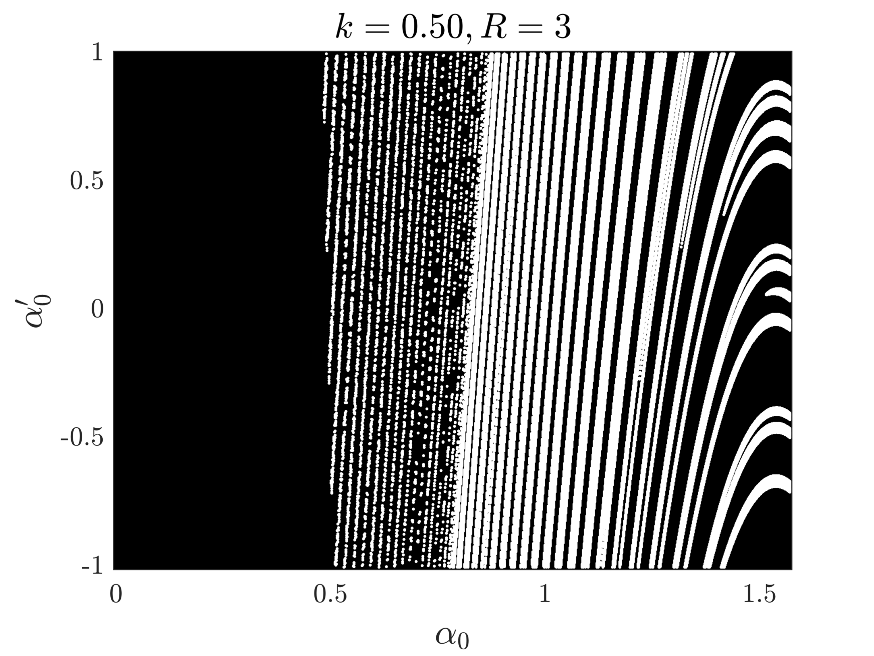}

\includegraphics[scale=0.35]{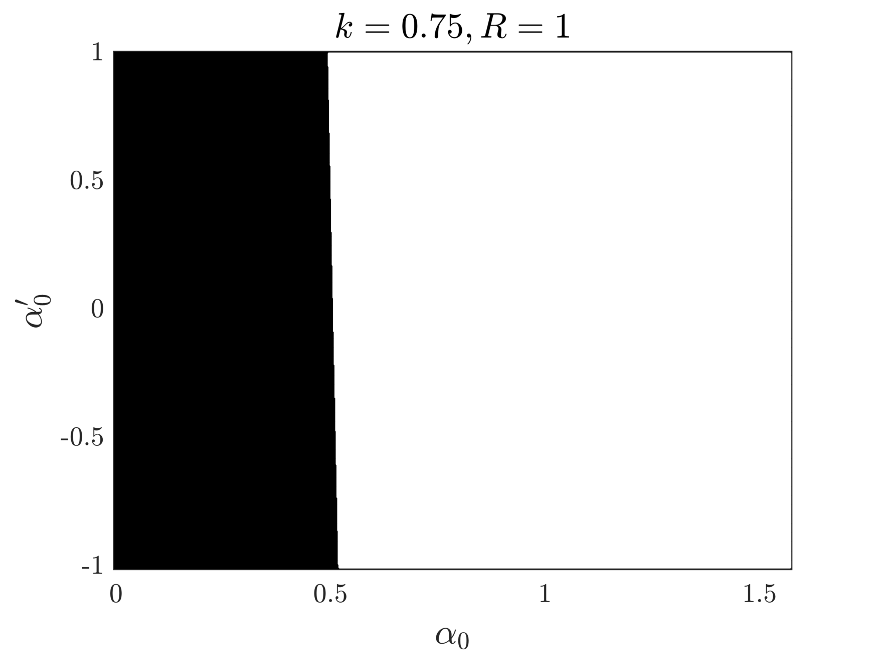} \hspace{-5mm}
\includegraphics[scale=0.35]{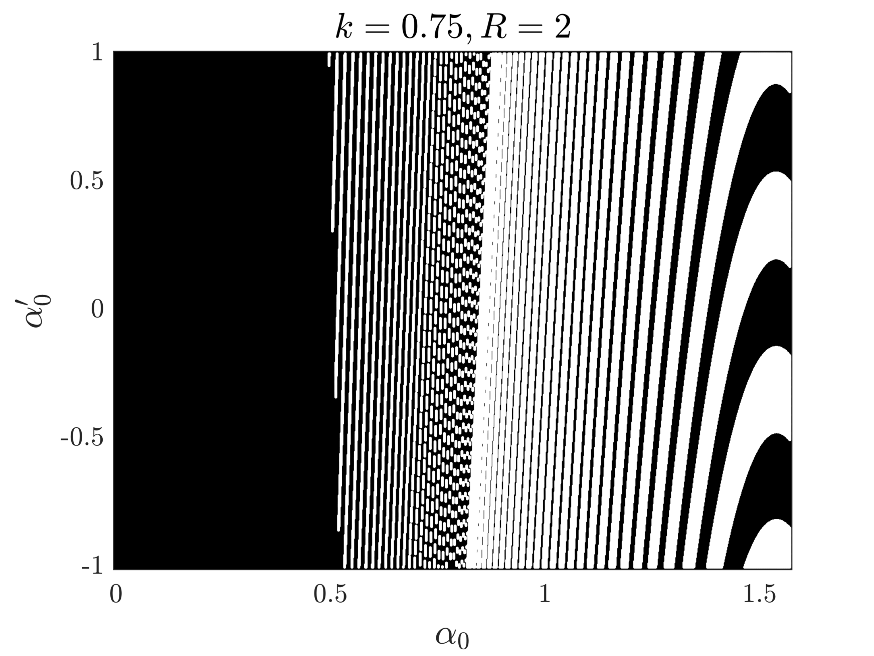} \hspace{-5mm}
\includegraphics[scale=0.35]{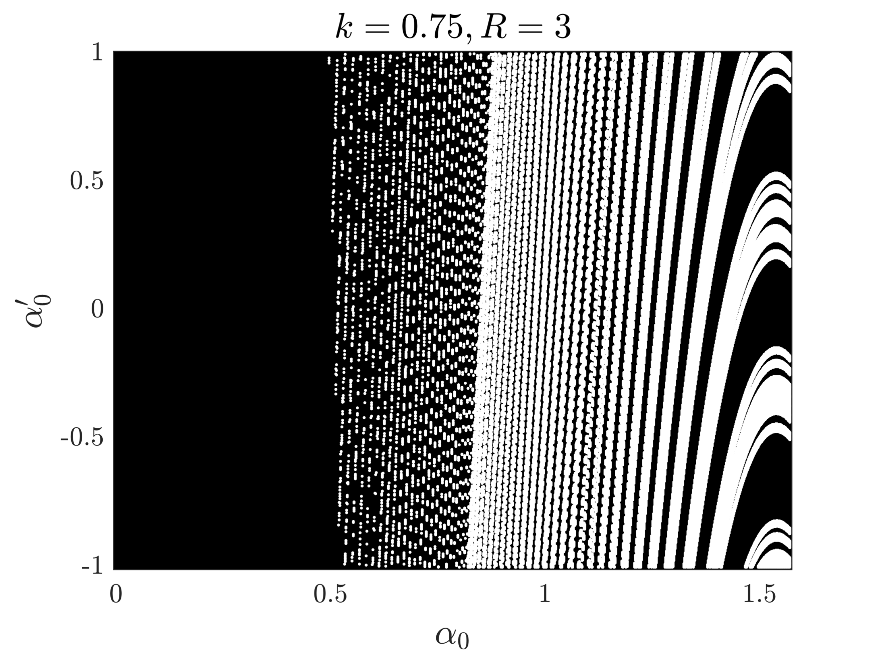}

\caption{Set of initial conditions $(\alpha_0,\alpha_0')$, represented in white color, for which the integrals $I_r < 0$ is satisfied up to the first (top row, $R=1$), second (central row, $R=2$) and third (bottom row, $R=3$) periapsis passages, for $e=0.9$ and $\kappa = 0.25$ (first column), $\kappa = 0.5$ (second row) and $\kappa = 0.75$ (right column). }
\label{fig:SOP_Intneg_k_05}
\end{figure}

In Fig. \ref{fig:SOP_Intneg_k_05} we plot in white color the initial conditions in the SOP with $e = 0.9$ such that the integrals $I_r$ for $r=0,\ldots,R-1$ are negative for $\kappa = 0.25$ (top row),  $\kappa = 0.5$ (center row) and $\kappa = 0.75$ (bottom row) up to one (left column, $R=1$), two (central column, $R=2$) and three (right column, $R=3$) periapsis passages.
For the computation of the initial conditions fulfilling the `in-phase integral' condition in the SOP, we select as initial angle $\alpha_0 \in [0,\pi/2)$ (this is because $\sin (2 \alpha_0) > 0$, and the gravity-gradient moment at the initial epoch $t_0 = -\pi$ with $f_0 = 0$ is negative). Then, for the detection of the initial conditions fulfilling the `in-phase integral' condition, we compute the integrals $I_r$, with $r=0,\ldots,R-1$ at each periapsis passages. If the orbit associated to an initial conditions is characterized by a negative value of $I_r$ for $r=0,\ldots,R-1$, then we store the initial datum. 
We notice that the regions of the initial conditions fulfilling the `in-phase integral' condition present a similar behavior as in the DM. In particular, at the second periapsis passage these regions are distributed in curved stripes. We remark that the length of these stripes for a fixed value of $\alpha_0$ is compatible with the length found for the DM (i.e., $1/4$). The main difference between the DM and the SOP is the absence of symmetry of the initial conditions regions satisfying the in-phase condition after three periapsis passages.
Moreover, we notice that the stripes are more curved around $\alpha_0 = \pi/2$ as the inertia ratio grows. As for the DM, for $R=2$ the regions of the initial conditions is distributed in curved stripes of length $1/4$ for a fixed value of $\alpha_0$. Moreover, after the third periapsis passage, the principal stripes (found after the second periapsis passage) are splitted in more stripes. It seems that the number of strip splits increases as the value of the inertia ratio.

\begin{figure}[htbp!]
	\centering
\includegraphics[scale=0.4]{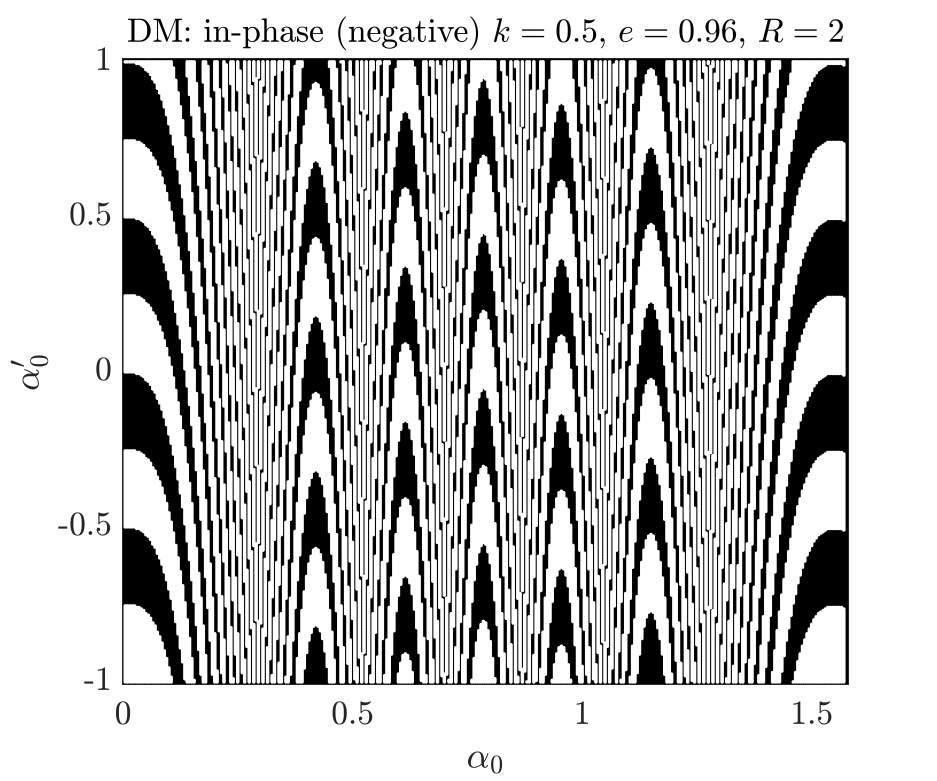}
	\includegraphics[scale=0.4]{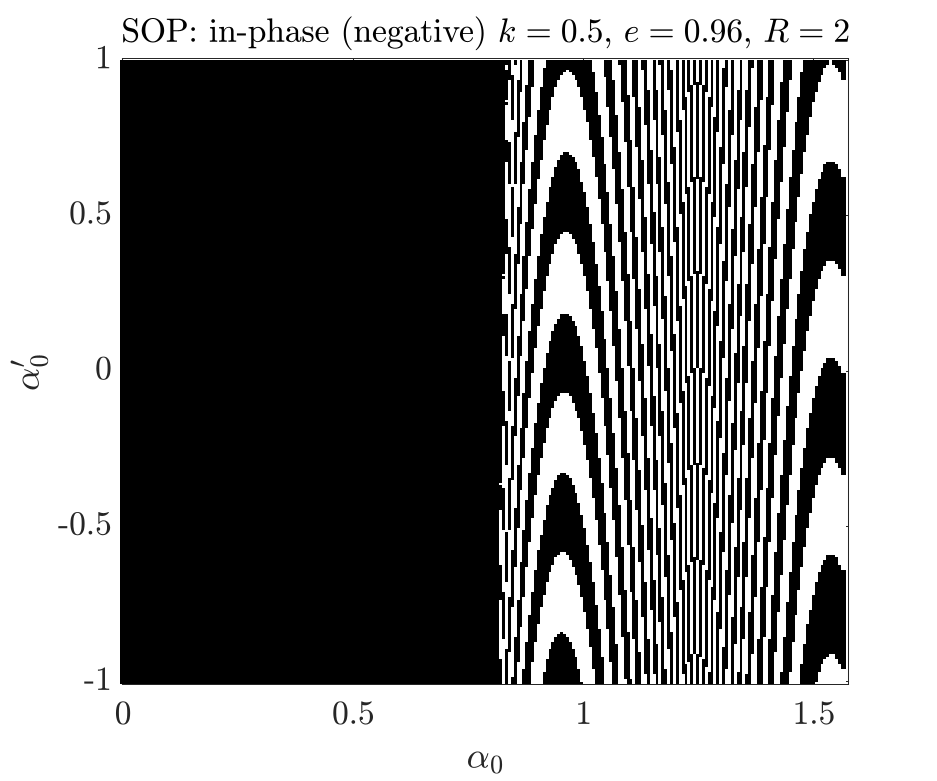}

\includegraphics[scale=0.4]{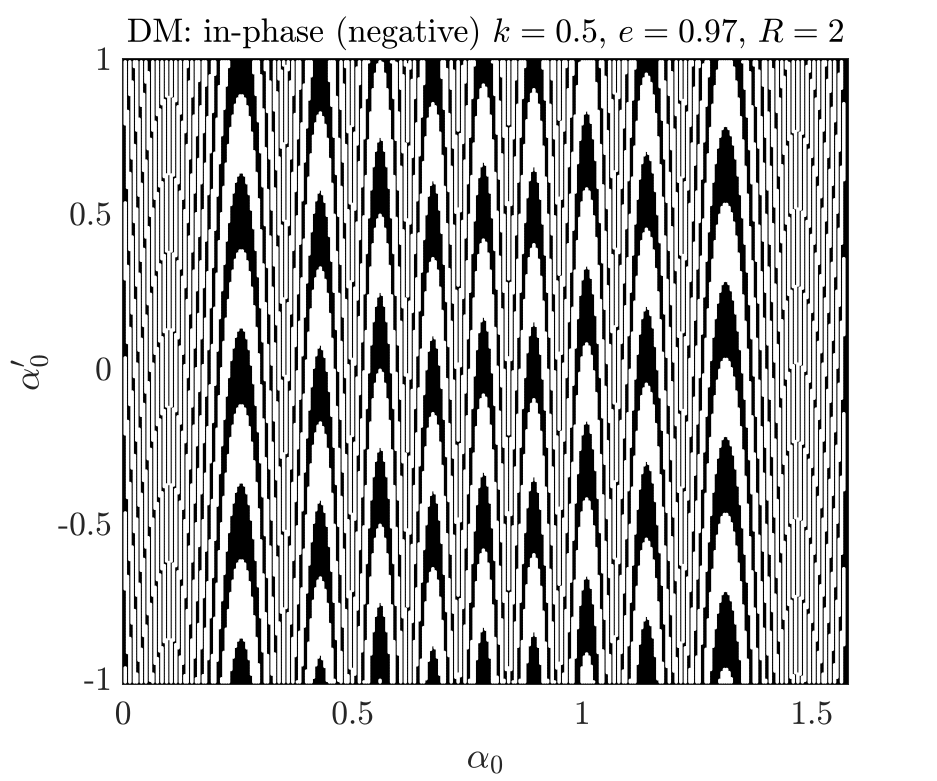}
	\includegraphics[scale=0.4]{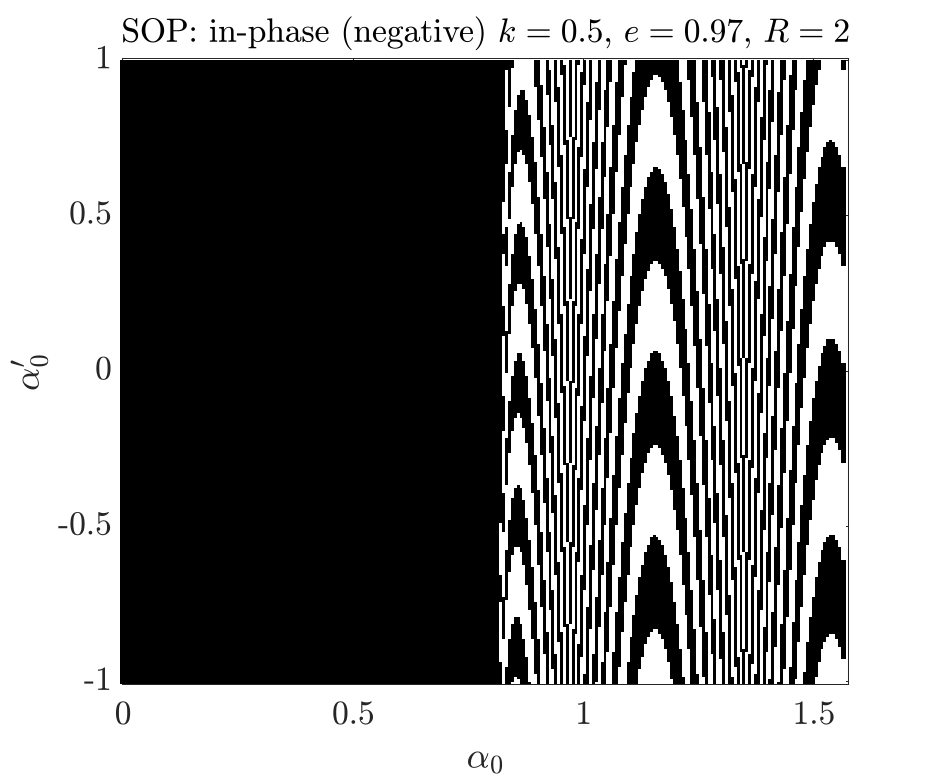}

\includegraphics[scale=0.4]{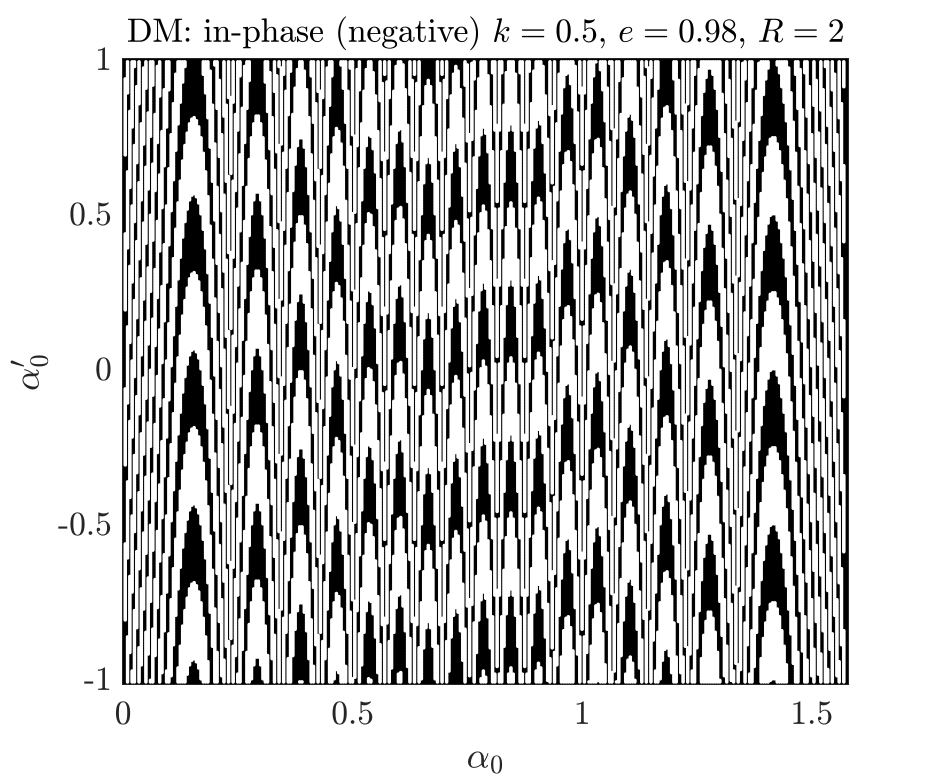}
	\includegraphics[scale=0.4]{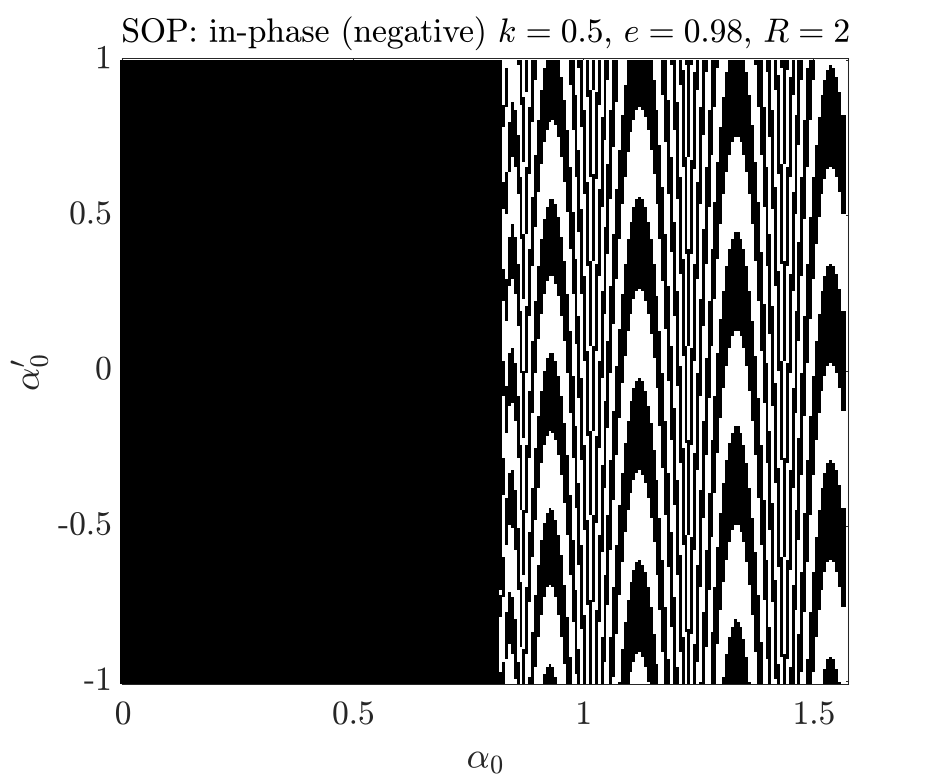}

\includegraphics[scale=0.4]{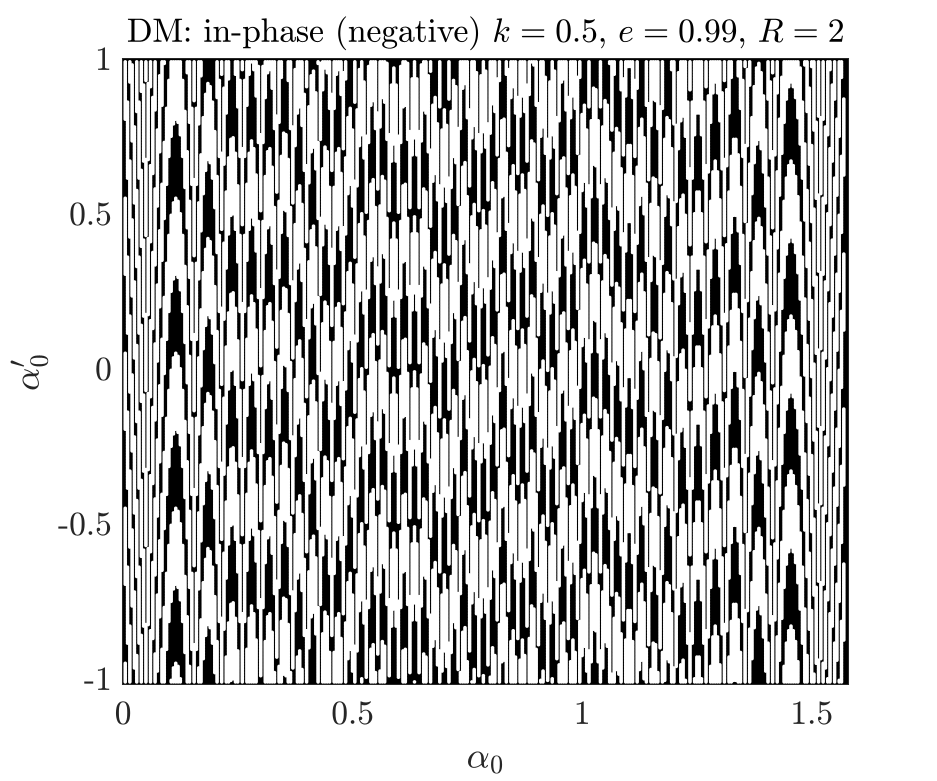}
	\includegraphics[scale=0.4]{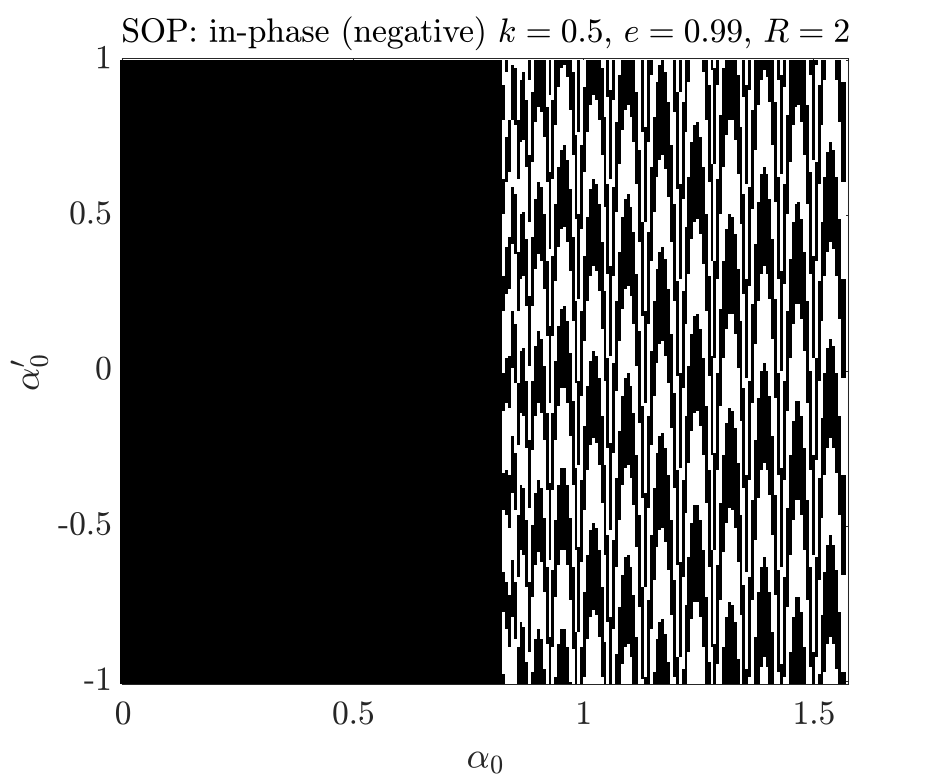}
\caption{Set of initial conditions $(\alpha_0,\alpha_0 ')$ for the discrete map (left column) and the spin-orbit problem (right column), represented in white color,  for which $\alpha_2  '- \alpha_1 ' < 0$ and $\alpha_1 ' -\alpha_0 ' < 0$ for $\kappa = 0.5$, and for the four values of the eccentricity $e=0.96$, $0.97$, $0.98$, and $0.99$ (from top to bottom row).}
\label{fig:inphaseneg_different_ecc_k05}
\end{figure}

Another method to identify initial conditions satisfying the in-phase condition in the SOP, is to verify if the angular velocity $\alpha'$ continues to grow or decrease, i.e., if $\alpha_r ' -\alpha_{r-1} ' < 0$ or $\alpha_r ' -\alpha_{r-1} ' > 0$ for $r =1,\ldots, R$. In Fig. \ref{fig:inphaseneg_different_ecc_k05} we show the initial conditions satisfying the in-phase condition (represented in white color)  with $\alpha_r ' -\alpha_{r-1} ' < 0$ in the DM (left column) and the SOP (right column)  for $r=1,2$, the inertia ratio $\kappa = 0.5$, and $e=0.96$, $0.97$, $0.98$, and $0.99$ (from top to bottom row).  The panels in the right column, show that the  interval of the initial conditions satisfying the in-phase condition present an offset of $\sim 0.83 \, \text{rad}$ with respect to the DM (left column). In particular, for increasing values of the eccentricity, the offset remain the same. Furthermore, even if the two models present differences, we notice that for increasing values of the eccentricity, the number of curved striped that are vertically grouped increase too. Furthermore, in both the DM and the SOP, these groups are displaced between them along the $\alpha_0'$ axis.

\begin{figure}[htbp!]
	\centering
\includegraphics[scale=0.4]{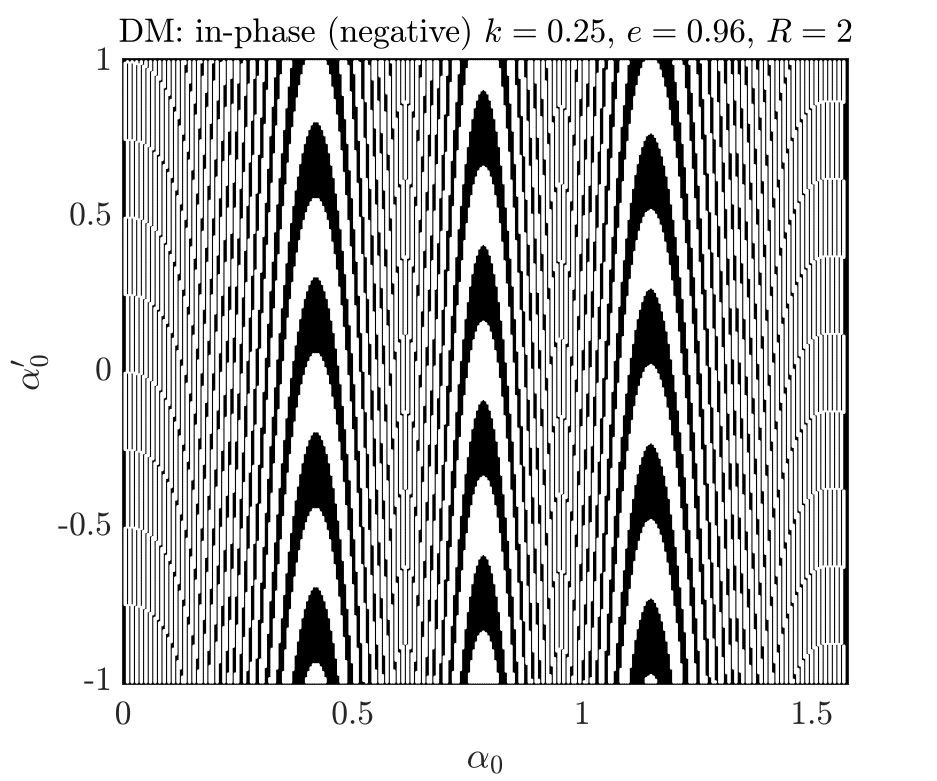}
	\includegraphics[scale=0.4]{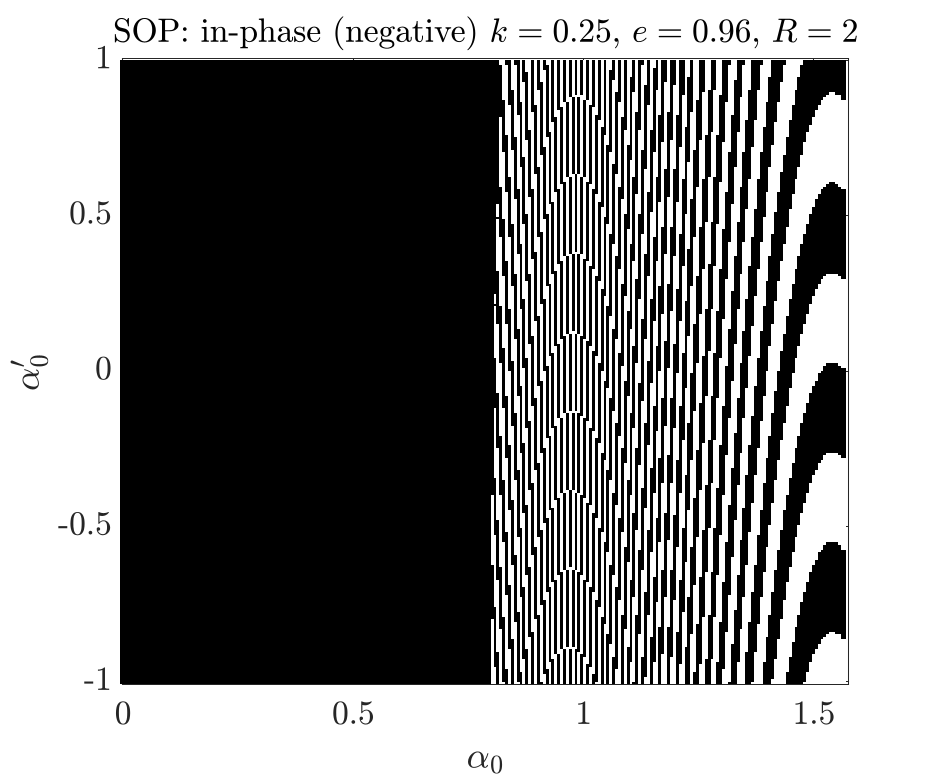}

\includegraphics[scale=0.4]{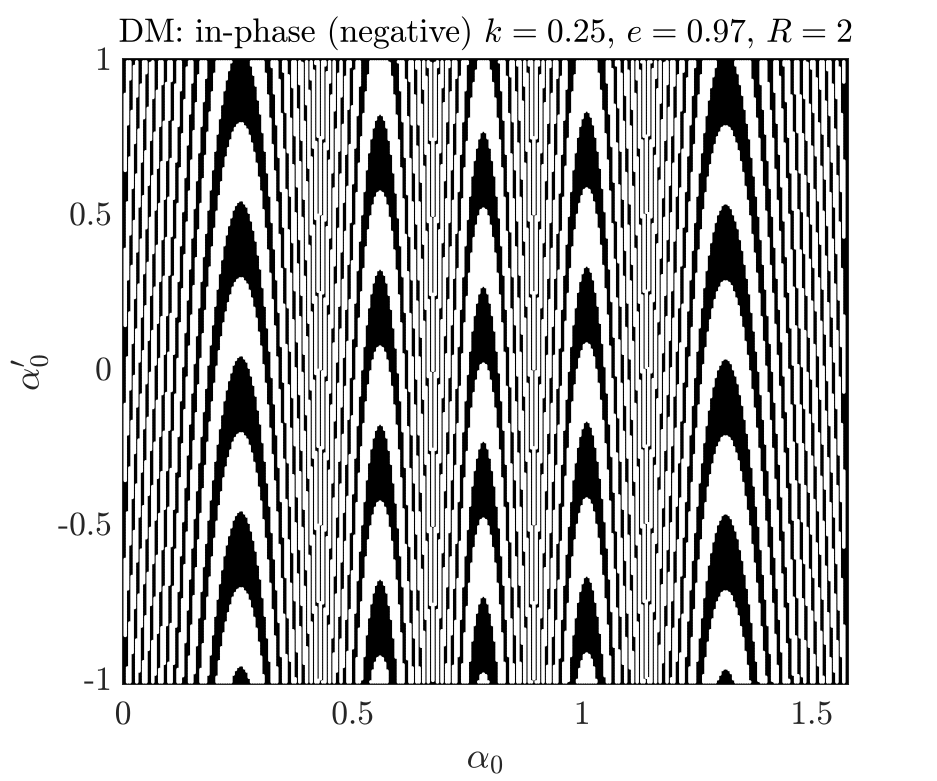}
	\includegraphics[scale=0.4]{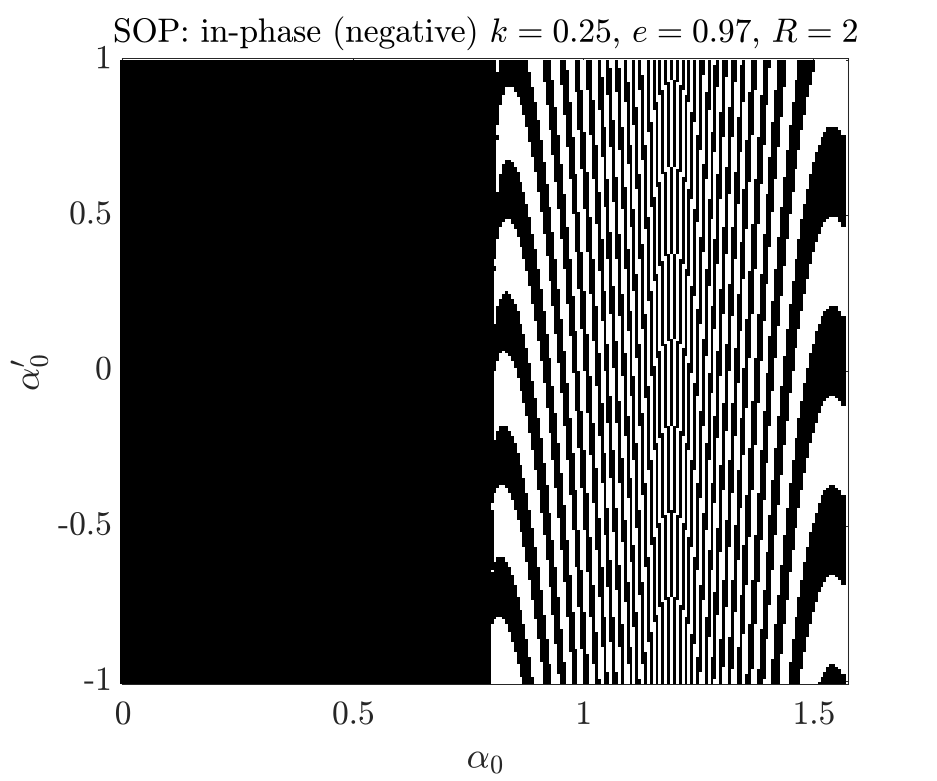}

\includegraphics[scale=0.4]{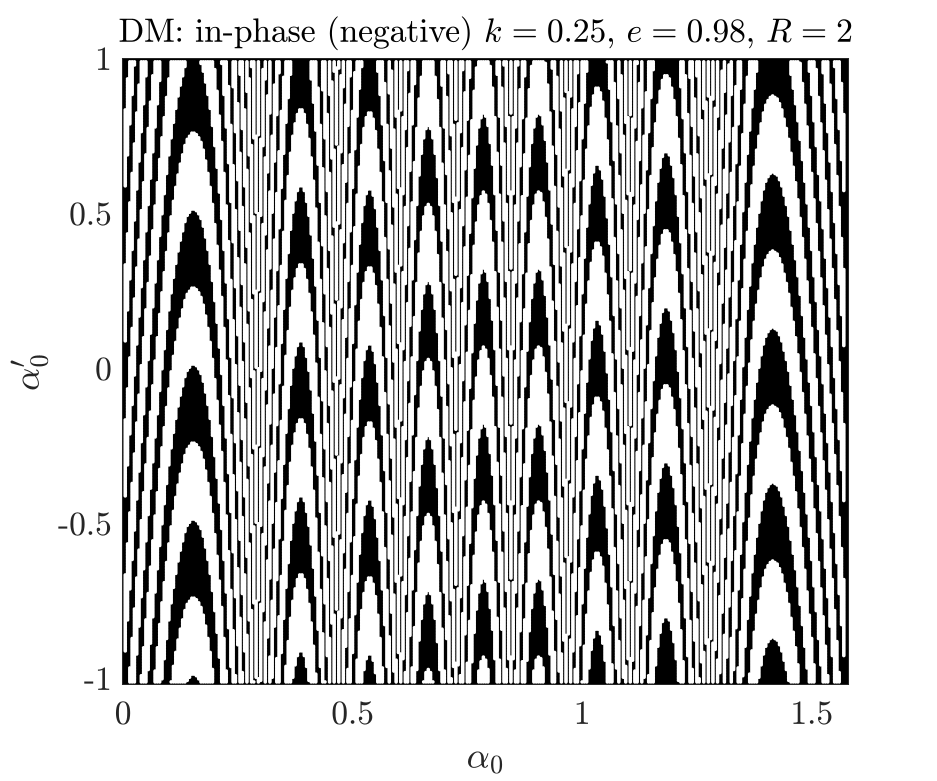}
	\includegraphics[scale=0.4]{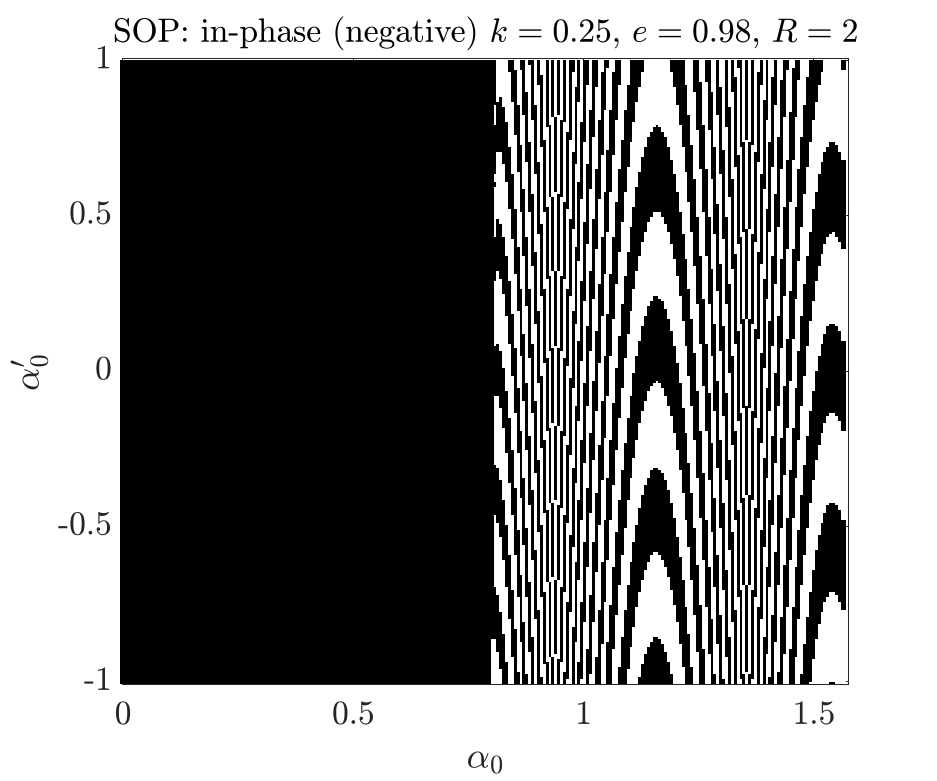}

\includegraphics[scale=0.4]{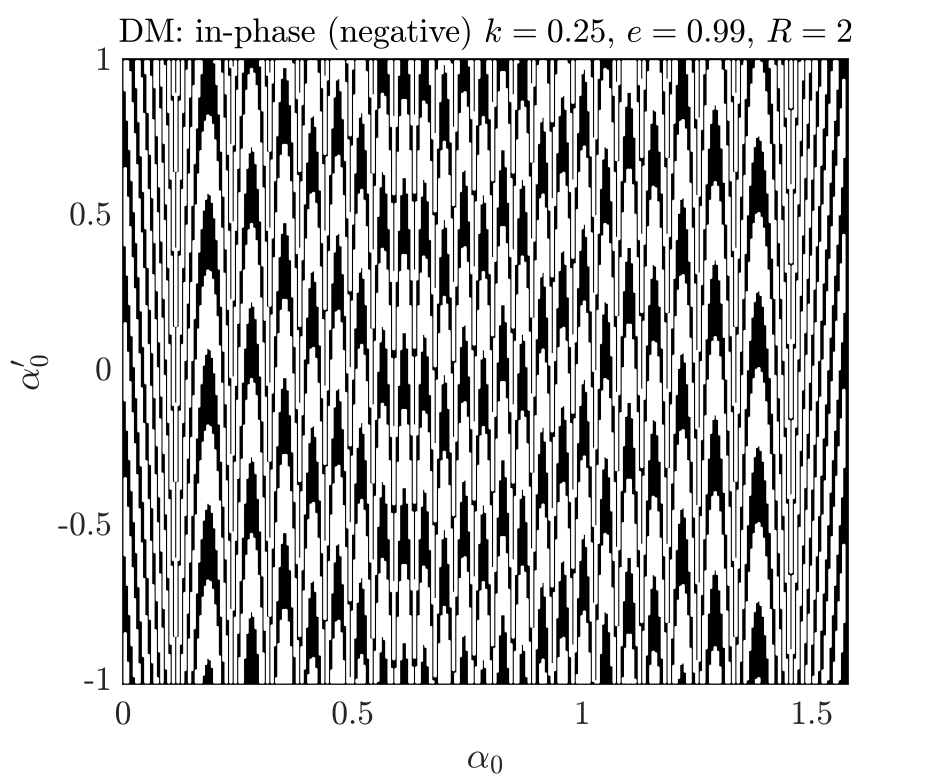}
	\includegraphics[scale=0.4]{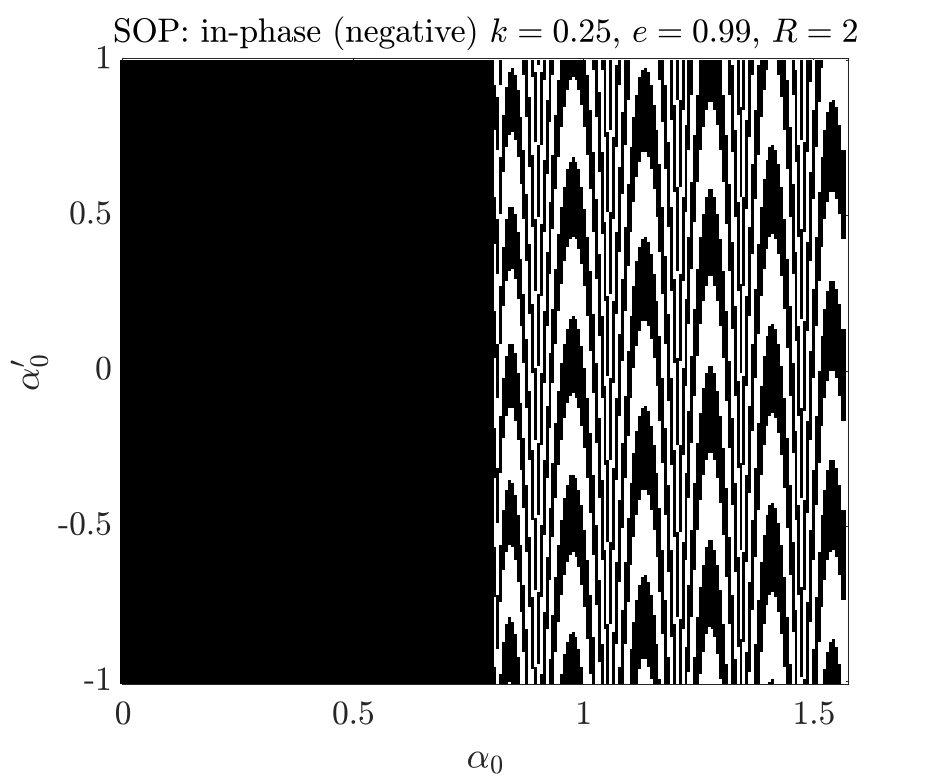}
\caption{Set of initial conditions $(\alpha_0,\alpha_0 ')$ for the discrete map (left column) and the spin-orbit problem (right column), represented in white color,  for which $\alpha_2  '- \alpha_1 ' < 0$ and $\alpha_1 ' -\alpha_0 ' < 0$ for $\kappa = 0.25$, and for the four values of the eccentricity $e=0.96$, $0.97$, $0.98$, and $0.99$ (from top to bottom row).}
\label{fig:inphaseneg_different_ecc_k025}
\end{figure}

\begin{figure}[htbp!]
	\centering
\includegraphics[scale=0.4]{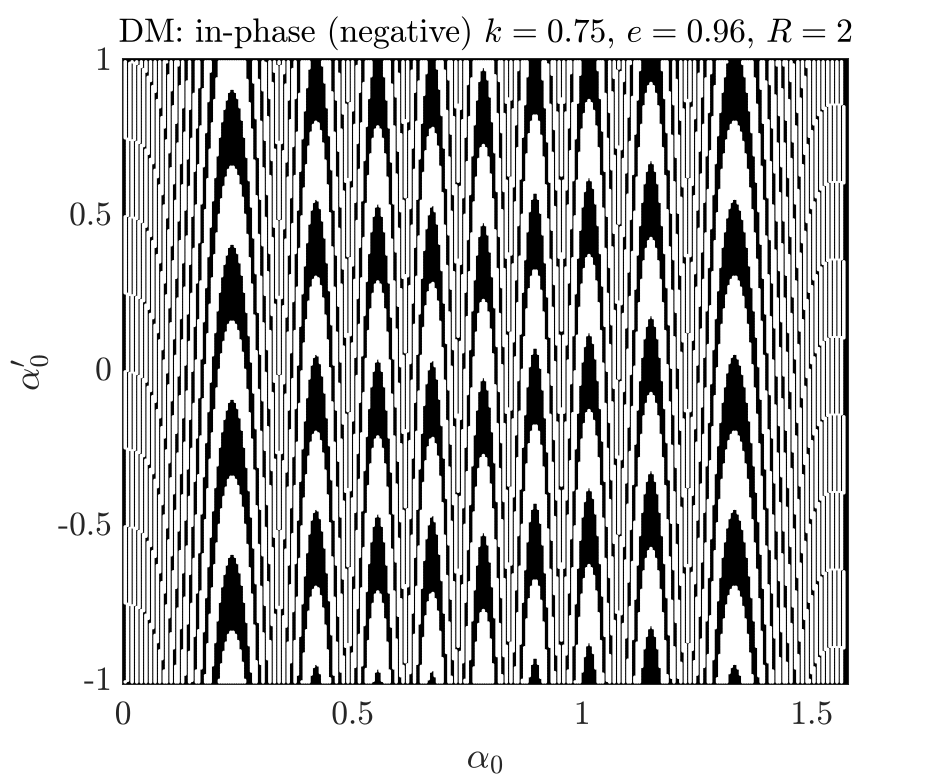}
	\includegraphics[scale=0.4]{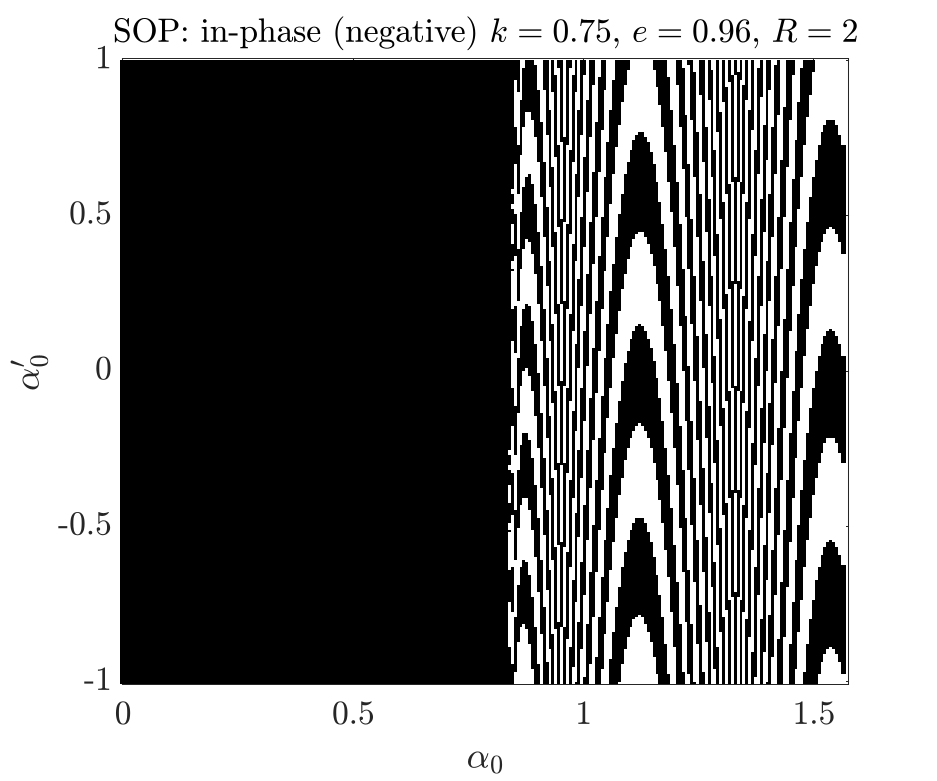}

\includegraphics[scale=0.4]{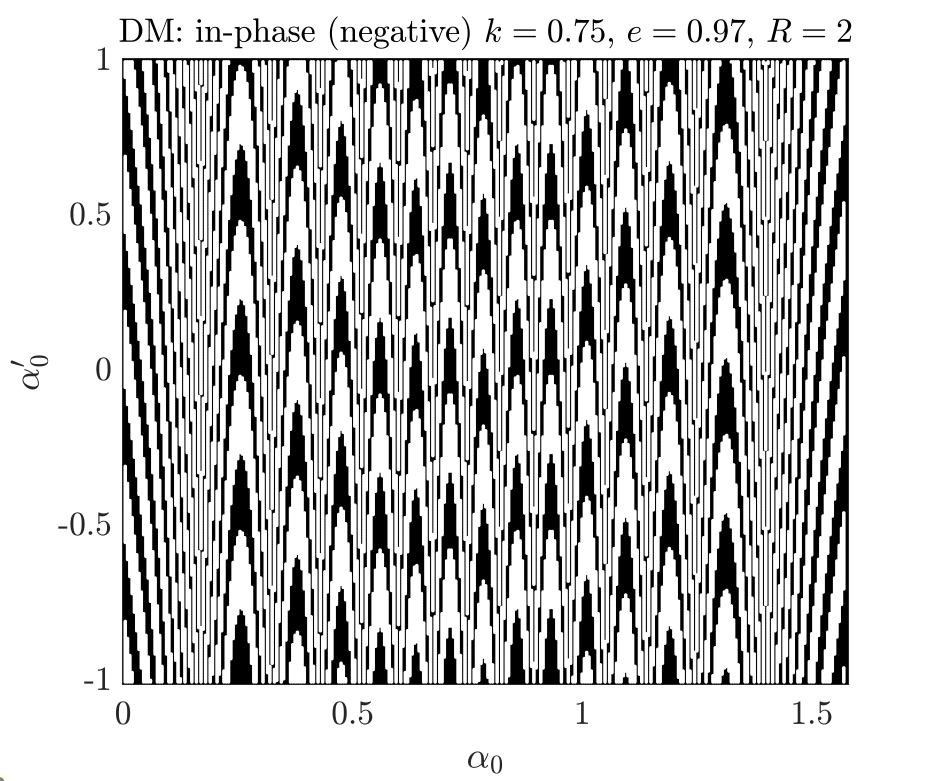}
	\includegraphics[scale=0.4]{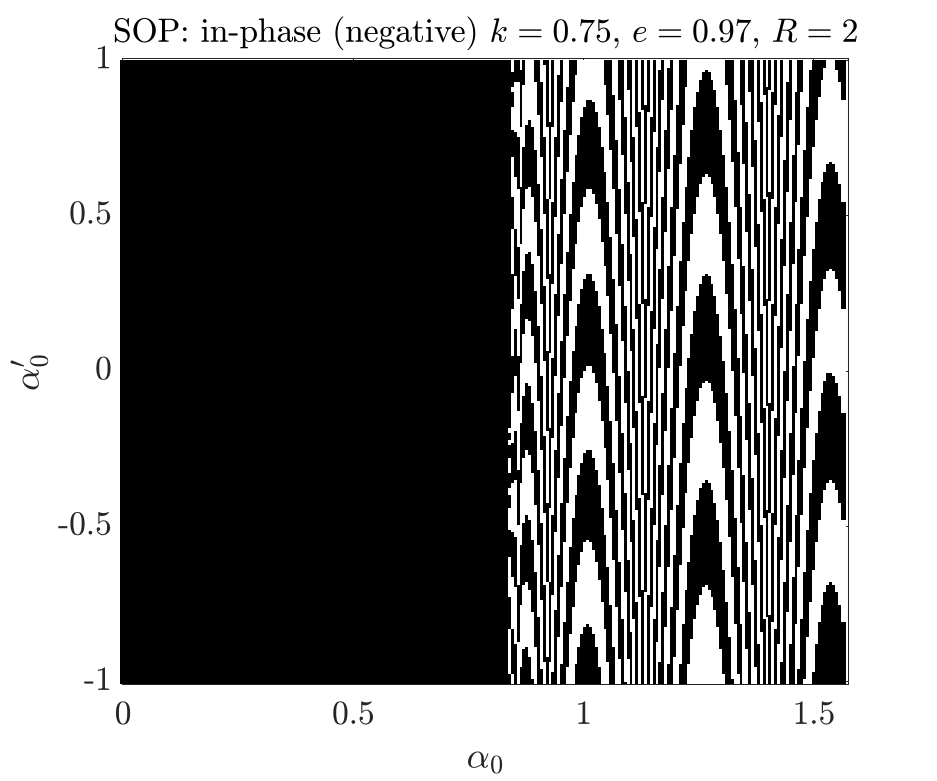}

\includegraphics[scale=0.4]{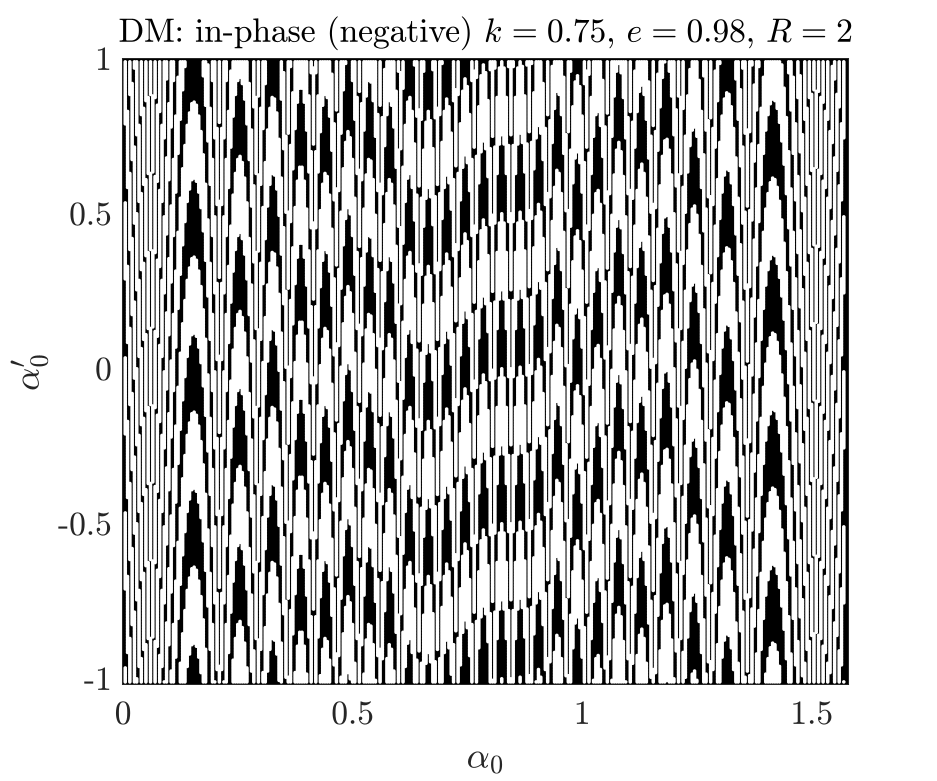}
	\includegraphics[scale=0.4]{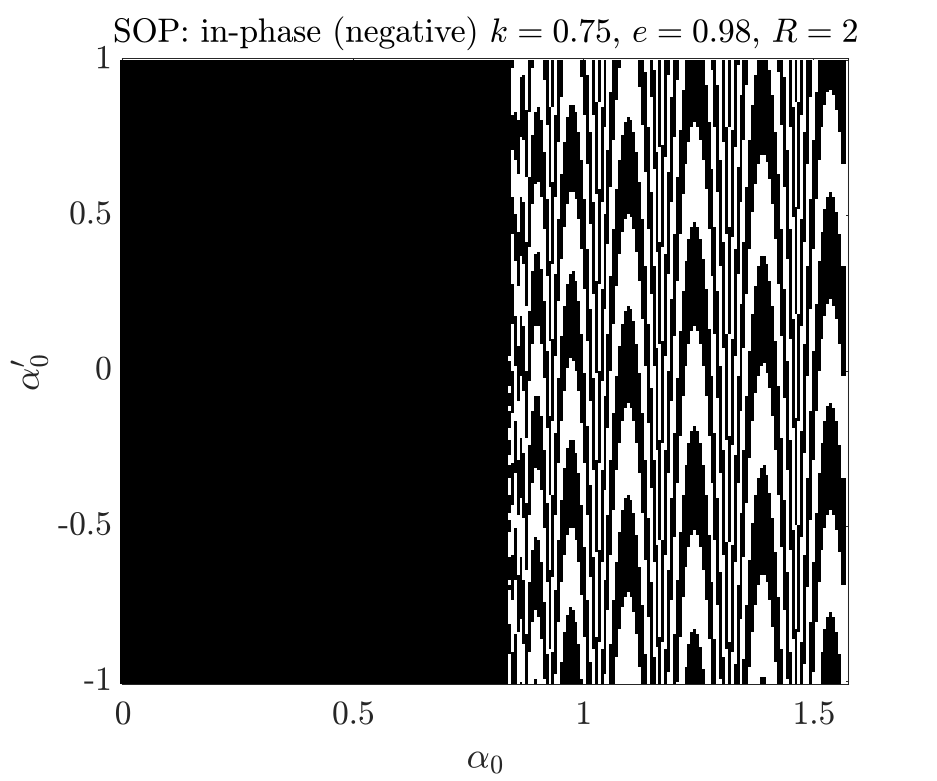}

\includegraphics[scale=0.4]{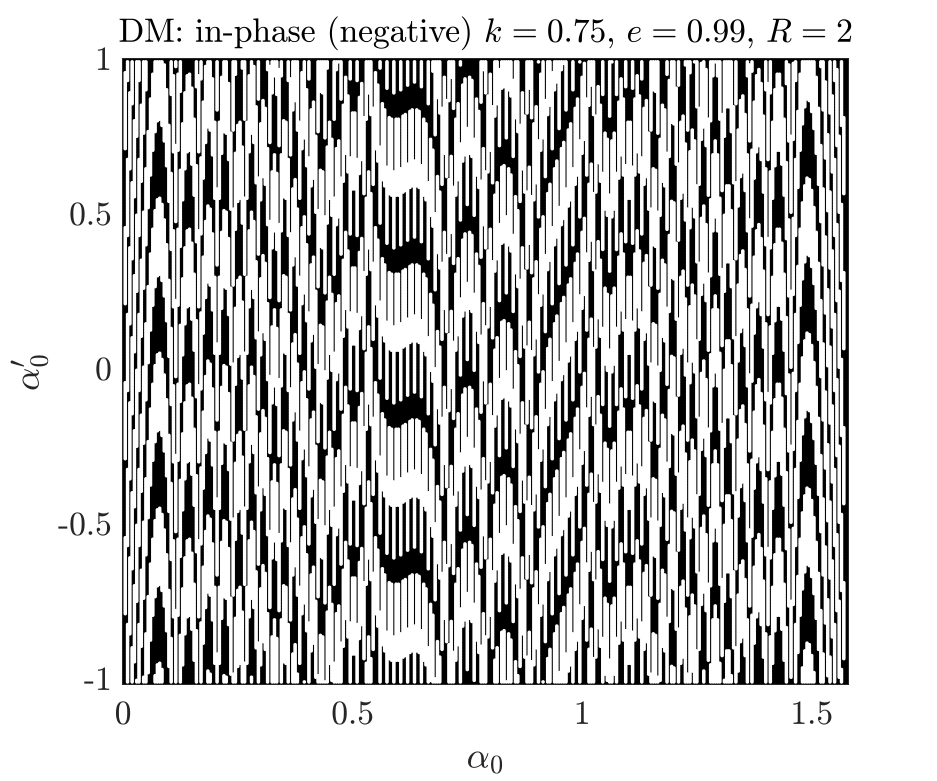}
	\includegraphics[scale=0.4]{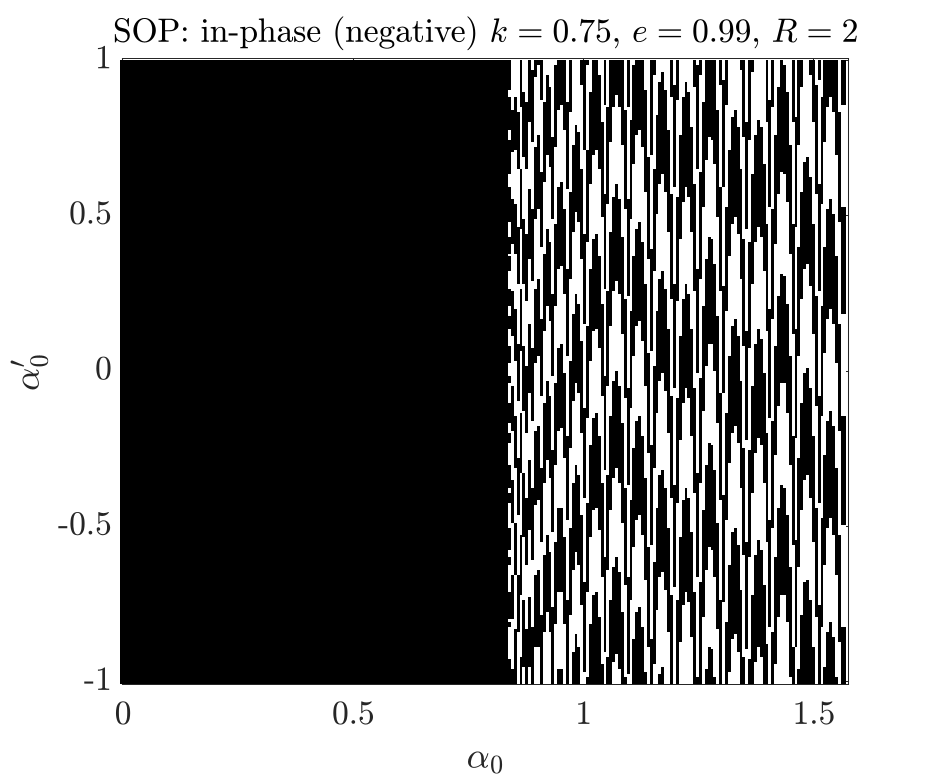}
\caption{Set of initial conditions $(\alpha_0,\alpha_0 ')$ for the discrete map (left column) and the spin-orbit problem (right column), represented in white color,  for which $\alpha_2  '- \alpha_1 ' < 0$ and $\alpha_1 ' -\alpha_0 ' < 0$ for $\kappa = 0.75$, and for the four values of the eccentricity $e=0.96$, $0.97$, $0.98$, and $0.99$ (from top to bottom row).}
\label{fig:inphaseneg_different_ecc_k075}
\end{figure}

In Figures  \ref{fig:inphaseneg_different_ecc_k025} and \ref{fig:inphaseneg_different_ecc_k075}, we plot the initial conditions $(\alpha_0,\alpha_0 ')$ (represented in white color), satisfying the in-phase condition for $\alpha_r ' -\alpha_{r-1} ' <0$ with $r=1,2$ for $\kappa = 0.25$ (Fig. \ref{fig:inphaseneg_different_ecc_k025}) and $\kappa = 0.75$ (Fig. \ref{fig:inphaseneg_different_ecc_k075}) in the discrete map (left column) and spin-orbit problem (right column).  From a comparison between the in-phase initial conditions in the SOP with $\kappa=0.25$ (Fig. \ref{fig:inphaseneg_different_ecc_k025}, right column) and $\kappa = 0.75$ (Fig. \ref{fig:inphaseneg_different_ecc_k025}, right column), we notice that the offset in $\alpha_0$ is $\sim \, 0.81 \, \text{rad}$ and $\sim 0.84 \, \text{rad}$. Hence, the offset amount is $\sim 0.03 \, \text{rad}$, i.e., $\sim 1.7 \, \text{deg}$. We remark that, for a given value of the inertia ratio $\kappa$, the offset along $\alpha_0$ for different eccentricity values remain the same.

Hence, in conclusion, from Figures \ref{fig:inphaseneg_different_ecc_k025}, \ref{fig:inphaseneg_different_ecc_k05}, and \ref{fig:inphaseneg_different_ecc_k075} we notice that: i) in the SOP, the offset of the interval along $\alpha_0$, where the in-phase initial conditions are identified, increases for increasing values of the inertia ratio $\kappa$.  In the DM, this interval along $\alpha_0$ is always the same, according to Eq. \eqref{eq:pos_cond}; ii) in the SOP, all the principal curved stripes have the same dimension. In the DM, the principal curved stripes have not the same dimension. In particular, the greatest ones are located nearby the extrema of the $\alpha_0$ range. Furthermore, even if the distribution is not the same, we stress that the number of the principal curved stripes along the $\alpha_0 '$ axis is the same for the DM and the SOP.  

\begin{figure}[ht!]
	\centering
	\includegraphics[scale=0.31]{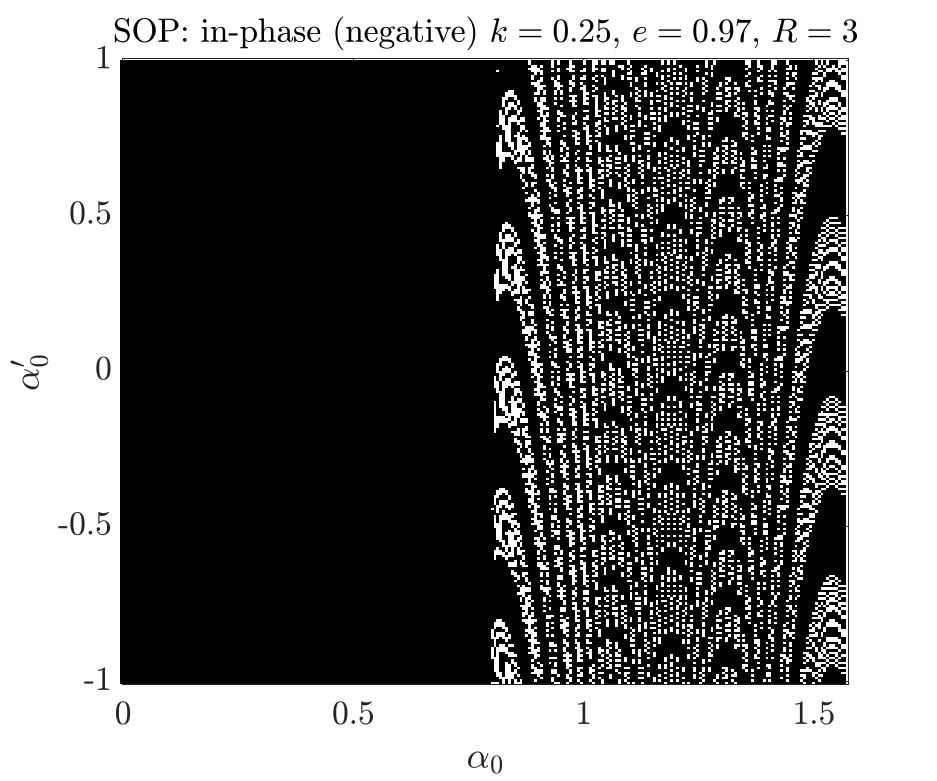}
	\includegraphics[scale=0.31]{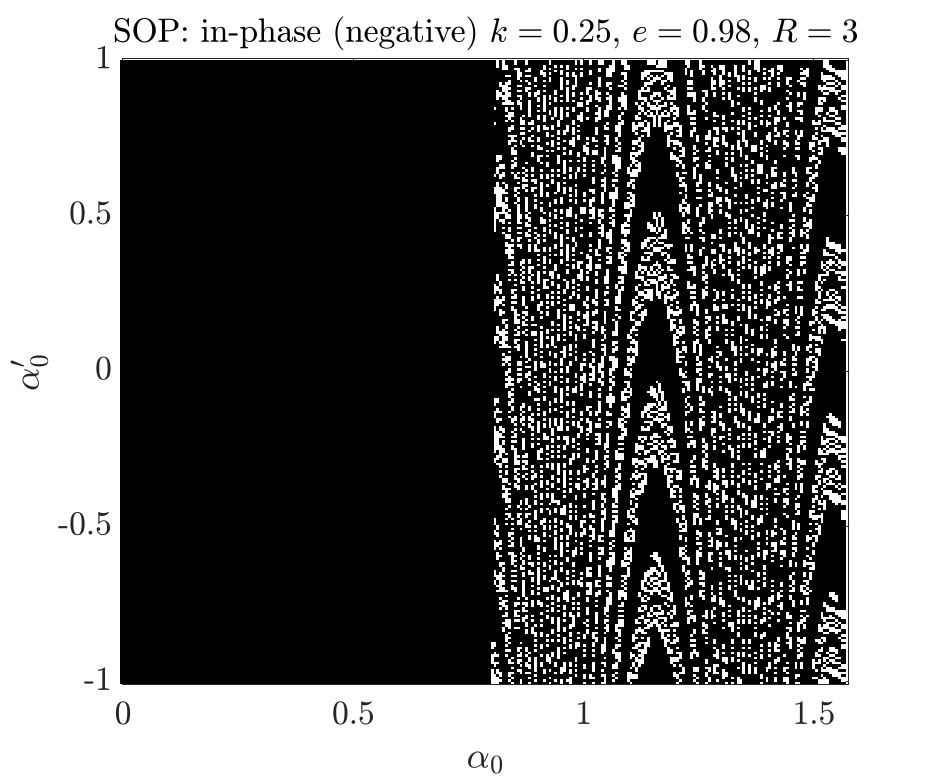}
	\includegraphics[scale=0.31]{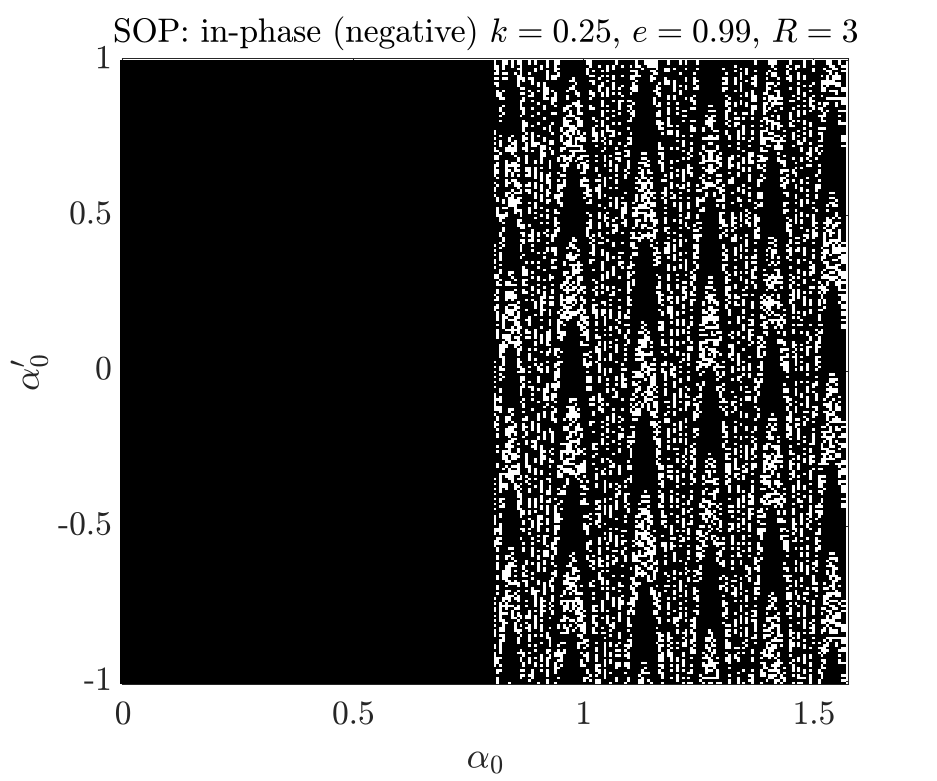}

\includegraphics[scale=0.31]{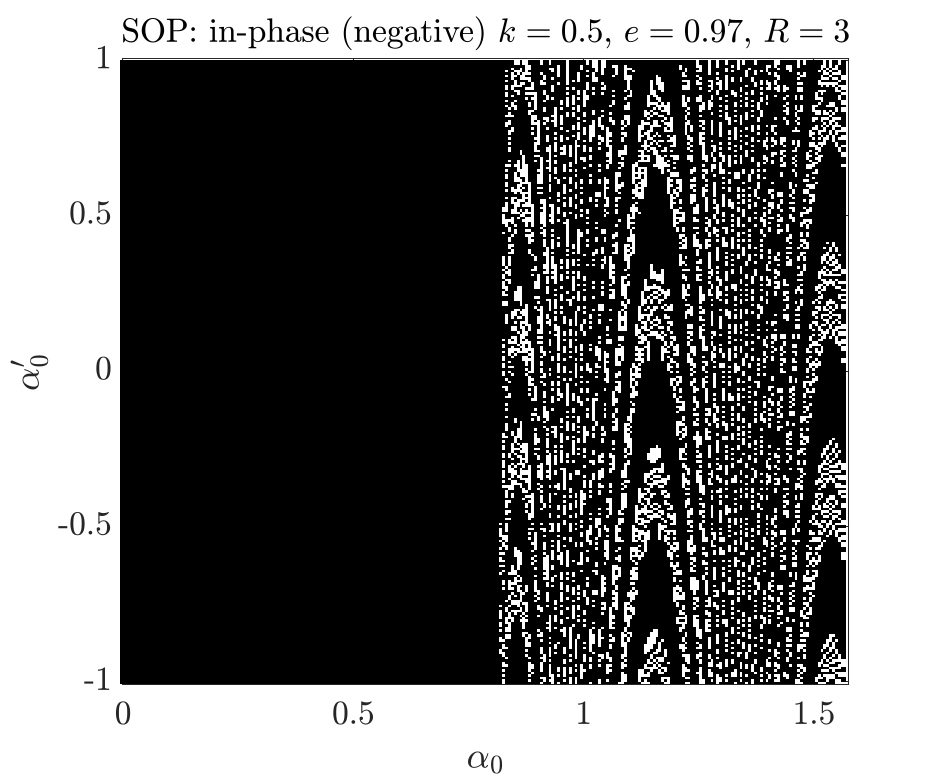}
	\includegraphics[scale=0.31]{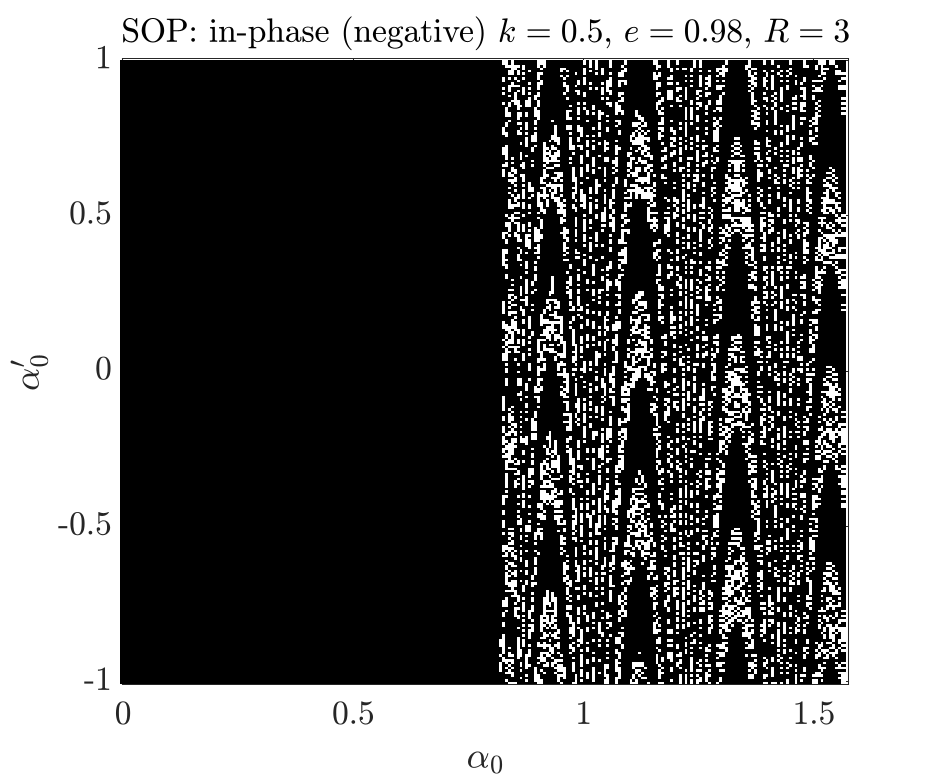}
	\includegraphics[scale=0.31]{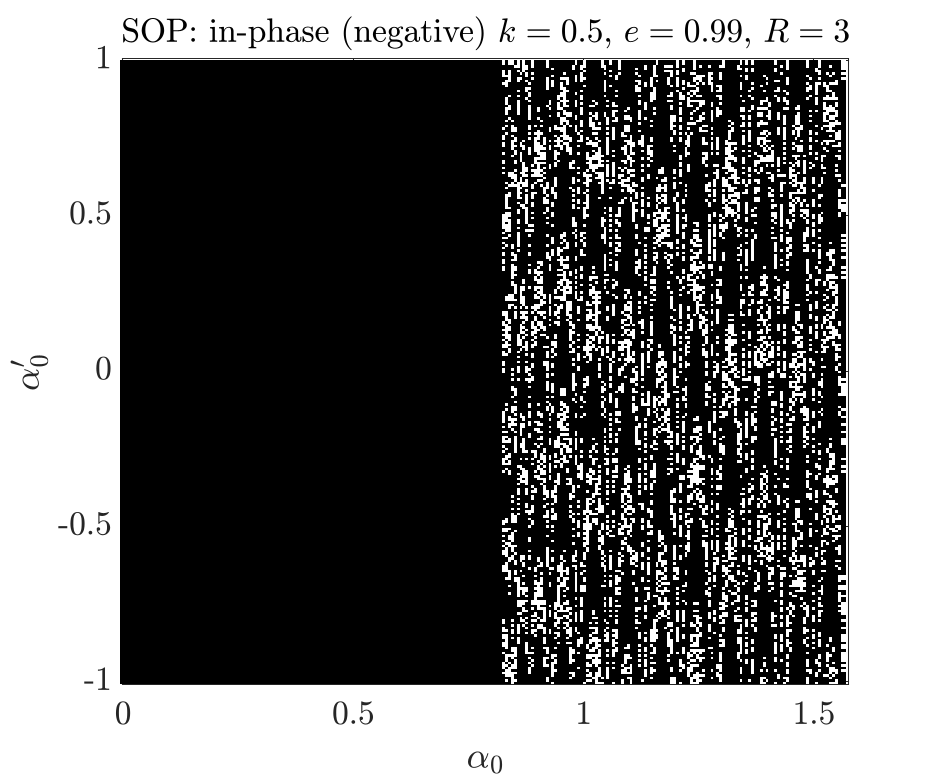}

\includegraphics[scale=0.31]{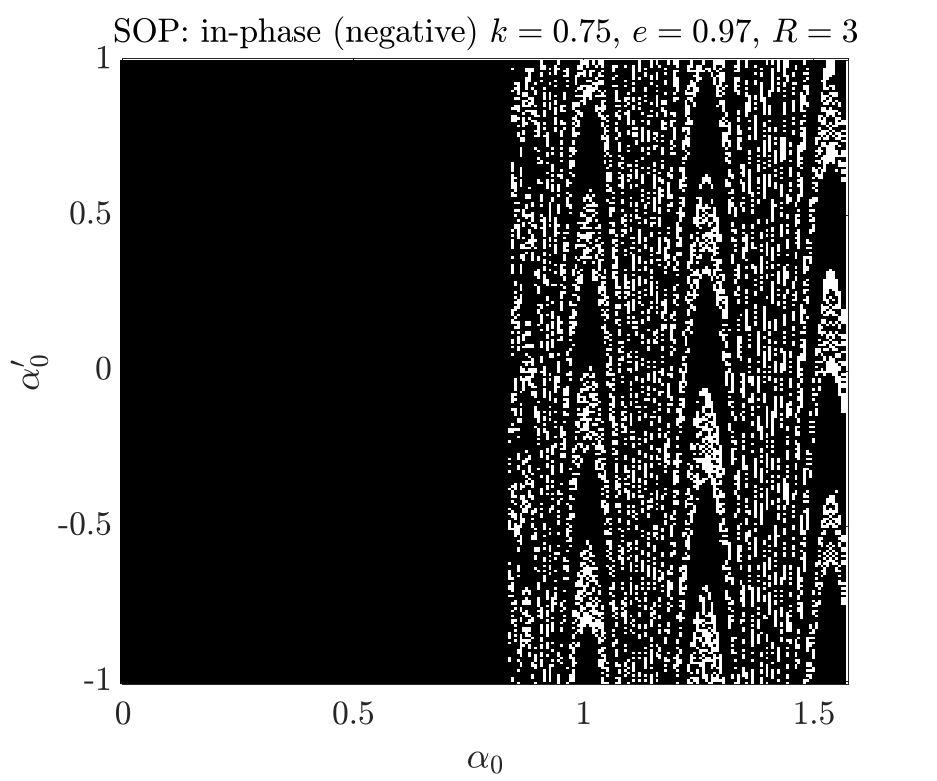}
	\includegraphics[scale=0.31]{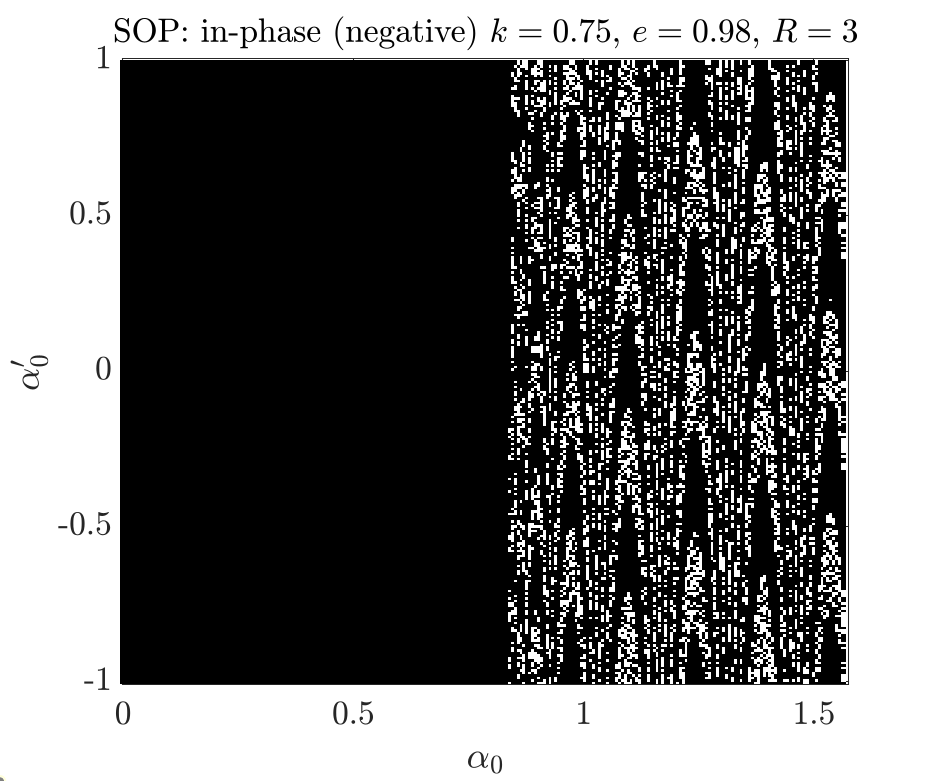}
	\includegraphics[scale=0.31]{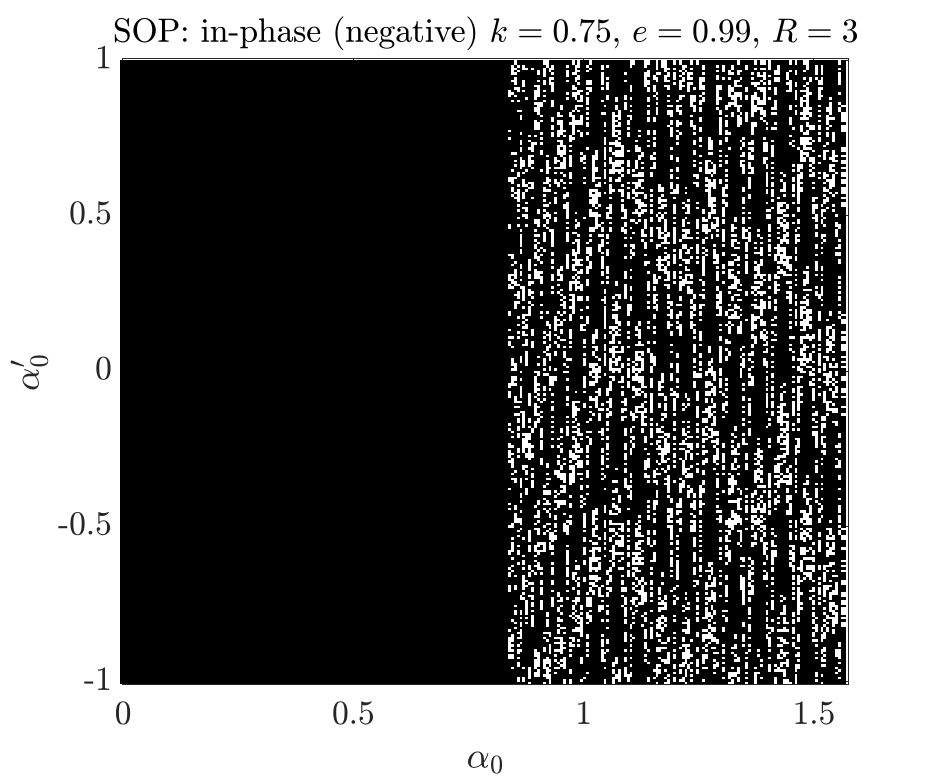}
	\caption{Set of initial conditions $(\alpha_0,\alpha_0')$ for the spin-orbit problem, represented in white color, for which $\alpha_r ' -\alpha_{r-1} < 0$ up to the first three periapsis passages, and for $\kappa = 0.25$ (top row), $\kappa = 0.5$ (central row) and $\kappa = 0.75$ (bottom row) , and for $e=0.97$ (left column), $e=0.98$ (central column) and $\kappa = 0.99$ (right column).}
	\label{fig:SOPinneg_R3}
\end{figure}

\begin{figure}[ht!]
	\centering
	\includegraphics[scale=0.31]{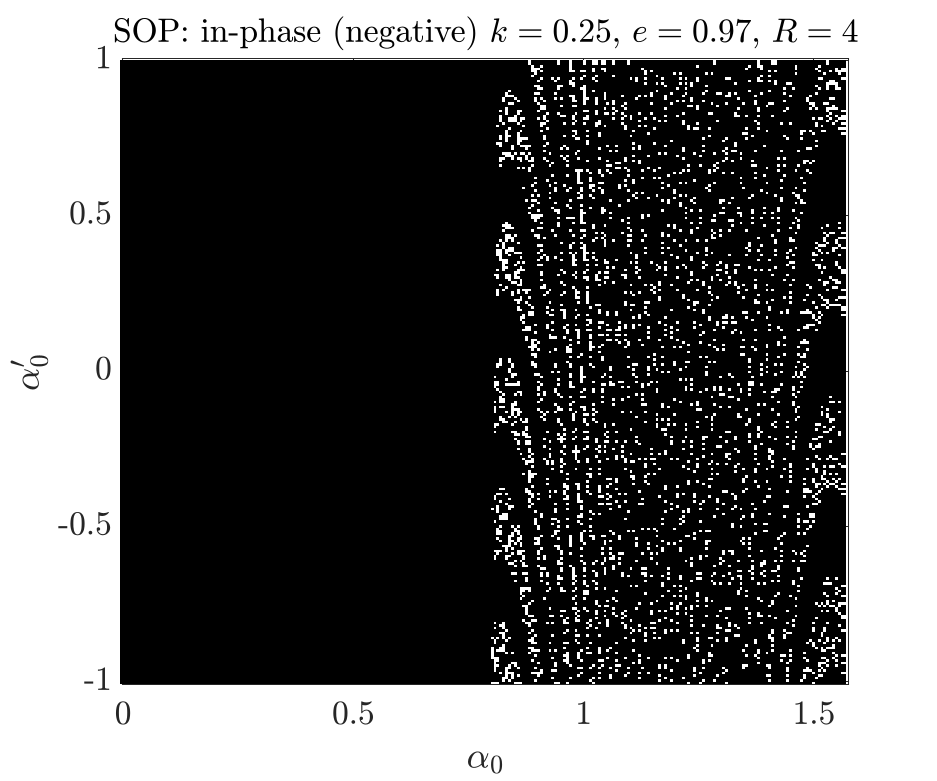}
	\includegraphics[scale=0.31]{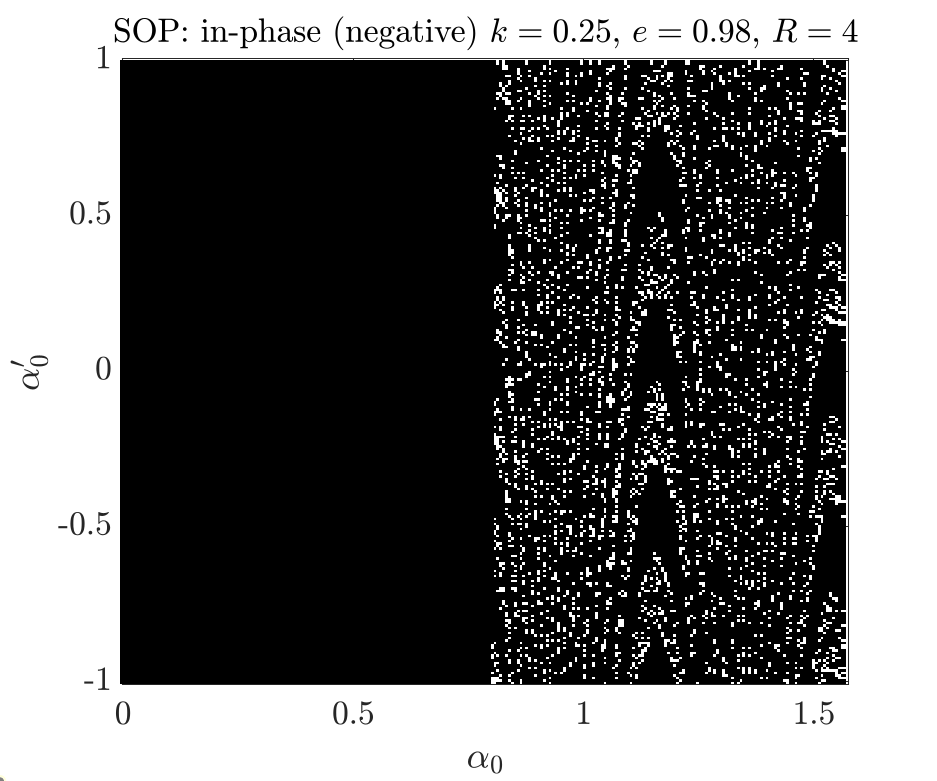}
	\includegraphics[scale=0.31]{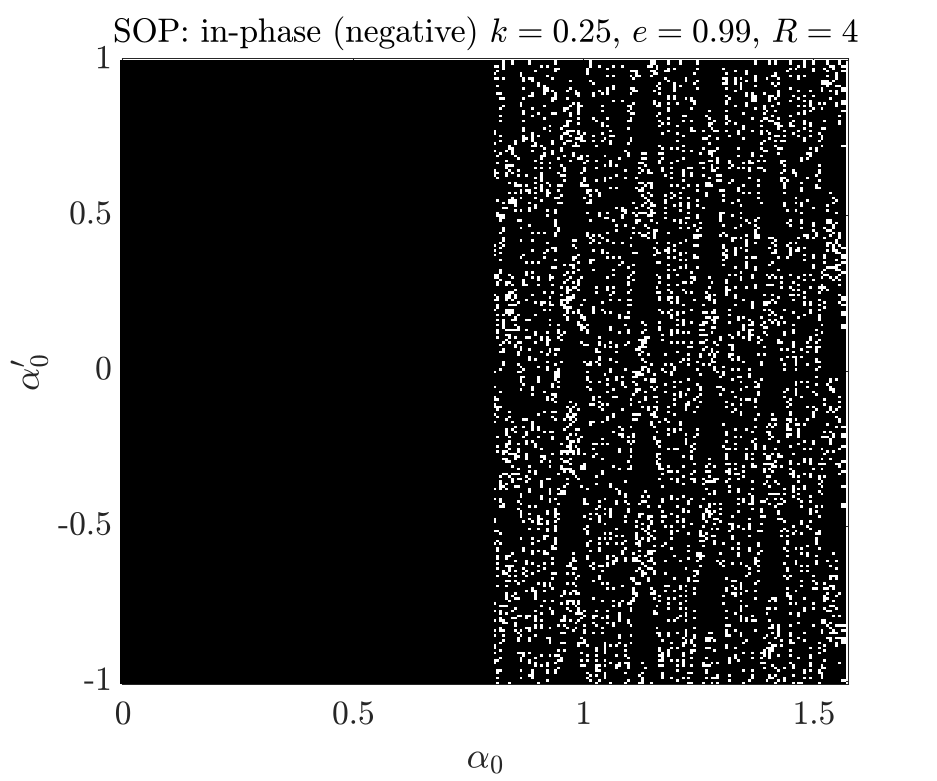}

\includegraphics[scale=0.31]{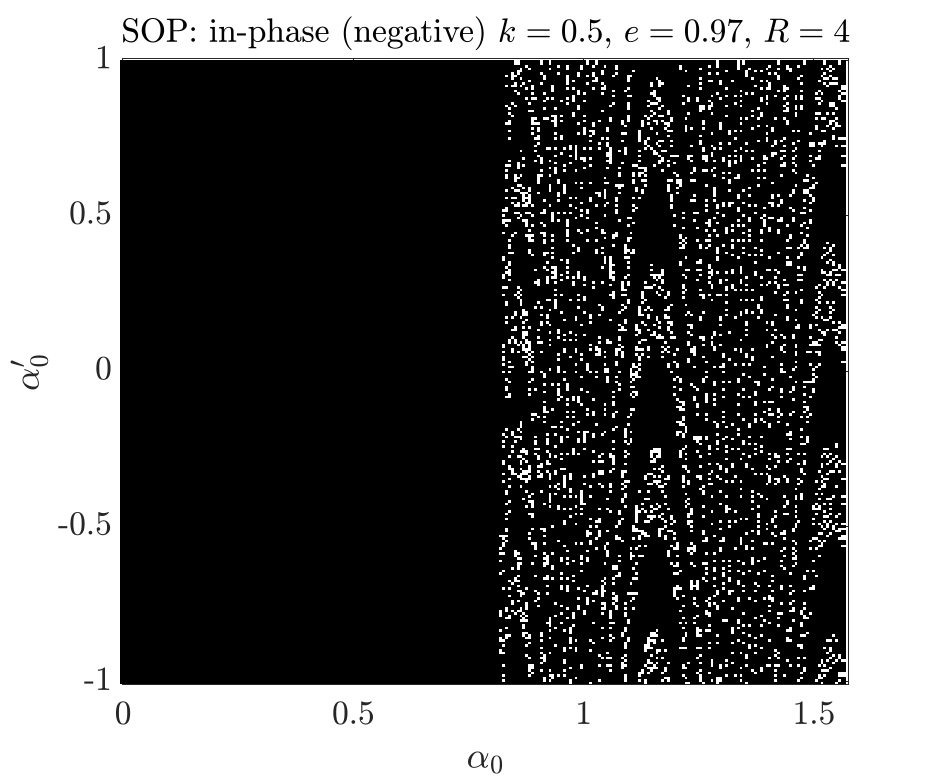}
	\includegraphics[scale=0.31]{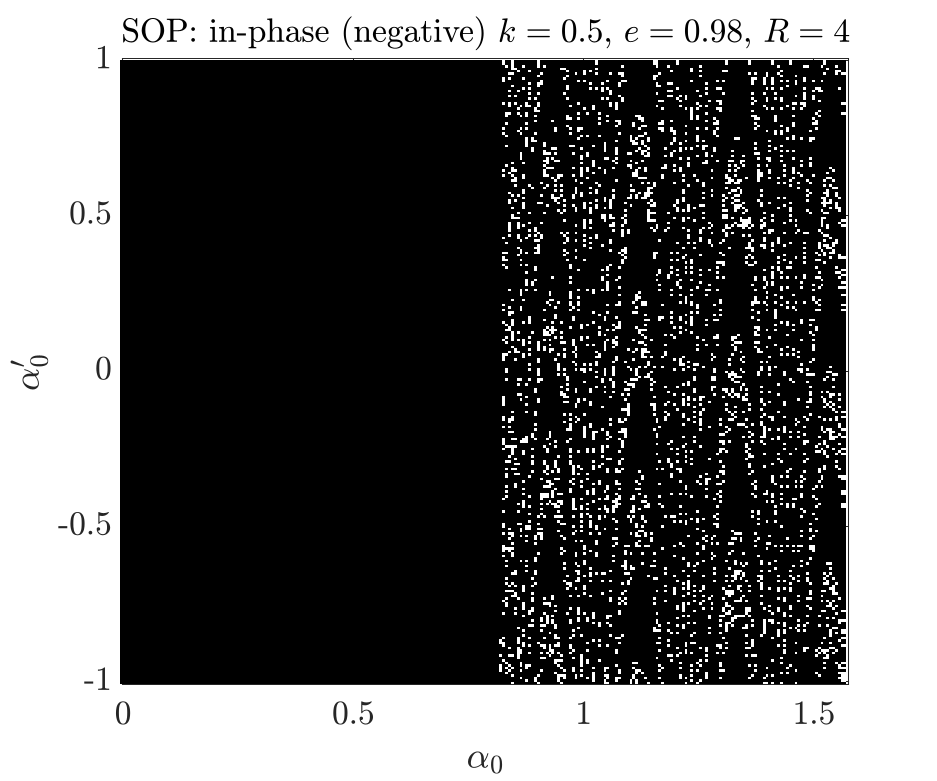}
	\includegraphics[scale=0.31]{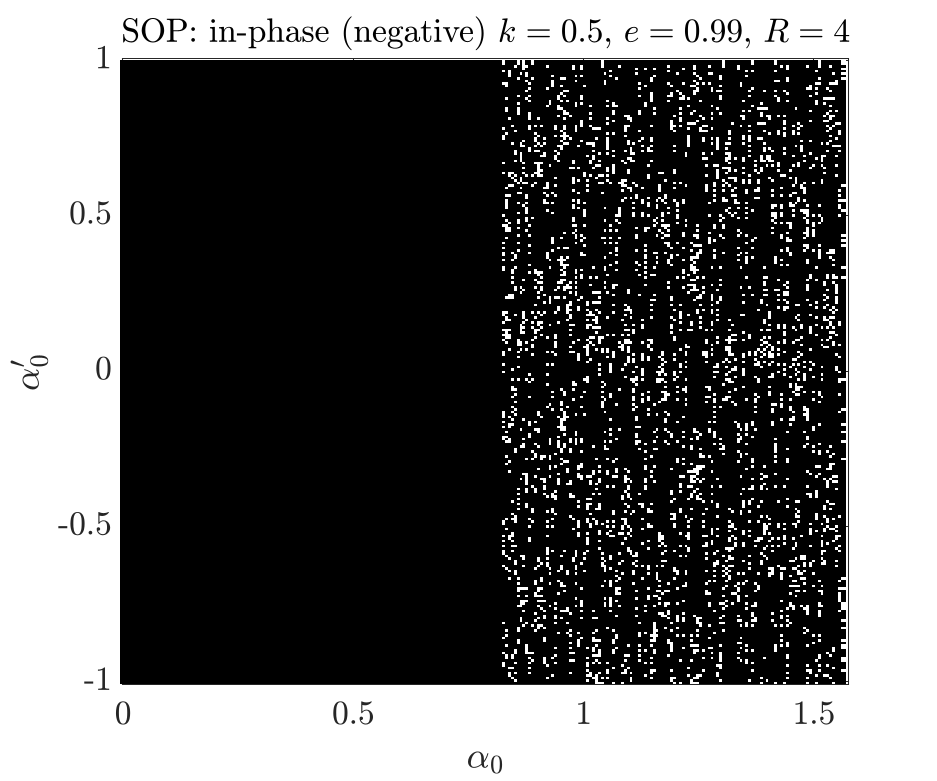}

\includegraphics[scale=0.31]{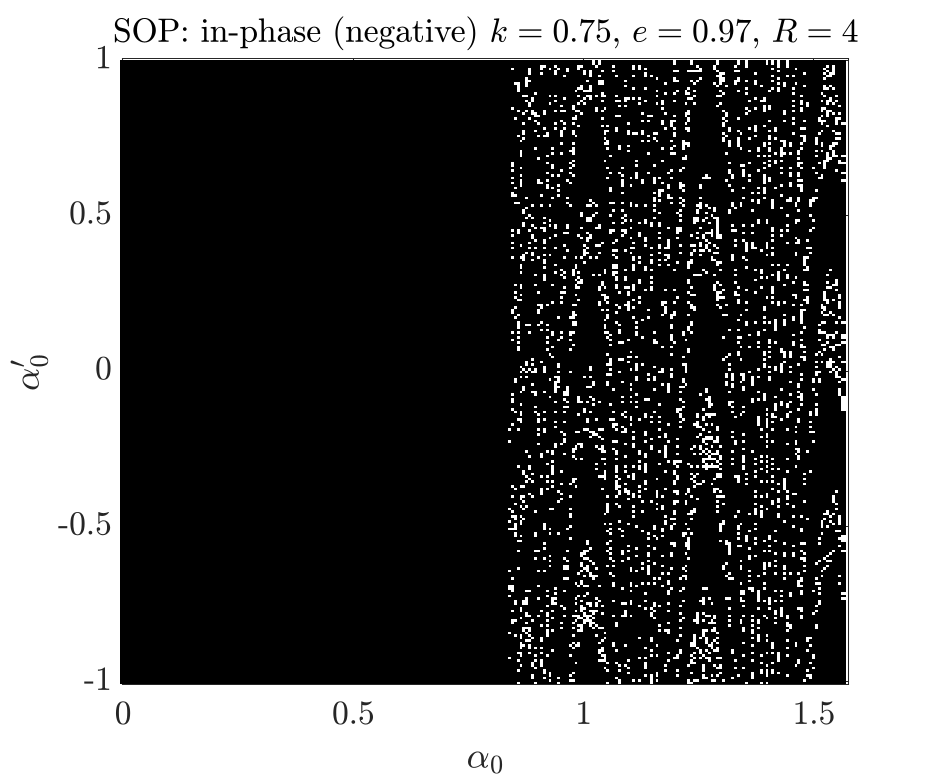}
	\includegraphics[scale=0.31]{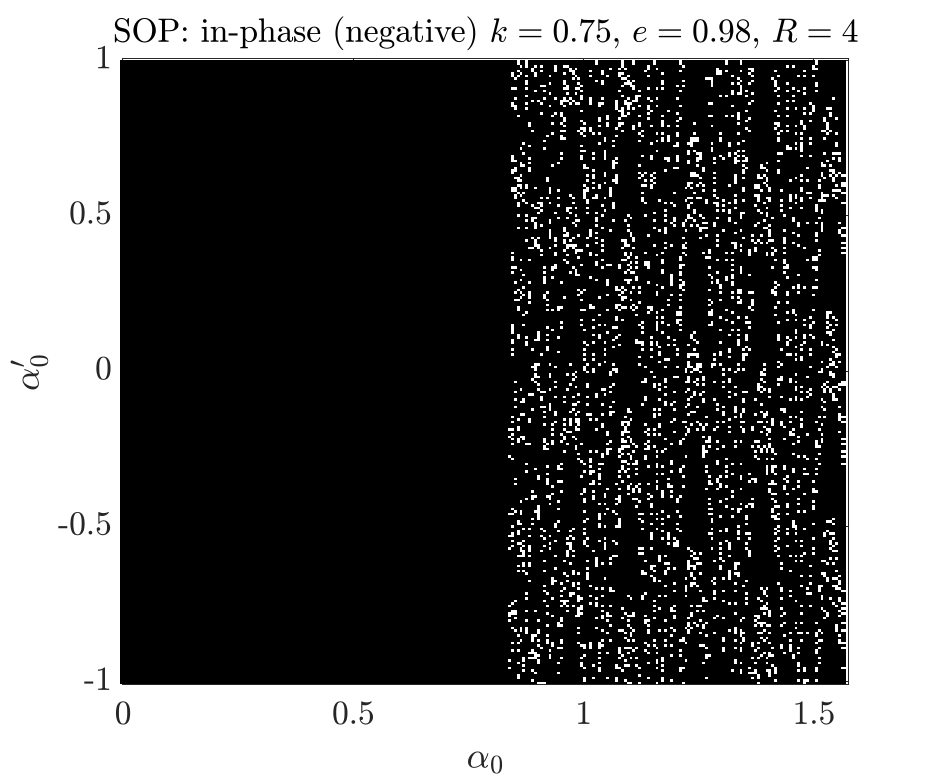}
	\includegraphics[scale=0.31]{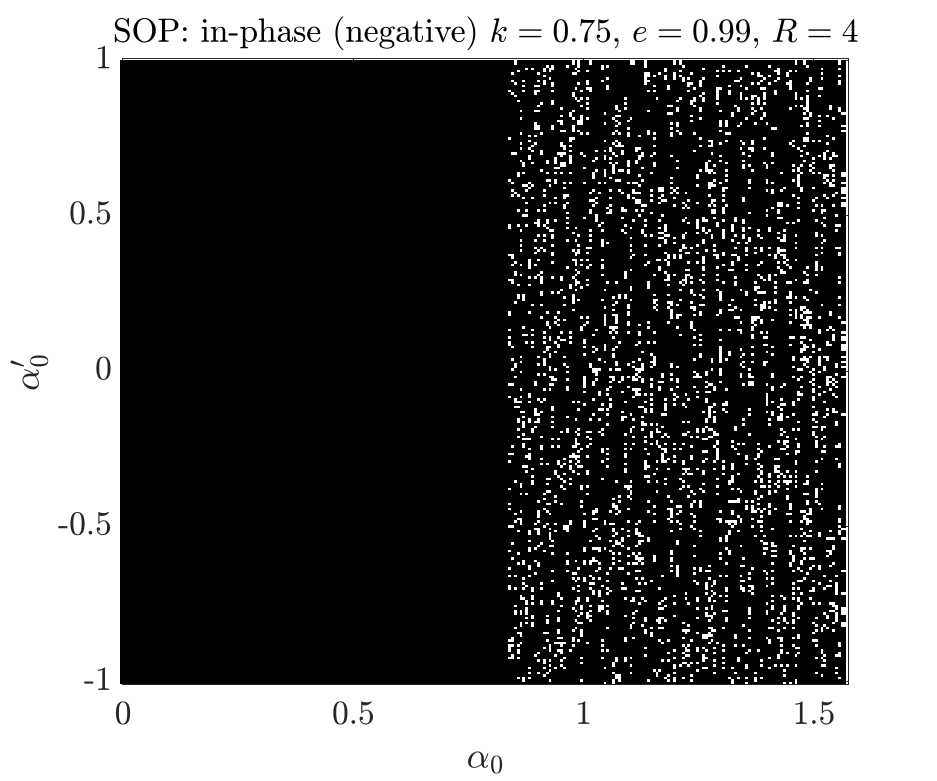}
	\caption{Set of initial conditions $(\alpha_0,\alpha_0')$ for the spin-orbit problem, represented in white color, for which $\alpha_r ' -\alpha_{r-1} < 0$ up to the first four periapsis passages, and for $\kappa = 0.25$ (top row), $\kappa = 0.5$ (central row) and $\kappa = 0.75$ (bottom row) , and for $e=0.97$ (left column), $e=0.98$ (central column) and $\kappa = 0.99$ (right column).}
	\label{fig:SOPinneg_R4}
\end{figure}

In Figure \ref{fig:SOPinneg_R3} and \ref{fig:SOPinneg_R4} we show the regions of the initial data satisfying the in-phase condition with $\alpha_r ' - \alpha_{r-1} ' < 0$ up to the first three (Fig. \ref{fig:SOPinneg_R3}) and four (Fig. \ref{fig:SOPinneg_R4}) periapsis passages. As in the DM, also in the SOP the regions of the initial data satisfying the in-phase condition appear more disrupted for increasing number of the periapsis passages.

All these simulations show that, even if there are differences in the motion prediction, the DM captures the phenomenological behaviour due to the impulsive nature of the gravity-gradient moment in the SOP for highly elliptical orbits. Hence, the introduction of the DM represents a first guess model that requires further development.

\subsection{Spin-orbit problem for highly elliptical orbits: application of the Fast Lyapunov Indicators}
\label{ssec:SOP_FLI}

Now we ask the following questions: is the in-phase condition (that, we recall, lead to an uncontrolled growth of the spacecraft angular velocity) source of chaos? Is the counterphase condition characterized by a small value of chaoticity? In order to obtain an answer we compute the chaos charts of the SOP for HEOs.

The chaos indicator method represents a powerful tool able to numerically compute the phase portrait. There exists several chaos indicators, such as the Finite Time Lyapunov Exponents (FTLE) \citep{tang1996} and the Mean Exponential Growth of Nearby Orbits (MEGNO) \citep{cincotta2000}. In this paper we use the Fast Lyapunov Indicators (FLI), that was originally introduced in \citet{froeschle1997_a,froeschle1997_b} in order to discriminate between regular and chaotic orbits of asteroids motions. In the last two decades, the method has been modified to compute solutions originated by partially hyperbolic equilibria, such as the center, stable and unstable manifolds tube of $L_1$ and $L_2$ \citep{guzzo2014,lega2016_a,guzzo2018}, the collisional manifolds of comet 67P/Churyumov Gerasimenko \citep{guzzo2015,guzzo2017}, as well as the detection of orbits of libration around the Sun--Earth $L_3$ points in a high-fidelity model \citep{scantamburlo2020}. In connection to the SOP, the investigation of chaotic motion is relevant since all the spin-orbit resonances are surrounded by chaotic separatrices \citep{chirikov1979,wisdom1984,wisdom1987}. In particular, \citet{wisdom1984} provides an estimation of the size of the chaotic separatrices for the $1$:$1$ spin-orbit resonances in terms of the eccentricity $e$, and inertia ratio $\kappa$. This estimation is linear in $e$ and exponential in $\sqrt{\kappa}$. In \citet{celletti2007} the investigation of the $1$:$1$, and $3$:$2$ spin-orbit resonances has been conducted via the maximum Lyapunov indicator \citep[the reader is referred to][for the definition of the maximum Lyapunov indicator]{wolf1985} for the Earth--Moon and Sun--Mercury systems.

Let us consider the system of ordinary differential equation
\begin{equation}
    \dot{\bm{\xi}} = F (\bm{\xi}), \quad \bm{\xi} \in \Omega \subset \mathbb{R}^n ,
    \label{eq:generic_Iorderode}
\end{equation}
where $\Omega$ is an open set, and $F:\Omega \rightarrow \mathbb{R}^n$ is the vector field. The variational equations associated to system \eqref{eq:generic_Iorderode} are
\begin{equation}
    \dot{\bm{\Xi}} = \left[ \frac{\partial F}{\partial \bm{\xi}} (\bm{\xi}(t)) \right] \bm{\Xi}, \quad \bm{\Xi} \in \mathbb{R}^{n} .
\end{equation}
Given the initial condition $\bm{\xi}_0 = \bm{\xi}(0) \in \Omega$, the initial tangent vector $\bm{\Xi}_0 = \bm{\Xi}(0) \in \mathbb{R}^n \backslash \{ \bm{0} \}$, and $\bm{\Xi}(t)$ the solution of the variational equation with $\bm{\xi}_0 $ and $\bm{\Xi}_0 $ as initial conditions, the characteristic Lyapunov exponent (CLE) is defined as
\begin{equation}
\text{CLE}(\bm{\xi}_0,\bm{\Xi}_0) \triangleq \lim _{t \rightarrow \infty} \frac{1}{t} \ln \left( \frac{\|\bm{\Xi} (t)\|}{\|\bm{\Xi}_0\|} \right).
\end{equation}

We emphasize that the CLE provide an estimation of the separation of trajectories with initial conditions close to $\bm{\xi}_0$ \citep[see, for example,][]{lyapunov_problem,cesari_book,benettin1980_a,benettin1980_b}. 

All the chaos indicators aim to compute the CLE in a short computational time. Specifically, the FLI at time $T$ of an initial condition $\bm{\xi}_0$ and initial tangent vector $\bm{\Xi}_0$ is defined as
\begin{equation}
    \text{FLI}(\bm{\xi}_0,\bm{\Xi}_0;[0,T]) \triangleq \max_{ t \in [0,T]} \ln \left( \frac{\|\bm{\Xi}(t)\|}{\|\bm{\Xi}_0\|} \right).
\end{equation}

Hence, given the state $\bm{\xi} = (f,\alpha,\omega_3)$ and the tangent vector $\bm{\Xi} = (\Xi_f,\Xi_{\alpha},\Xi_{\omega_3})$, the variational equations of the SOP are
\begin{equation}
    \begin{split}
        f' &= \frac{(1+e \cos f)^2}{(1-e^2)^{3/2}} \cr
        \alpha ' & = \omega_3 \cr
        \omega_3 ' & = \overline{A}(f) \sin (2(f-\alpha)) \cr 
        \Xi_f ' & = -2 e \sin f \frac{1+e \cos f}{(1-e^2)^{(3/2)}} \Xi_f \cr
        \Xi_{\alpha} ' & = \Xi_{\omega_3} \cr
        \Xi_{\omega_3} ' & = \frac{d}{df}\left( \overline{A} \sin (2f-2\alpha) \right) \Xi_f - 3 \kappa \frac{(1+e \cos f)^3}{(1-e^2)^3} \cos (2(f-\alpha)) \Xi_{\alpha}
    \end{split},
\end{equation}
where
\begin{equation}
    \frac{d}{df}\left( \overline{A} \sin (2f-2\alpha) \right) = 3 \kappa \frac{(1+e \cos f)^2}{(1-e^2)^3} \left[ (1+e\cos f) \cos(2(f-\alpha)) - \frac{3}{2} e \sin f \sin (2(f-\alpha))\right].
\end{equation}

\begin{figure}[h!]
    \centering
    \includegraphics[scale=0.56]{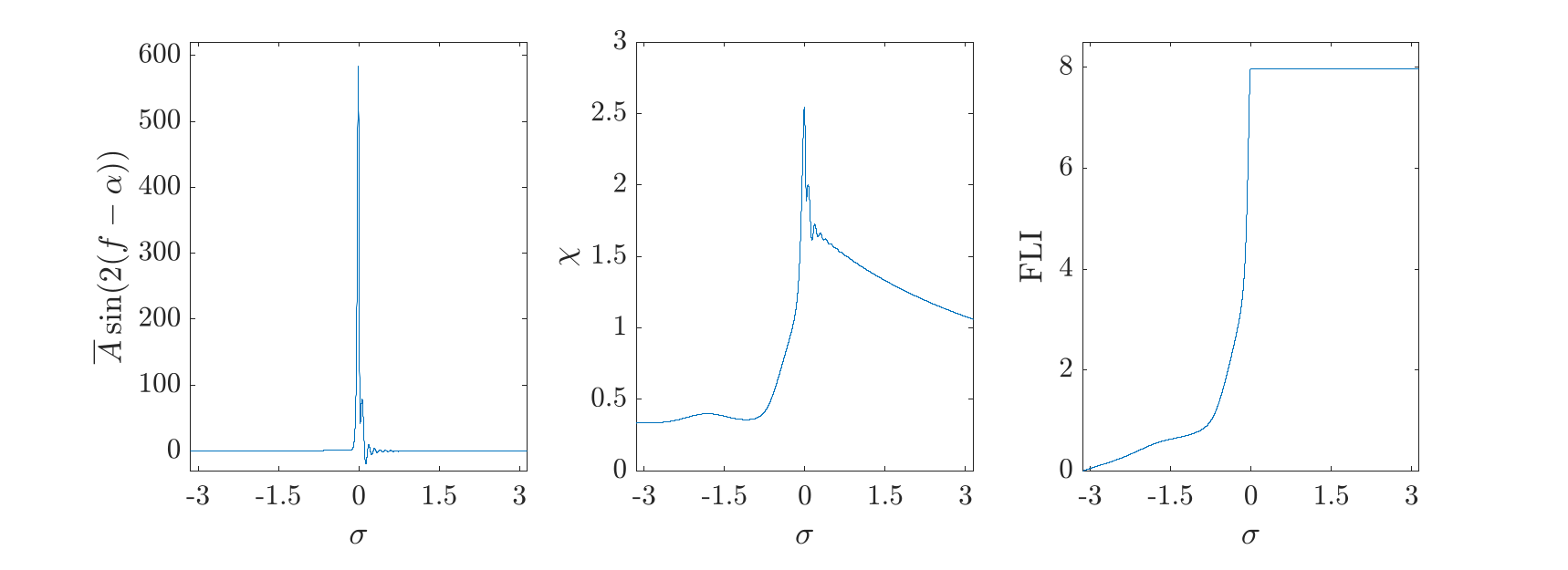}

    \includegraphics[scale=0.56]{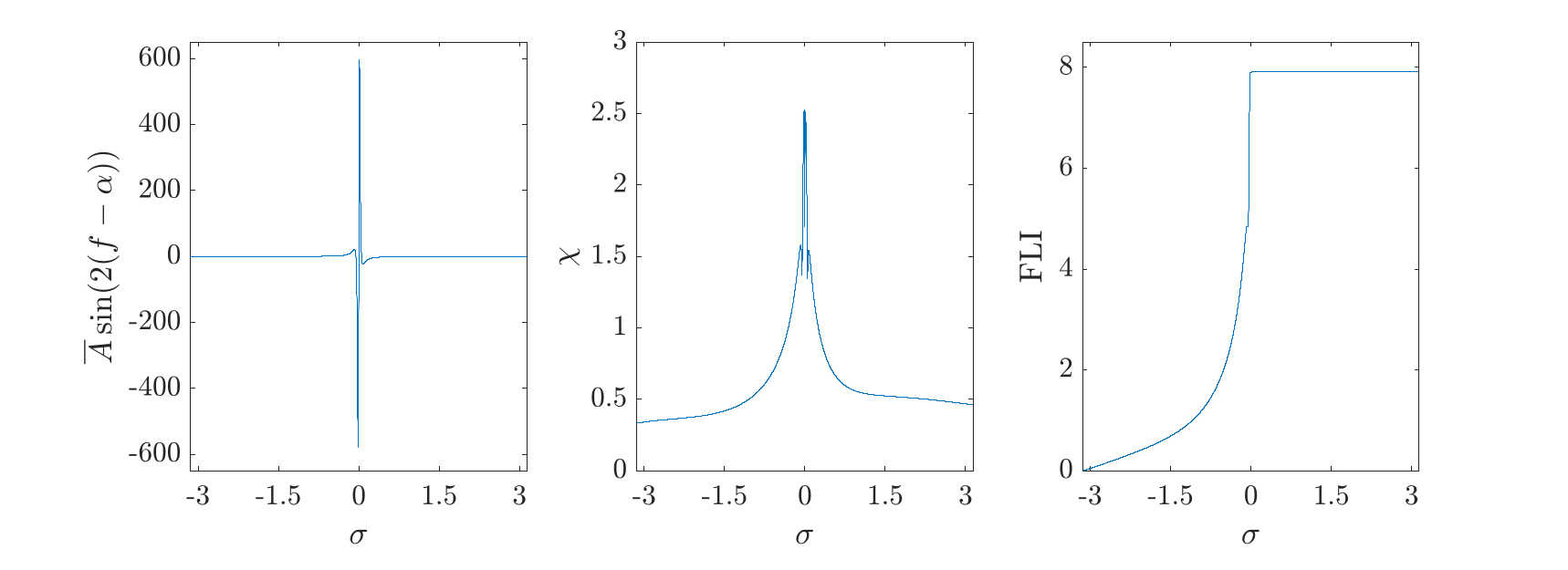}

    \includegraphics[scale=0.56]{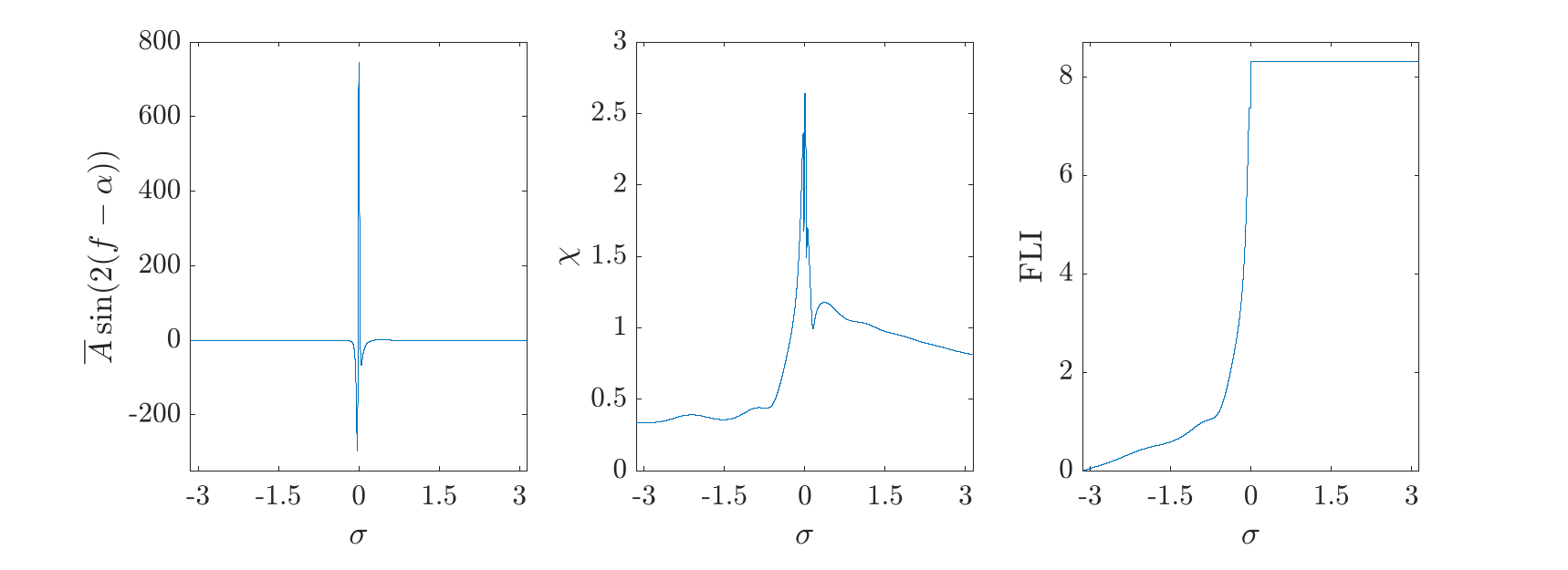}
    \caption{$\chi$ and FLI evolution for the same orbits plotted in Fig. \ref{fig:spinorbit_solution_example}. }
    \label{fig:spinorbit_FLI_example}
\end{figure}

In Fig. \ref{fig:spinorbit_FLI_example} we plot the evolution of
\begin{equation}
    \chi(\sigma) = \frac{1}{\sigma-\sigma_0}\ln \left( \frac{\| \bm{\Xi} (\sigma)\|}{\|\bm{\Xi}_0\|} \right)
\end{equation}
(central panels of each row), and the evolution of the FLI (right panels of each row) for the trajectories depicted in Fig. \ref{fig:spinorbit_solution_example}, where $\sigma_0 = -\pi$ is the initial epoch and we choose $(1/\sqrt{3},1/\sqrt{3},1/\sqrt{3})$ as initial tangent vector (for convenience, we plot also the gravity-gradient moment associated to the trajectories in the left panels of each row). We notice that for all of the three trajectories, $\chi$ and the FLI grow up to the periapsis passage, and then their value stabilizes. This is compatible with the classical behaviour of the FLI evolution \citep[for further details about the effects of close encounters on the FLI, the reader is referred to][]{guzzo2017,guzzo2023}, since at the periapsis passage the trajectory is close to a singularity and so there is a loss of precision in the numerical integration. We remark that this behaviour is not completely equal to the FLI evolution for close encounters analyzed in \citet{guzzo2017}, since for gravity-gradient moment a change of sign may appear, while not for the orbital dynamics.
We notice that the evolution of $\chi$ is affected by short and successive peaks close to the periapsis passage, and that these successive peaks may grow if the gravity-gradient moment can be approximated by two or more Dirac pulses with opposite sign. These behaviour is reported also in the FLI evolution (during the periapsis passage there are very short time-span in which the FLI seems to stabilize).
We stress that the highest FLI value between these three cases is reached by the orbit whose gravity-gradient moment is characterized by two pulses with opposite sign and not equal amplitude.

\begin{figure}[h!]
    \centering
    \includegraphics[scale=0.6]{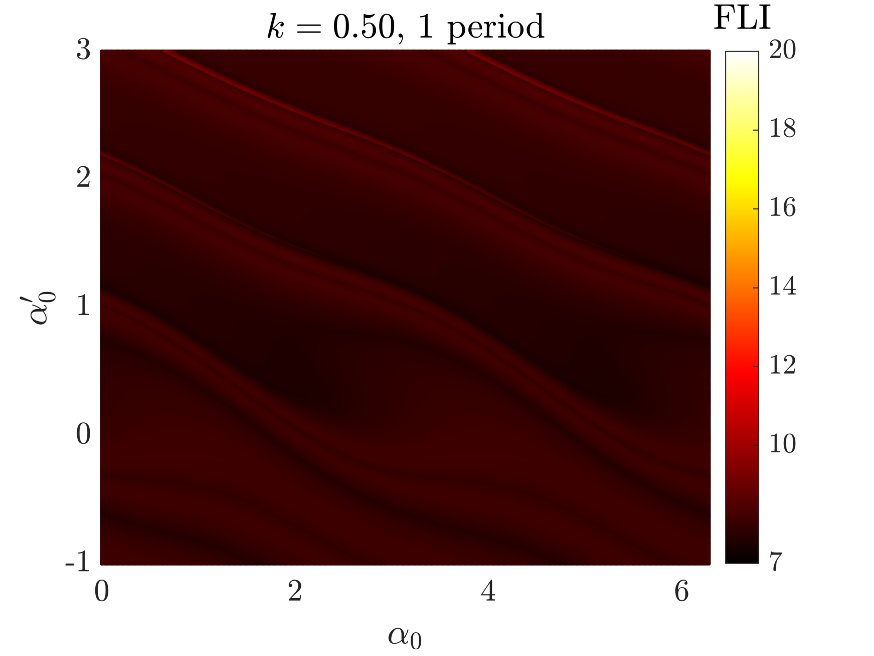}
    
    \includegraphics[scale=0.6]{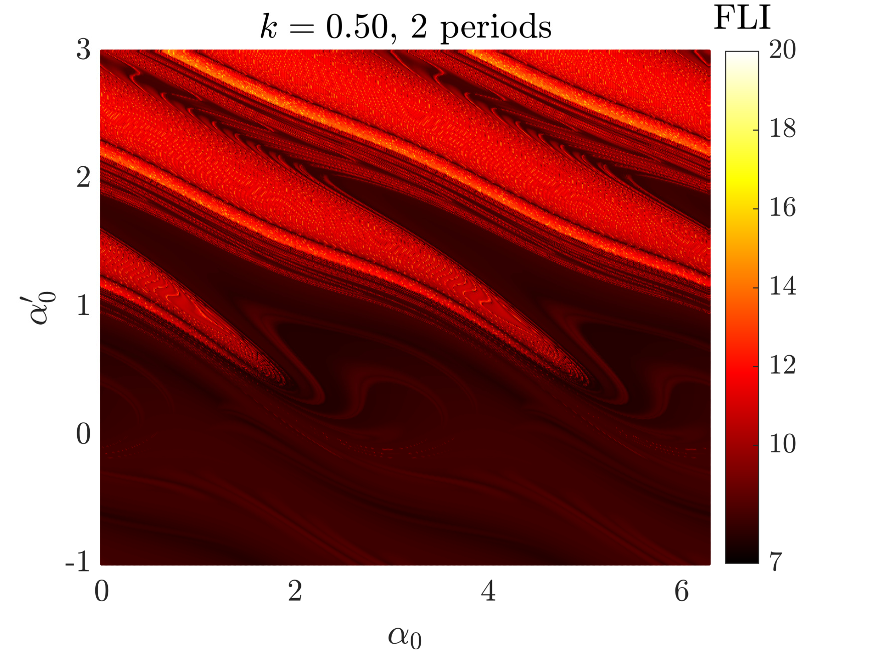}

    \includegraphics[scale=0.6]{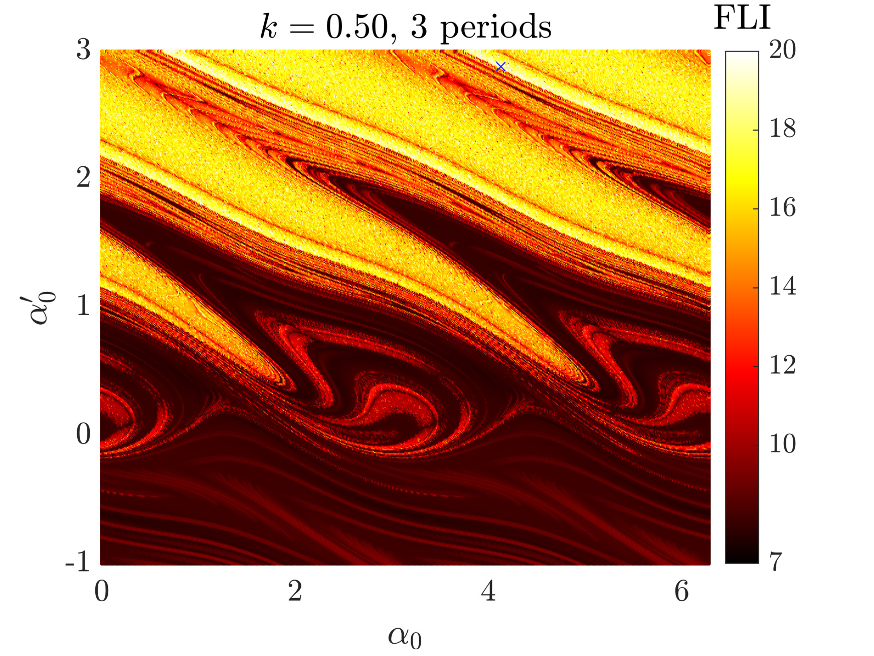}
    \caption{FLI charts for one orbit (top panel), two orbits (central panel), and three orbits (bottom panel) for the SOP with $e = 0.9$ and $\kappa = 0.5$. For the computation of the FLI charts we consider a grid of $500\times 500$ initial conditions in the interval $\alpha_0 \in [0,2\pi]$ and $\alpha_0 ' \in [-1,3]$. For all the three FLI charts, we restrict the colorbar in the range $[7,20]$. }
    \label{fig:SOP_FLImaps_ecc_09_kappa_05}
\end{figure}

In Fig. \ref{fig:SOP_FLImaps_ecc_09_kappa_05} we plot the values of the FLI for the SOP with $e = 0.9$ and $\kappa=0.5$ after one (top panel), two (central panel), and three (bottom panel) orbital periods. As initial conditions, we consider $f_0 = -\pi$ as initial true anomaly, and a grid of evenly space points for $\alpha_0 \in [0,2\pi], \alpha_0 ' \in [-1,3]$. For all the three FLI charts, we cut-off the FLI values in the range $[7,20]$ (hence, for a FLI value that is smaller than $7$, the initial datum is colored in black; if the FLI value is greater than $20$, the initial datum is colored in white). We notice that after the second periapsis passage, the highest values of the FLI (yellow and white initial conditions) are distributed in regions where $\alpha_0 '$ is positive.

We stress that the FLI maps of Fig. \ref{fig:SOP_FLImaps_ecc_09_kappa_05} are not similar to the in-phase charts of Fig. \ref{fig:SOP_Intneg_k_05}. This is because the regions of initial conditions fulfilling the in-phase condition are not necessary chaotic.

\begin{figure}[h!]
    \centering
    \includegraphics[scale=0.56]{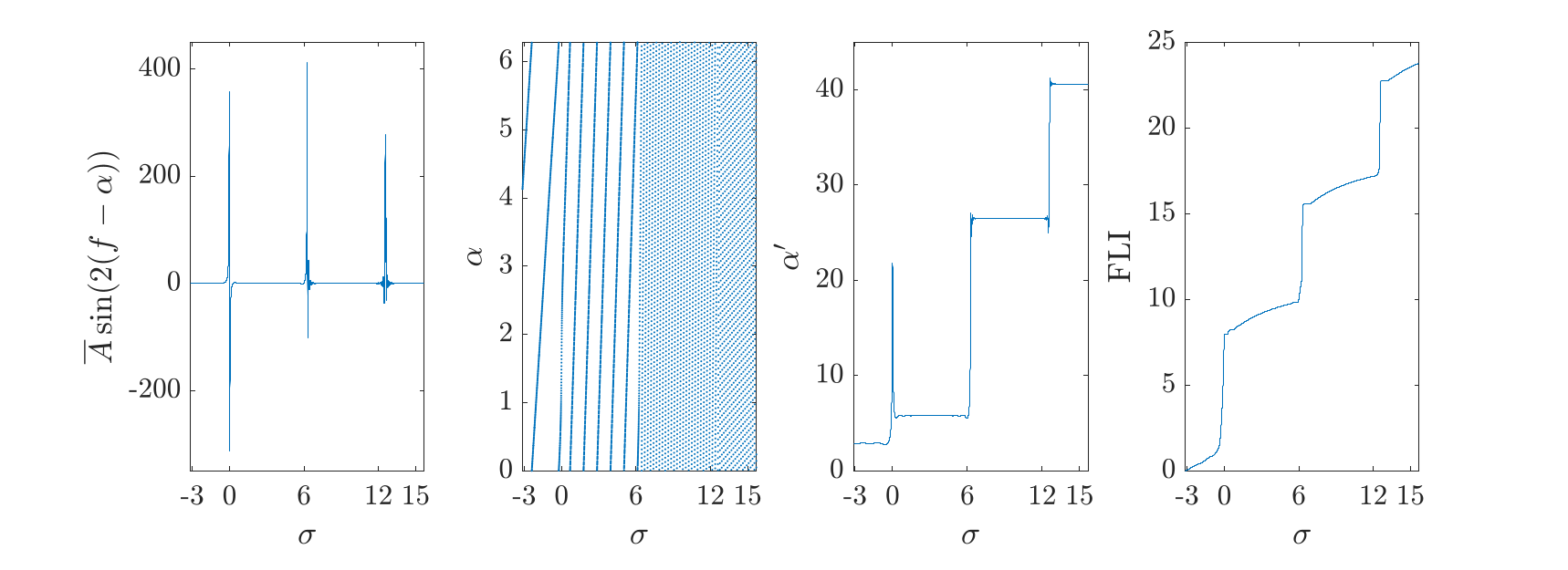}
    
    \caption{Orbit associated to the initial datum characterized by the maximum FLI after three orbital periods for $\kappa = 0.5$ (blue point in the bottom panel of Fig. \ref{fig:SOP_FLImaps_ecc_09_kappa_05}).}
    \label{fig:SOP_orbFLImax_ecc_09_kappa_05}
\end{figure}

In the bottom panel of Fig. \ref{fig:SOP_FLImaps_ecc_09_kappa_05} we plot with a blue cross point the initial datum characterized by the maximum FLI after $3$ orbital periods. We plot the orbit evolution of such a initial datum in Fig. \ref{fig:SOP_orbFLImax_ecc_09_kappa_05}. We notice that the evolution of the FLI for this orbit is different from those depicted in Fig. \ref{fig:spinorbit_FLI_example}; as a matter of fact the FLI increases not only when the satellite is approaching to the periapsis passage. This implies that the source of chaos for this orbit is not only due to the periapsis passage (hence, to the numerical errors in the vicinity to the singularity), but also to the evolution of the dynamical system. This suggests that such a trajectory is originated by a orbit with an hyperbolic nature (for example, a resonance). 

\begin{figure}
\centering
\includegraphics[scale=0.5]{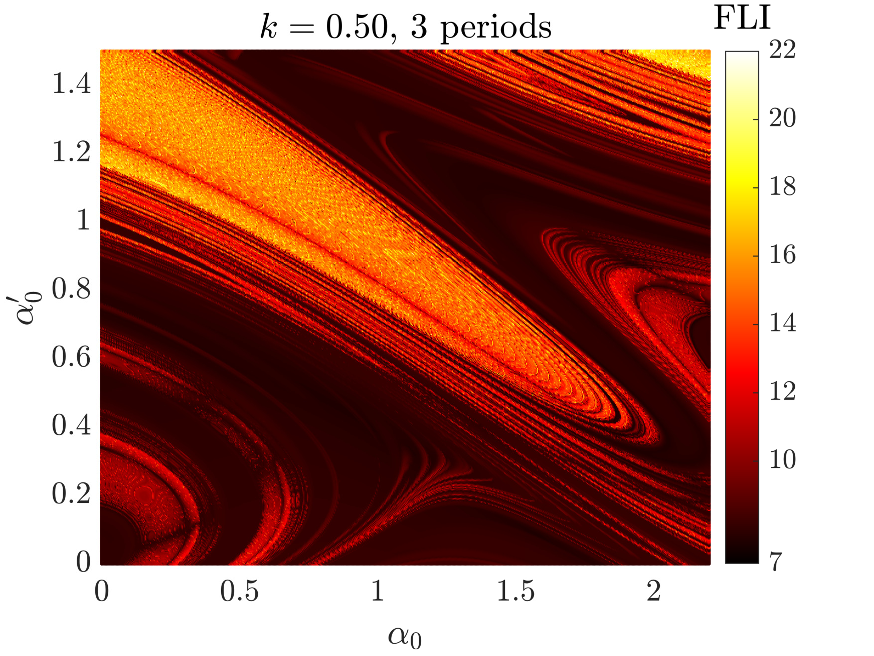}
\includegraphics[scale=0.5]{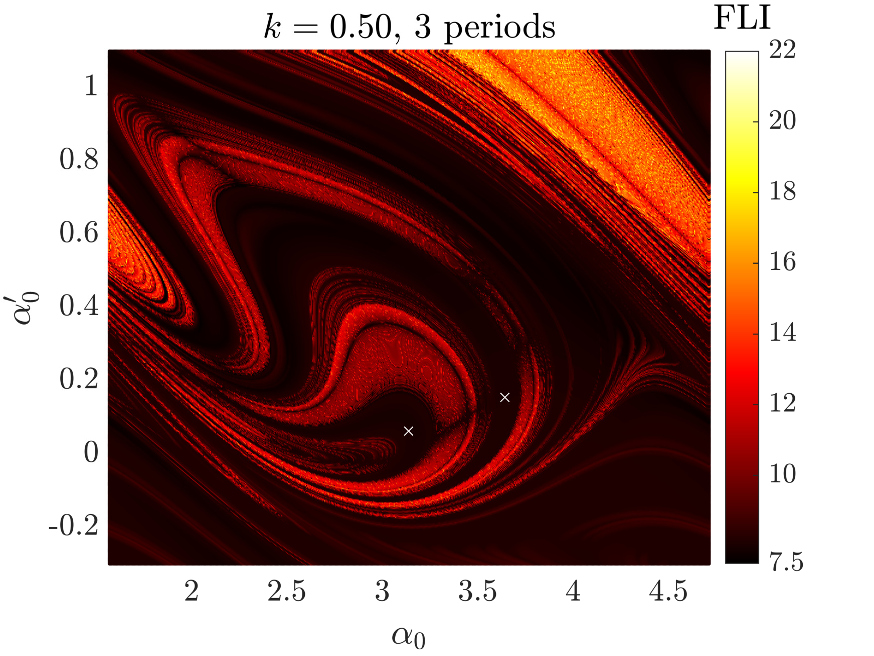}
\caption{Zoom-in of the FLI map for the SOP with $e=0.9$ and $\kappa = 0.5$.}
\label{fig:SOP_FLI_ecc_09_zooms}
\end{figure}

In Fig. \ref{fig:SOP_FLI_ecc_09_zooms} we plot two zoomed in charts of the FLI map for the SOP with $e=0.9$ and $\kappa = 0.5$ after three periapsis passages around two regions. The left panel of Fig. \ref{fig:SOP_FLI_ecc_09_zooms} shows the creation of lobes that are accumulating and bending over, creating the zone with the highest FLI values. In the right panel of Fig. \ref{fig:SOP_FLI_ecc_09_zooms} the highest values of the FLI are distributed in curves that curl in, and are sharply surrounded by regions with small values of the FLI.

\begin{figure}[h!]
\centering
\includegraphics[scale=0.56]{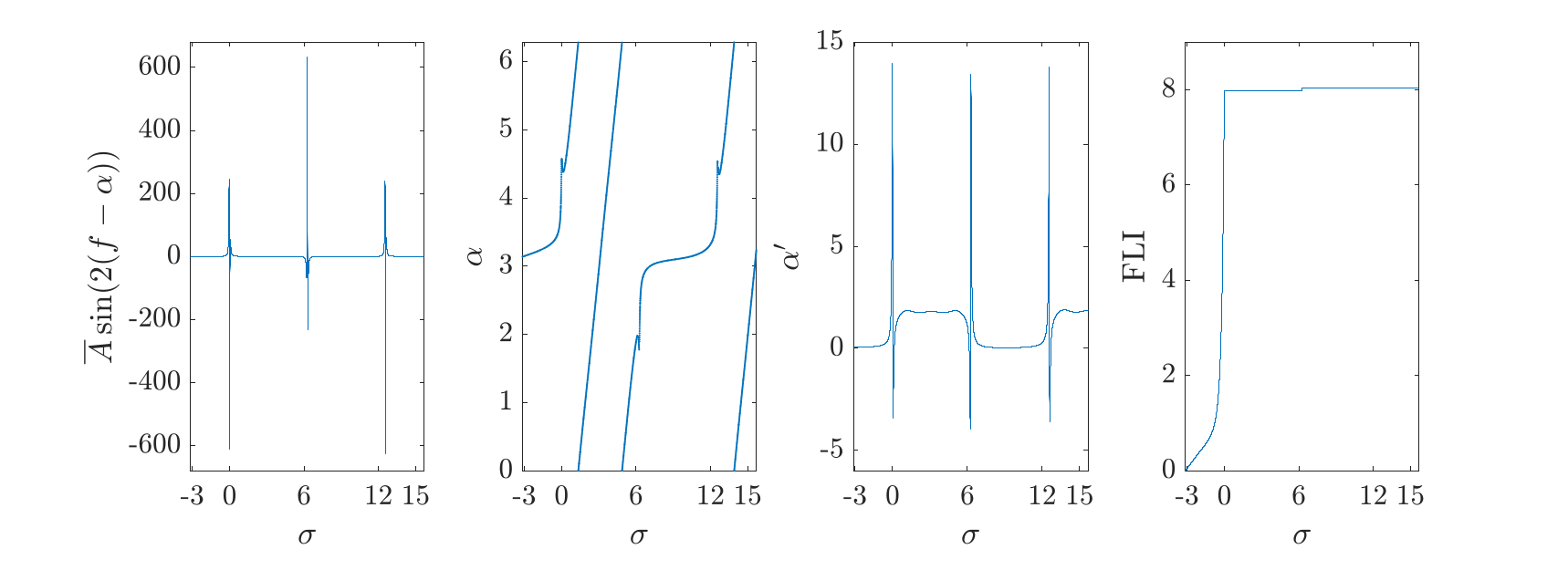}

\includegraphics[scale=0.56]{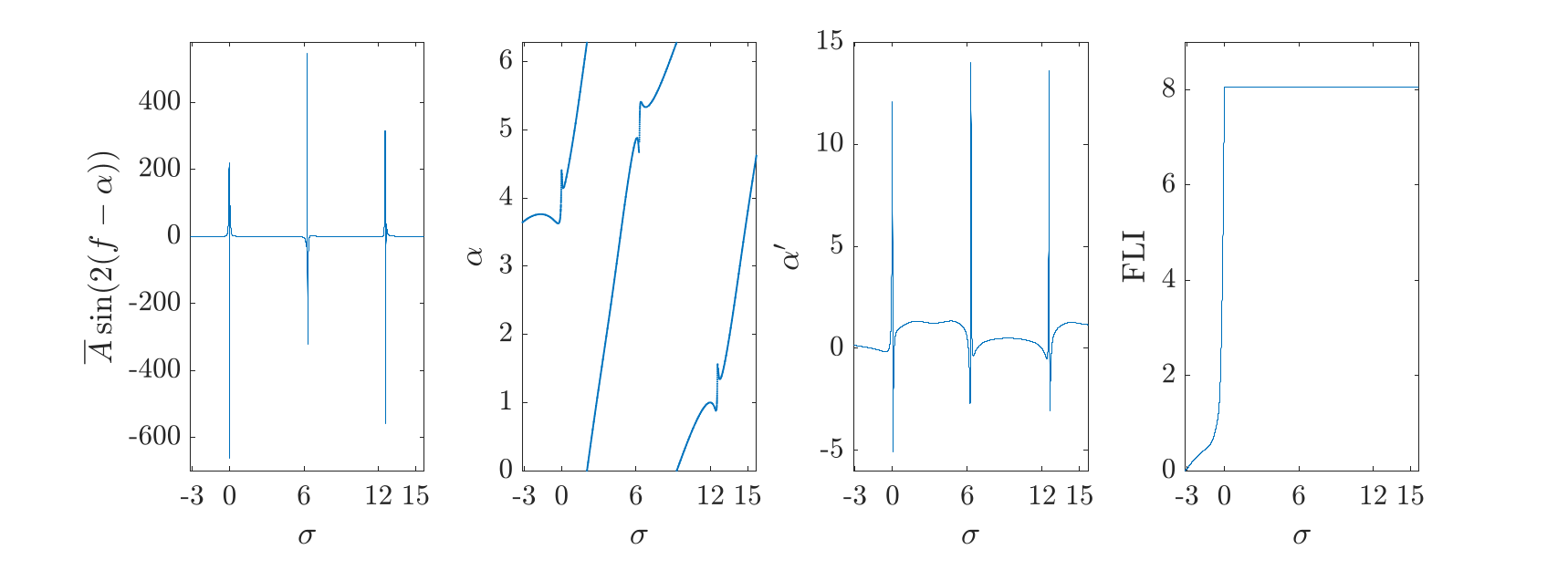}
\caption{Evolution of the initial conditions represented by a white cross point in the right panel of Fig. \ref{fig:SOP_FLI_ecc_09_zooms}.}
\label{fig:SOP_smallFLI_whitecolor_ecc_09_kappa_05}
\end{figure}

Always in the right panel of Fig. \ref{fig:SOP_FLI_ecc_09_zooms}, we select two initial conditions (depicted by a white cross point) with a small value of FLI; we plot their evolution in Fig. \ref{fig:SOP_smallFLI_whitecolor_ecc_09_kappa_05}. It is interesting to notice that the orbits characterized by a small value of the FLI are those with the evolution of the angular velocity that is compatible with the counterphase condition, i.e., the behaviour of the angular velocity can be described by a gravity-gradient evolution fulfilling the counterphase condition. Moreover, we stress that for both the two orbits, the FLI increases during the first periapsis passage, but then it remains almost constant. These investigations suggest that if periodic orbits fullfilling the counterphase condition exists, then their initial condition have to be searched in the regions characterized by the smallest values of the FLI. 

\begin{figure}[htbp!]
\centering
\includegraphics[scale=0.5]{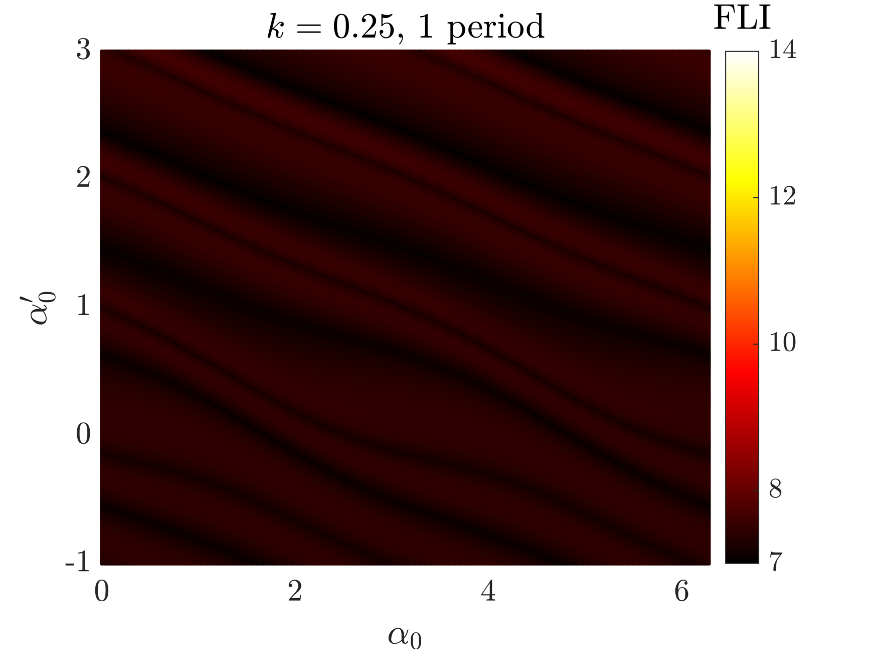}
\includegraphics[scale=0.5]{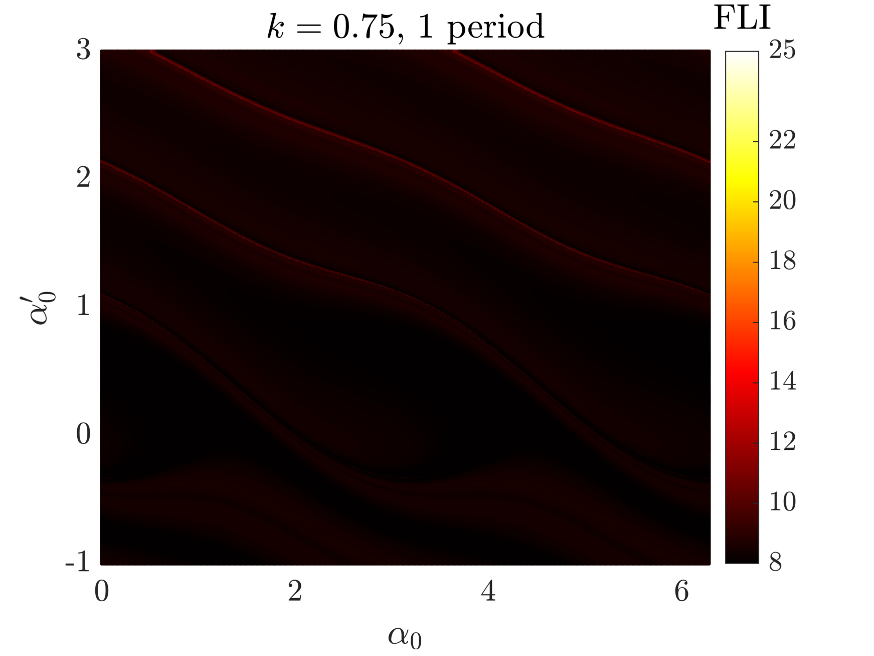}

\includegraphics[scale=0.5]{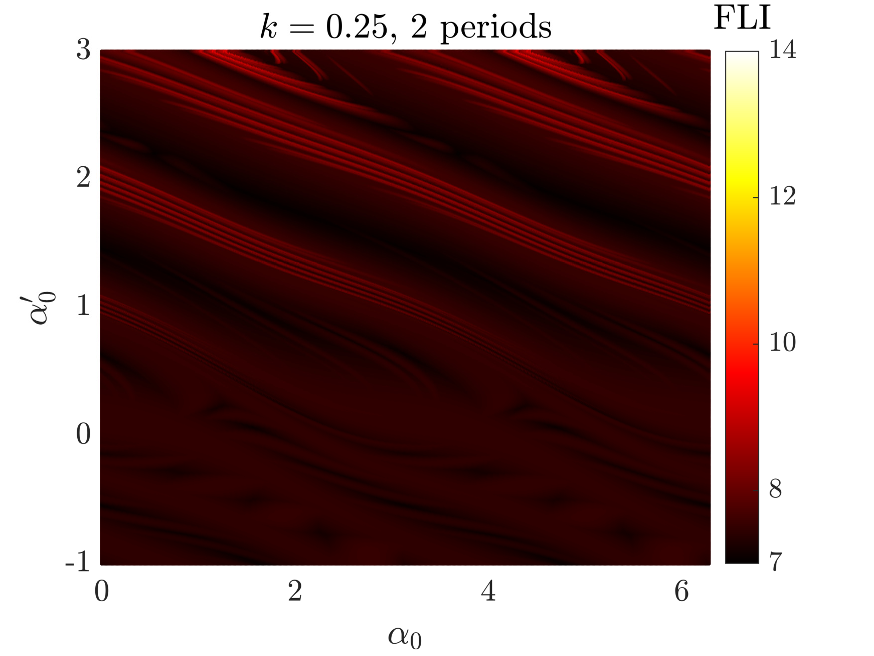}
\includegraphics[scale=0.5]{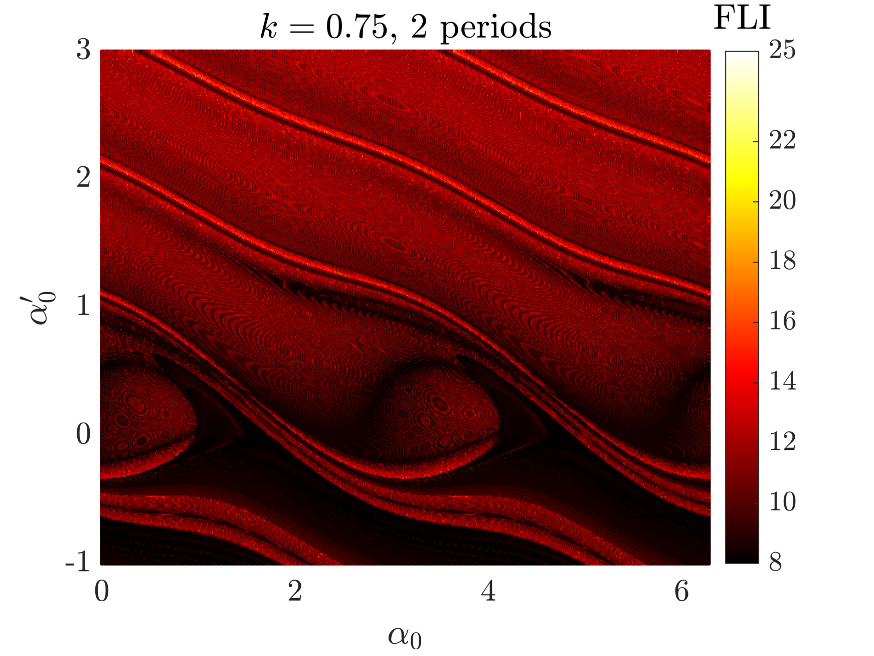}

\includegraphics[scale=0.5]{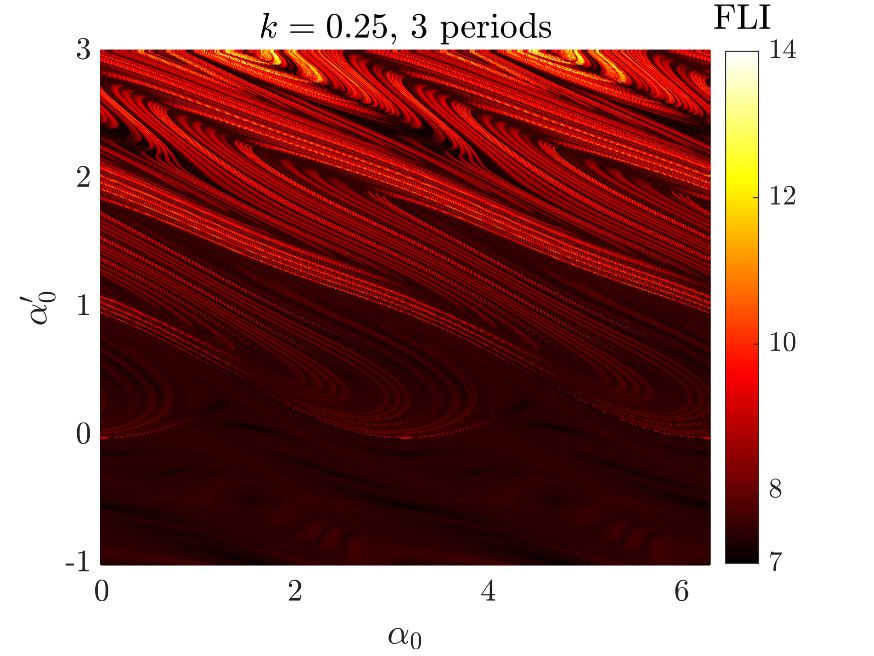}
\includegraphics[scale=0.5]{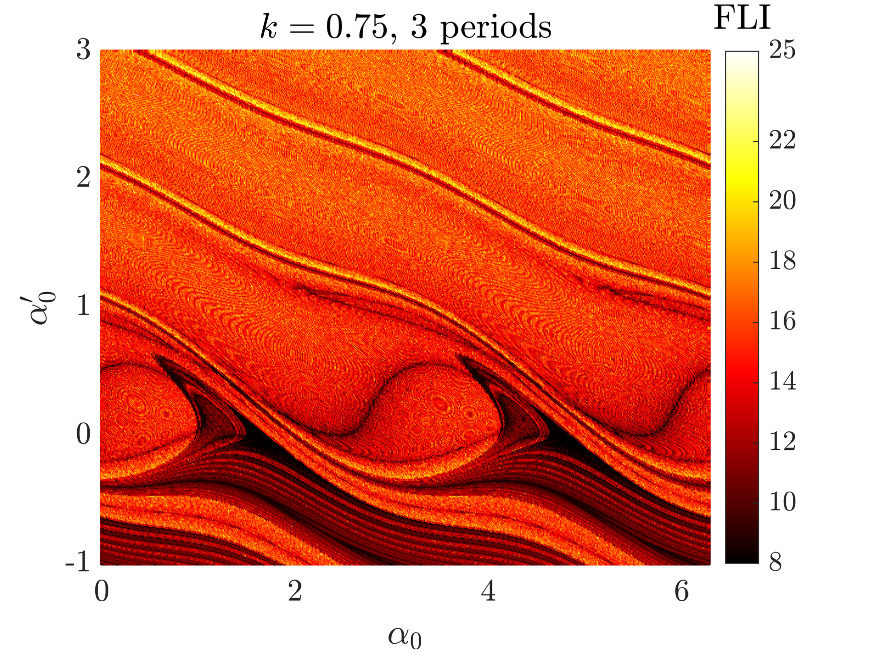}

\caption{FLI maps of the SOP for $e = 0.9$ and after one (top row), two (central row), and three (bottom row) orbital periods. For the computation of the FLI charts we consider a grid of $500\times 500$ initial conditions in the interval $\alpha_0 \in [0,2\pi]$ and $\alpha_0 ' \in [-1,3]$. For all the three FLI maps related to the inertia ratio $\kappa = 0.25$ (left column), we restrict the colorbar in the range $[7,14]$. For all the three FLI maps related to $\kappa = 0.75$ (right column), we restrict the colorbar in the range $[8,25]$.}
\label{fig:FLI_ecc_09_kappa_025_075}
\end{figure}

In Fig. \ref{fig:FLI_ecc_09_kappa_025_075} we plot the FLI maps for the SOP with $e = 0.9$ and $\kappa = 0.25$ (left column) and $\kappa = 0.75$ (right column) at one (top row), two (central row), and three (bottom row) orbital periods. As for $\kappa = 0.5$, the higher FLI values are located in the regions with $\alpha_0 ' > 0$. In particular, the highest values of the FLI are distributed in stripes that are diagonal in the $\alpha_0-\alpha_0'$ plane. Moreover, the regions characterized by the highest values of the FLI are sharply detected for $\kappa = 0.75$, while the distinction between the highest and smallest FLI values regions is more difficult for $\kappa = 0.25$. This is due to the fact that for smaller values of $\kappa$, the amplitudes of the gravity-gradient moment reduce (until it vanishes for $\kappa \rightarrow 0$), but the impulsive behaviour is still present. 

\section{Conclusion}

In this paper we investigate the spin-orbit coupling effects in highly elliptical orbits. This analysis is of interest in connection to attitude motion analysis for space mission, since it enhances our understanding of the natural rotational dynamics (i.e., without the action of external controls) of an artificial satellite in highly elliptical orbits, such as the Near-Rectilinear Halo Orbits. 
We start our investigation by analysing the gravity-gradient moment for elliptical orbits with high eccentricity; we then prove that gravity-gradient moment plays the role of an instantaneous excitation acting at the periapsis passage that may induce the satellite to rotate faster and faster. 
Moreover, at each periapsis passage the gravity-gradient moment can be approximated by a Dirac pulses. Thanks to this approximation, we are able to derive a recursive discrete map describing the spin-angle and velocity at each periapsis passage. Even if there are differences between the recursive map and the spin-orbit problem, the map we propose is a first guess model capturing phenomenological behaviours of the spin-orbit problem for highly elliptical orbits, and for this reason requires further and future developments. The examination of this discrete map is relevant as a preliminary study of the spin-orbit problem for highly elliptical orbits; for example, with the recursive discrete map we can provide a rigorous definition of the attitude initial configurations that can induce an increasingly fast angular velocity of the satellite.
As a justification of the discrete map introduction, we then compare the results of the recursive map with those found in the spin-orbit problem.
Finally, the computation of the FLI charts reveals that there does not exist a correlation between the in-phase initial conditions and chaoticity.

\paragraph{Acknowledgments}
E. Scantamburlo and M. Romano acknowledge the project ``Advanced Space System Engineering to Address Broad Societal Benefits – Starting Grant" funded by a contract between Politecnico di Torino and Compagnia di San Paolo (CSP) 2019/2021 within the call “Attrazione e retention di docenti di qualità” (P.I. Prof. Marcello Romano).

\bibliographystyle{plainnat}
\bibliography{listbib.bib}

\begin{appendices}
\section{Fixed points of the discrete map}\label{app:fixedpoints}

We stress that the fixed points $(\alpha^{\ast},\alpha^{\prime \ast})$ of the recursive map \eqref{eq:recursivemapnew} are 
\begin{equation}
(\alpha^{\ast},\alpha^{\prime \ast} ) = \left(\ell_{\alpha} \frac{\pi}{2}, \ell_{\alpha^{\prime}} \right) , \qquad \ell_{\alpha},\ell_{\alpha^{\prime}} \in \mathbb{Z}.
\end{equation}
For $r \geq 2$, the Jacobian associated to the map calculated at a fixed point is 
\begin{equation}
JF_{r\geq 2} = \begin{bmatrix}
1 & 2 \pi \\
-2K \cos (2 \alpha ^{\ast} + \alpha^{\prime \ast} 4 \pi) & 1 - 4 K \pi \cos (2 \alpha ^{\ast} + \alpha^{\prime \ast} 4 \pi)
\end{bmatrix} ,
\end{equation}
whose eigenvalues are
\begin{equation}
\lambda_{1,2} = 1-2K \pi \cos (2 \alpha ^{\ast} + \alpha^{\prime \ast} 4 \pi) \mp  2 \sqrt{K \pi \cos (2 \alpha ^{\ast} + \alpha^{\prime \ast} 4 \pi) (K \pi \cos (2 \alpha ^{\ast} + \alpha^{\prime \ast} 4 \pi) -1)}.
\end{equation}

For $(\alpha^{\ast},\alpha^{\prime \ast}) = (2 \ell \pi/2,\ell_{\alpha'})$, with $\ell \in \mathbb{Z}$,  the eigenvalues are given by
\begin{equation}
\lambda_{1,2} = 1 - 2K \pi \mp 2 \sqrt{K \pi (K \pi -1)}.
\end{equation}
Hence $\lambda_{1,2} \in \mathbb{R}$ if $K > 1/\pi$ (we recall that, by definition, $K$ is positive). Furthermore, for $K > 1/\pi$, we notice that the fixed point $(\alpha^{\ast},\alpha^{\prime \ast}) = (2 \ell \pi/2,\ell_{\alpha'})$ is unstable since $|\lambda_1| > 1$, while for  $0 < K < 1/\pi$, we have $|\lambda_1|=|\lambda_2| = 1$.

For $(\alpha^{\ast},\alpha^{\prime \ast}) = ((2 \ell +1)\pi/2 ,\ell_{\alpha'})$ with $\ell \in \mathbb{Z}$,  the eigenvalues are given by
\begin{equation}
\lambda_{1,2} = 1 + 2K \pi \mp 2 \sqrt{K \pi (K \pi +1)}.
\end{equation}
Then $\lambda_{1,2} \in \mathbb{R}$ for $K> 0$. In particular, $|\lambda_1| < 1$, and $|\lambda_2| > 1$ for $ K > 0$. Hence, the fixed points are unstable.

\end{appendices}

\end{document}